\documentclass{ws-ijmpd}
\usepackage{cite}
\usepackage{hyperref}
\usepackage{color}

\def\be{\begin{equation}}
\def\ee{\end{equation}}
\newcommand{\bea}{\begin{eqnarray}}
\newcommand{\eea}{\end{eqnarray}}

\begin{document}

\markboth{McDonough, Hill, Ivanov, La Posta, and Toomey}{Observational Constraints on Early Dark Energy}

\catchline{}{}{}{}{}
%
%%%%%%%%%%%%%%%%%%%%%%%%%%%%%%%%%%%%%%%%%%%%%%%%%%%%%%%%%%%%%%%%%%%%

% \begin{flushright}
% MIT-CTP/5618
% \end{flushright}

\title{\bf OBSERVATIONAL CONSTRAINTS ON EARLY DARK ENERGY}
%INSTRUCTIONS FOR TYPESETTING
%MANUSCRIPTS\footnote{For the title, try not to use more than 3 lines.
%Typeset the title in 10~pt Times roman, uppercase and boldface.} 

\author{EVAN MCDONOUGH}
\address{Department of Physics,
University of Winnipeg, Winnipeg MB, R3B 2E9, Canada}

\author{J.~COLIN HILL}
\address{Department of Physics, Columbia University, New York, NY, USA 10027}

\author{MIKHAIL M.~IVANOV}
\address{Center for Theoretical Physics, Massachusetts Institute of Technology,\\ Cambridge, MA 02139, USA}

\author{ADRIEN LA POSTA}
\address{Universit\'{e} Paris-Saclay, CNRS/IN2P3, IJCLab, 91405 Orsay, France}

\author{MICHAEL W.~TOOMEY}
\address{Center for Theoretical Physics, Massachusetts Institute of Technology,\\ Cambridge, MA 02139, USA}

% \author{FIRST AUTHOR\footnote{Typeset names in
% 8~pt roman, uppercase. Use the footnote to indicate the
% present or permanent address of the author.}}

% \address{University Department, University Name, Address\\
% City, State ZIP/Zone,
% Country\footnote{State completely without abbreviations, the
% affiliation and mailing address, including country. Typeset in
% 8~pt Times italic.}\\
% first\_author@university.edu}

% \author{SECOND AUTHOR}

% \address{Group, Laboratory, Address\\
% City, State ZIP/Zone, Country\\
% second\_author@group.com}

\newcommand{\jch}[1]{{\textcolor{magenta}{{[JCH: #1]}}}}

\maketitle

%\begin{history}
%\received{Day Month Year}
%\revised{Day Month Year}
%\end{history}

\begin{abstract}
We review and update constraints on the Early Dark Energy (EDE) model from cosmological data sets, in particular \emph{Planck} PR3 and PR4 cosmic microwave background (CMB) data and large-scale structure (LSS) data sets including galaxy clustering and weak lensing data from the Dark Energy Survey, Subaru Hyper Suprime-Cam, and KiDS+VIKING-450, as well as BOSS/eBOSS galaxy clustering  and Lyman-$\alpha$ forest data. We detail the fit to CMB data, and perform the first analyses of EDE using the \texttt{CAMSPEC}  and \texttt{Hillipop} likelihoods for {\it Planck} CMB data, rather than \texttt{Plik}, both of which yield a tighter upper bound on the allowed EDE fraction than that found with \texttt{Plik}.  We then supplement CMB data with large-scale structure data in a series of new analyses. All these analyses are concordant in their Bayesian preference for $\Lambda$CDM over EDE, as indicated by marginalized posterior distributions. We perform a series of tests of the impact of priors in these results, and compare with frequentist analyses based on the profile likelihood, finding qualitative agreement with the Bayesian results. All these tests suggest prior volume effects are not a determining factor in analyses of EDE. This work provides both a review of existing constraints and several new analyses.
% The abstract should summarize the context, content
%and conclusions of the paper in less than 200 words. It should
%not contain any references or displayed equations. Typeset the
%abstract in 8 pt Times roman with baselineskip of 10~pt, making
%an indentation of 1.5 pica on the left and right margins.
\end{abstract}

%\keywords{Keyword1; keyword2; keyword3.}

%\ccode{PACS numbers:}

\ccode{MIT-CTP/5618}

\tableofcontents

%%%%%%%%%%%%%%%%%%%%%%%%%%%%%%%%%%%%%%%%%%%%%%%%%%%%%%%%%%%%%%%%%%%%%%
%%%%%%%%%%%%%%%%%%%%%%%%%%%%%%%%%%%%%%%%%%%%%%%%%%%%%%%%%%%%%%%%%%%%%%
\section{Introduction}
%%%%%%%%%%%%%%%%%%%%%%%%%%%%%%%%%%%%%%%%%%%%%%%%%%%%%%%%%%%%%%%%%%%%%%
%%%%%%%%%%%%%%%%%%%%%%%%%%%%%%%%%%%%%%%%%%%%%%%%%%%%%%%%%%%%%%%%%%%%%%

The Hubble tension has emerged as one of the leading challenges to the $\Lambda$ Cold Dark Matter ($\Lambda$CDM) model of cosmology. That is, direct measurements of the local expansion rate today, $H_0$, disagree with $\Lambda$CDM predictions inferred from early-Universe probes. While most acute between the \emph{Planck} CMB data and SH0ES Cepheid-variable-calibrated cosmic distance ladder $H_0$ measurement, which are discrepant at a statistical significance of $5\sigma$ \cite{Riess:2021jrx}, the Hubble tension persists across a wide array of probes \cite{Verde:2019ivm}. This has spurred on an active theory effort to find new models to restore concordance, though there still remains the possibility that the resolution lies in systematic errors in one or both of the data sets \cite{Rigault:2014kaa,NearbySupernovaFactory:2018qkd,Addison:2017fdm,CSP:2018rag,Jones:2018vbn,Efstathiou:2020wxn,Brout:2020msh,Mortsell:2021nzg,Mortsell:2021tcx,Freedman:2021ahq,Garnavich:2022hef,Kenworthy:2022jdh,Riess:2022mme,Feeney:2017sgx,Breuval:2020trd,Javanmardi:2021viq,Wojtak:2022bct}.  Note, however, that one can completely discard CMB data in the early-Universe approach and instead use baryon acoustic oscillation, Big Bang Nucleosynthesis, and weak lensing data to constrain $H_0$ via the sound horizon scale (e.g.,~\cite{Abbott:2017smn}), obtaining results fully consistent with those of \emph{Planck} and other CMB data sets, strongly suggesting that systematics are not at play in the early-Universe inference.

Early Dark Energy \cite{Karwal:2016vyq,Poulin:2018cxd,Agrawal:2019lmo,Lin:2019qug} (EDE) is a prominent proposal to resolve the Hubble tension, using new physics in the very early Universe to bring the $H_0$ inferred from \emph{Planck} CMB data into agreement with the cosmic distance ladder measurement of SH0ES. This is a daunting task given the high precision of \emph{Planck} 2018 CMB data: for example, the angular size of the sound horizon $\theta_s$, imprinted in the acoustic peaks of the CMB spectra, is measured to $0.03\%$. The EDE approach is to compensate a larger $H_0$, and hence smaller Hubble radius $r_H=c/H_0$, with a smaller sound horizon at last scattering $r_s$, in order to leave $\theta_s$ unchanged  \cite{Bernal:2016gxb,Addison:2017fdm,Aylor:2018drw,Schoneberg:2019wmt, Evslin:2017qdn}. This is achieved by introducing a new component of the Universe, that evolves as dark energy at early times (``Early Dark Energy'') in order to shrink $r_s$, but decays rapidly after matter-radiation equality in order to sequester its cosmological effects. Many EDE-like models have been proposed to realize this phenomenology; see e.g.,~Refs.~\cite{Smith:2019ihp,Agrawal:2019lmo,Lin:2019qug,Alexander:2019rsc,Sakstein:2019fmf,Das:2020wfe,Niedermann:2019olb,Niedermann:2020dwg,Niedermann:2021vgd,Ye:2020btb,Berghaus:2019cls,Freese:2021rjq,Braglia:2020bym,Sabla:2021nfy,Sabla:2022xzj,Gomez-Valent:2021cbe,Moss:2021obd,Guendelman:2022cop,Karwal:2021vpk,McDonough:2021pdg,Wang:2022nap,Alexander:2022own,McDonough:2022pku,Nakagawa:2022knn,Gomez-Valent:2022bku,MohseniSadjadi:2022pfz,Kojima:2022fgo,Rudelius:2022gyu,Oikonomou:2020qah,
Tian:2021omz,Maziashvili:2021mbm}.

While EDE was first proposed in \cite{Karwal:2016vyq},\footnote{Note that the constraints derived on EDE in Ref.~\cite{Karwal:2016vyq} erroneously concluded that the scenario was not viable to resolve the $H_0$ tension, due to a mismodeling of the linear-perturbation equations of motion (as pointed out in Ref.~\cite{Poulin:2018cxd}).} the first analysis of constraints on EDE from \emph{Planck} data alone, i.e., not combined with other data sets, was performed in \cite{Hill:2020osr}. This analysis, using the \texttt{Plik} likelihoood for \emph{Planck} 2018 temperature and polarization anisotropies, confirmed the intuition of earlier works, namely the parameter compensations that underlie the ability of EDE to fit the CMB data even for relatively large values of $H_0$. In particular, the EDE model introduces a degeneracy between $H_0$ and the peak fraction of energy density in the EDE component $f_{\rm EDE}$: larger values of $H_0$ are correlated with larger values of $f_{\rm EDE}$, with $f_{\rm EDE} \simeq 0.1$ near the epoch of matter-radiation equality to come into agreement with SH0ES and hence resolve the Hubble tension. Movement along this degeneracy direction towards larger $H_0$ is accompanied by compensating shifts in other $\Lambda$CDM parameters, in particular an increase in the physical dark matter density $\Omega_c h^2$ and an increase in the spectral index $n_s$, and correspondingly an increase in the parameter $S_8 \equiv \sigma_8 (\Omega_m/0.3)^{0.5}$. These compensations can be understood in terms of conventional CMB physics, such as the driving of acoustic oscillations and the early integrated Sachs-Wolfe effect (see, e.g., Refs.~\cite{Lin:2019qug,Hill:2020osr}).

The changes in these $\Lambda$CDM parameters result in non-negligible modifications to the cosmology that persist well after EDE has decayed away. For example, the $z=0$ matter power spectrum $P(k)$ is increased by ${\cal O}(10\%)$ at $k=1 ~h~{\rm Mpc}^{-1}$ \cite{Hill:2020osr}.  This presents a challenge to maintaining consistency with other cosmological data sets, in particular large-scale structure data sets, which span a wide range of  wavenumber and redshift where the EDE-induced parameter shifts leave a sizeable imprint. This challenge is a general feature of early-Universe resolutions to the Hubble tension, as emphasized in Refs.~\cite{Jedamzik:2020zmd,Lin:2021sfs}.

%\\

In this review we present and update constraints on the canonical axion-like EDE model from a variety of cosmological data sets and data set combinations.  We examine in detail the constraints from \emph{Planck} 2018 (PR3) and \emph{Planck} NPIPE (PR4) CMB data, including constraints from alternative choices of the likelihood, namely the \texttt{Plik} likelihood for {\it Planck} PR3 data, the \texttt{CamSpec} likelihood for both PR3 and PR4 data, and the \texttt{HiLLiPoP} likelihood for PR4 data. We then include large-scale structure data, including galaxy clustering and weak lensing data from the Dark Energy Survey, weak lensing from Subaru Hyper Suprime-Cam and KiDS+VIKING-450, BAO and full-shape data from BOSS DR12 galaxy
power spectrum and bispectrum likelihood, 
and Lyman-$\alpha$ forest data from  eBOSS \cite{Chabanier:2018rga} and XQ-100 \cite{Irsic:2017sop}. In each of these cases we find tight upper bounds on $f_{\rm EDE}$ that are well below the benchmark value required to resolve the Hubble tension.

We also examine in detail the role of priors, and so-called `prior-volume effects,' (see e.g. \cite{Planck:2013nga}) in Bayesian constraints on EDE, and the implications of this for the status of EDE as a concordant cosmological model. We perform a multipronged analysis of priors in the EDE scenario:
\begin{enumerate}
    \item {\it Impact of SH0ES $H_0$ prior}: The MCMC analysis including $S_8$ data from DES, HSC, and KV-450, is performed with a {\it prior on $H_0$ given by the SH0ES measurement,} which is then removed in post-processing of the samples, to ensure adequate exploration of the $H_0$-tension resolving region of EDE parameter space.
    \item {\it Impact of particle-physics-motivated priors}: We compute the effective priors on the axion decay constant $f$ and mass $m$ implied by uniform priors on the derived parameters $f_{\rm EDE}$ and the critical redshift $z_c$. These effective parameters are almost exclusively used in EDE analyses in order to minimize the computational expense of the MCMC. We find that the {\it computational-expense-motivated priors on $f_{\rm EDE} $ and $z_c$ imply a prior distribution for the decay constant $f$ that is peaked at the Planck scale} $f \approx M_{pl}$, in conflict with priors from particle physics, such as the Weak Gravity Conjecture \cite{Arkani-Hamed:2006emk, Rudelius:2015xta, Brown:2015iha, Hebecker:2015zss}. We repeat the MCMC analysis of \emph{Planck} (\texttt{Plik}) with uniform priors on $f$ and $\log_{10}m$ and an find even tighter upper bound on $f_{\rm EDE}$ than that resulting from uniform priors on $f_{\rm EDE}$ and $z_c$.
    \item {\it Mean Likelihood Profile:} We extract the {\it mean likelihood profile} for $f_{\rm EDE}$ from MCMC chains of CMB+BOSS and CMB+BOSS+$S_8$. The mean likelihood profile of the samples (see \cite{LewisBridle2002} for a detailed discussion) by construction suppresses marginalization bias effects, and for this reason has been used in CMB analyses in e.g.~\cite{LewisBridle2002,Montier:2014bra,Martin:2010kz,Audren:2012wb,Planck2015likelihood}. Comparing the mean likelihood profile from our analysis we find that it closely resembles the corresponding 1d marginalized posteriors: both are monotonically decreasing functions of $f_{\rm EDE}$. This implies that the posterior is {\it not} driven by prior volume effects, but rather the likelihood. 
    \item {\it EDE away from the $\Lambda$CDM limit:} To mitigate the degeneracies of $z_c$ and $\theta_i$ in the $\Lambda$CDM limit $f_{\rm EDE}\rightarrow 0$, we repeat our analysis of BOSS and $S_8$ data with a lower bound $f_{\rm EDE}>0.04$ imposed as a {\it prior} on the model. The analysis results in an upper bound $f_{\rm EDE}<0.08$ at 95\% C.L., again less than the benchmark $f_{\rm EDE}=0.1$ needed to address the Hubble tension.
    \item {\it Profile Likelihood:}  As a final investigation into the impact of priors on Bayesian parameter constraints in EDE, we compare the frequentist analysis performed in \cite{Herold:2021ksg,Herold:2022iib} based on the profile likelihood, namely the maximum likelihood at fixed values of $f_{\rm EDE}$, to the Bayesian analyses performed here. We find qualitative agreement between the two approaches, namely a strong preference for $\Lambda$CDM over EDE when data are included from \emph{Planck} 2018, BOSS DR12, and $S_8$ data from DES.

\end{enumerate}

This work provides both a review of existing constraints and several new analyses, including the first analyses of EDE with  \texttt{CAMSPEC}  and \texttt{Hillipop} likelihoods for Planck CMB data, an EFT analysis including new BOSS power spectrum 
and bispectrum 
measurements, 
eBOSS emission line galaxies, and analyses of the role of priors as discussed above. The structure of this work is as follows: In Sec.~\ref{sec:EDEreview} we review the EDE model and the proposed EDE resolution of the Hubble tension, followed by a review of constraints for \emph{Planck} PR3 CMB data, and newly obtained constraints from \emph{Planck} PR4 data, in Sec.~\ref{sec:constraintsCMB}. In Sec.~\ref{sec:EDExLSS} we review large-scale structure (LSS) phenomenology in the EDE model, before proceeding to review LSS constraints on the model in Secs.~\ref{sec:constraintsWL}, \ref{sec:constraintsBOSS}, and \ref{sec:constraintsLymanalpha}. We then discuss the role of priors in these analyses in Sec.~\ref{sec:priors}, and conclude in Sec.~\ref{sec:Discussion} with a summary of the results and directions for future work.

%%%%%%%%%%%%%%%%%%%%%%%%%%%%%%%%%%%%%%%%%%%%%%%%%%%%%%%%%%%%%%%%%%%%%%
%%%%%%%%%%%%%%%%%%%%%%%%%%%%%%%%%%%%%%%%%%%%%%%%%%%%%%%%%%%%%%%%%%%%%%
%\clearpage
\section{Early Dark Energy}
\label{sec:EDEreview}
%%%%%%%%%%%%%%%%%%%%%%%%%%%%%%%%%%%%%%%%%%%%%%%%%%%%%%%%%%%%%%%%%%%%%%
%%%%%%%%%%%%%%%%%%%%%%%%%%%%%%%%%%%%%%%%%%%%%%%%%%%%%%%%%%%%%%%%%%%%%%

%%The EDE approach

The insight for early-Universe solutions to the Hubble tension is to recognize that CMB data do not directly measure length scales, such as the CMB damping scale $r_d$, the sound horizon $r_s$, or the Hubble length $H_0 ^{-1}$, but instead angles on the sky formed by ratios of these length scales. In particular, the angular extent of the sound horizon,
\begin{equation}
\label{eq:thetas}
    \theta_s = \frac{r_s (z_*)}{D_A(z_*)}\,,
\end{equation}
is constrained to 0.03\% precision in the \emph{Planck} 2018 analysis,  $100 \theta_s = 1.0411 \pm 0.0003$ \cite{Aghanim:2018eyx}. The primacy of the angular sound horizon is manifest in the acoustic peaks of the CMB temperature anisotropy angular power spectrum, where $\theta_s$ corresponds to the position (multipole number) of the first peak, and the spacing between subsequent peaks.

The numerator and denominator in the above are the comoving sound horizon,
\begin{equation}
\label{eq:rs}
   r_s(z_*) = \int _{z_*} ^{\infty} \frac{{\rm d} z}{H(z)} c_s(z) ,
\end{equation}
and the angular diameter distance to last scattering,
\begin{equation}
D_A (z_*) = \int _0 ^{z_*} {\rm d}z \frac{1}{H(z)}\,,
\end{equation}
respectively, where $z_*$ is the redshift of last scattering.

The tight CMB constraint on $\theta_s$ can accommodate a relative increase in $H_0$ if there is a commensurate increase in $H(z>z_*)$: The former leads to a decrease in $D_A$ and the latter a decrease in $r_s$, leaving their ratio $\theta_s$ fixed. In order for this increase in $H(z>z_*)$ to have any appreciable impact on $r_s$, the effect must be active within the last decade of redshift preceding last scattering \cite{Knox:2019rjx}. A simple physical model to realize this possibility is a new dark-energy-like component in the Universe, which, present for $z \gtrsim z_*$, rapidly decays to leave the underlying $\Lambda$CDM cosmology otherwise unchanged. This scenario is termed ``Early Dark Energy''  \cite{Karwal:2016vyq}, and will be the focus of this review.

\subsection{The Canonical EDE Model}

The original EDE proposal \cite{Karwal:2016vyq} modelled the EDE field with a phenomenological fluid description described by the evolution of energy density fraction and a characteristic redshift.
A concrete field theory model of EDE was proposed in \cite{Poulin:2018cxd}
utilizing the class of scalar potentials proposed in \cite{Kamionkowski:2014zda} that generalized the conventional axion potential $V \propto (1 - \cos (\varphi/f))$ by introducing an exponent, as
\be
V(\varphi) = V_0 \left[ 1- \cos\left(\frac{\varphi}{f}\right)\right]^n \,\, , \,\, V_0 \equiv m^2 f^2 .
\label{eq:EDE_V}
\ee
The potential is characterized by a mass scale $m$, periodicity (`decay constant') $f$, and an exponent $n$.  Ref.~\cite{Poulin:2018cxd} considered $n=1,2,3$ and $n\gg1$, with the conclusion that $n=3$ is the best fit to data, though this preference is weak. The $n=3$ model has been widely studied in the literature, and we therefore define the `canonical' model as the potential in Eq.~\eqref{eq:EDE_V} with
\begin{equation}
    n=3 \quad \quad ({\rm canonical\,\,EDE}) \,,
\end{equation}
which will be the focus of much of this work.

The background dynamics of EDE can be understood directly from the Klein-Gordon equation for a canonical scalar field in an expanding background,
\begin{equation}
    \ddot{\varphi}+3 H \dot{\varphi} + \frac{dV}{d\varphi}=0\,.
\end{equation}
where $\dot{}$ denotes a derivative with respect to cosmic time.
If the potential $V(\varphi)$ is sufficiently flat, at early times the field will be held in place by the Hubble friction term $3 H \dot{\varphi}$, and the field will effectively act as a dark energy component of the Universe. As the Universe expands and the Hubble parameter decreases, eventually the field is released from Hubble drag and will roll down the potential. For a quadratic potential $V(\varphi) = \frac{1}{2} m^2 \varphi^2 $ this occurs at a critical redshift $z_c$ when $H(z_c) \simeq m$. 

The critical redshift $z_c$ can be tuned to the desired value simply by tuning the mass $m$: An EDE component which becomes dynamical around $z_{*}$ (in order to have an appreciable reduction of the sound horizon $r_s$) requires a scalar field mass of $m \sim H(z_{\rm eq})\sim 10^{-28}$ eV. The magnitude of the Hubble tension ($\approx 10\%$), determines the energy density fraction of the EDE component,
\begin{equation}
    f_{\rm EDE}(z)  \equiv \frac{\rho_{\rm EDE}(z)}{ \rho_{\rm tot}(z)}\, 
\end{equation}
to be $\approx 10\%$ at $z=z_c$. Combined with the natural initial field value $\varphi_i \sim f$, this determines the decay constant in Eq.~\eqref{eq:EDE_V} as $f\lesssim M_{pl}$. Indeed, the best-fit parameter values in the fit to varied data sets are $f\sim 0.2 M_{pl}$ and $m\sim 5 \times 10^{-28} $ eV; see, e.g., Ref.~\cite{McDonough:2021pdg}.

\begin{figure}[!t]
    \centering
    \includegraphics[width=\linewidth]{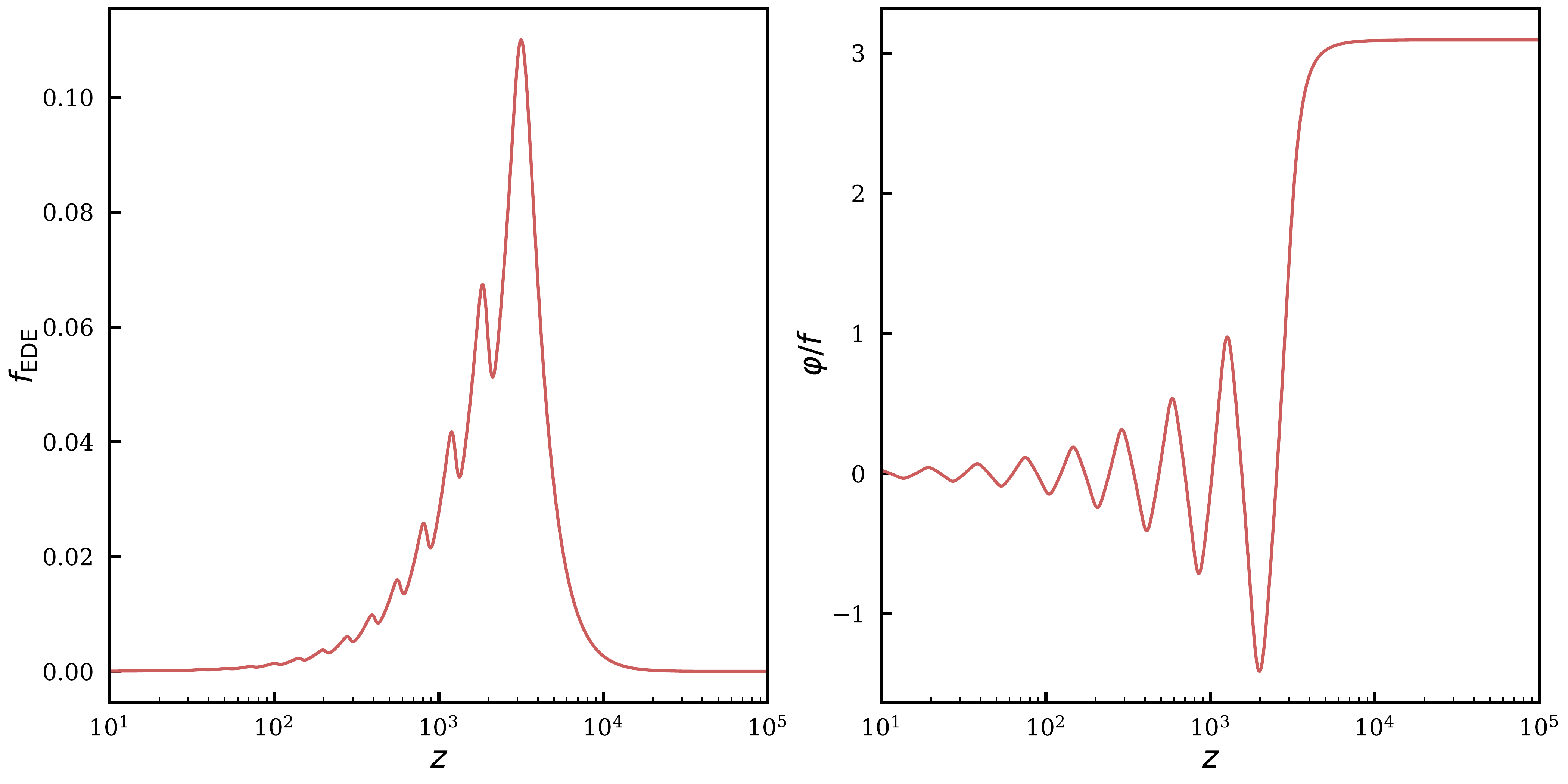}
    \caption{Evolution of the fraction of the total energy density (left panel) and scalar field (right panel) in the canonical EDE model  with parameters that increase $H_0$. Here $f_{\rm EDE} = 0.11, z_c = 3500,$ and $ \theta_i = 3.10$ for the effective cosmology parameters, or equivalently in terms of particle physics parameters, $m = 6.0 \times 10^{-28} {\rm~eV}, f = 2.7 \times 10^{26} {\rm~eV}$. }
    \label{fig:fede_plot}
\end{figure}

Following the release from Hubble drag, the energy density in the EDE component will redshift away as the field oscillates about the minimum of the potential $V(\varphi)$, or else runs away to $\infty$ as in, e.g., Acoustic Dark Energy \cite{Lin:2019qug,Lin:2020jcb}, the axion-dilaton model \cite{Alexander:2019rsc}, and a subset of the examples in Ref.~\cite{Agrawal:2019lmo}. The decay of the EDE, encoded in the equation of state $w$ as $\rho(t) \propto a(t)^{-3(1+w)}$, is controlled by the shape of the potential. For a potential with a minimum that is locally $V(\varphi)\propto \varphi^{2n}$, the energy density of an oscillating scalar field will redshift with time-averaged equation of state~\cite{Turner:1983he}
\begin{equation}
    \langle w \rangle = \frac{n-1}{n+1}
\end{equation}
where the angular brackets denote a time-average over field oscillations. The exponent $n$ in Eq.~\eqref{eq:EDE_V} can therefore be used to control the decay of the EDE. This in turn controls the change to the Silk damping scale $r_d$ of the CMB, which in turn drives the preference for $n=3$ \cite{Poulin:2018cxd}. 

A fiducial example is shown in Fig.~\ref{fig:fede_plot}, where we show the evolution of the scalar field $\varphi$ along with the energy density fraction $f_{\rm EDE}(z)$. This example features $f_{\rm EDE}(z_c) \sim 10\%$ and $z_c \sim 10^3$ relevant to the Hubble tension, realized by Eq.~\eqref{eq:EDE_V} with $m=6.0\times 10^{-28}$ eV, $f=2.6\times 10^{26}$ eV, $\varphi_i/f = 3.10$ , and $n=3$. This plot is generated using the code {\tt class\_ede} \cite{Hill:2020osr}. This code is capable of inferring $m$ and $f$ from a user-specified $f_{\rm EDE}$ and $z_c$, enabling efficient exploration of cosmological constraints on the model; we return to this point in Sec.~\ref{sec:priors}. The code and explanatory Python notebooks are publicly available.\footnote{\url{https://github.com/mwt5345/class_ede}}

As the Hubble tension has grown in statistical significance, many EDE-like models have been proposed \cite{Smith:2019ihp,Agrawal:2019lmo,Lin:2019qug,Alexander:2019rsc,Sakstein:2019fmf,Das:2020wfe,Niedermann:2019olb,Niedermann:2020dwg,Niedermann:2021vgd,Ye:2020btb,Berghaus:2019cls,Freese:2021rjq,Braglia:2020bym,Sabla:2021nfy,Sabla:2022xzj,Gomez-Valent:2021cbe,Moss:2021obd,Guendelman:2022cop,Karwal:2021vpk,McDonough:2021pdg,Wang:2022nap,Alexander:2022own,McDonough:2022pku,Nakagawa:2022knn,Gomez-Valent:2022bku,MohseniSadjadi:2022pfz,Kojima:2022fgo,Rudelius:2022gyu,Oikonomou:2020qah,
Tian:2021omz,Maziashvili:2021mbm}, including models which consider modifying the kinetic or potential terms for the EDE scalar \cite{Alexander:2019rsc,Alexander:2022own}, models that reproduce the EDE background dynamics via interactions amongst multiple fields \cite{Alexander:2022own,Niedermann:2019olb,Niedermann:2020dwg}, models that aim to resolve the apparent required fine-tuning of $z_c$ \cite{Lin:2022phm,Sabla:2021nfy,Sakstein:2019fmf,CarrilloGonzalez:2020oac}, and models which aim to resolve tensions with LSS data \cite{McDonough:2021pdg,Allali:2021azp,Alexander:2022own,Berghaus:2019cls,Berghaus:2022cwf,Clark:2021hlo,Cruz:2023lmn}. Each of these requires a dedicated analysis to determine the relative fit to data and any overall preference for the model in comparison to $\Lambda$CDM and the canonical $n=3$ EDE model in Eq.~\eqref{eq:EDE_V}.

%%%%%%%%%%%%%%%%%%%%%%%%%%%%%%%%%%%%%%%%%%%%%%%%%%%%%%%%%%%%%%%%%%%%%%
%%%%%%%%%%%%%%%%%%%%%%%%%%%%%%%%%%%%%%%%%%%%%%%%%%%%%%%%%%%%%%%%%%%%%%
%\clearpage
\subsection{Resolution of the Planck-SH0ES $H_0$ Tension}

\label{sec:EDECMB}

\begin{figure}%[h!]
    \centering
    \includegraphics[width=\linewidth]{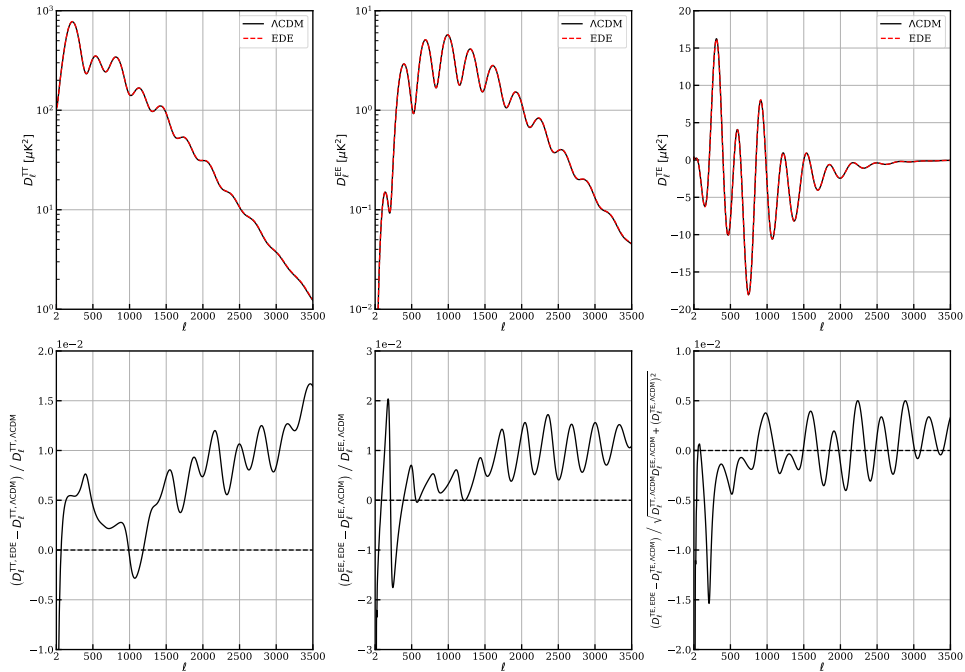}
    \caption{CMB power spectrum for TT (left panel), EE (middle panel), and TE (right panel) in the canonical EDE model (red, dashed) and $\Lambda$CDM (black, solid), with $H_0 = 71.15$ km/s/Mpc and $H_0 = 68.07$ km/s/Mpc, respectively, including the corresponding fractional difference between the two models (bottom). We normalize the fractional difference for TT and EE to the $\Lambda$CDM spectra, while we normalize TE by the variance to account for the zero crossings in this spectrum. 
    The model parameters used here correspond to the best-fit results from \cite{Hill:2020osr} in the fit to primary CMB (TT+TE+EE), CMB lensing, RSD, BAO, SNIa, and SH0ES data.
   }
    \label{fig:CMB_TT}
\end{figure}

The Early Dark Energy model delivers on its promise to accommodate an $H_0$ value compatible with SH0ES while remaining compatible with CMB data from \emph{Planck} 2018. This can be appreciated from Fig.~\ref{fig:CMB_TT} where we show the CMB temperature and polarization power spectra for a fiducial $\Lambda$CDM cosmology and EDE cosmology with $H_0=68$ and $71.5$ km/s/Mpc respectively, along with the fractional difference, indicating percent-level changes across the range of $\ell$ probed by \emph{Planck} and beyond.

\begin{figure}[h!]
\centering
\includegraphics[trim={400 0 400 0}, clip,width=\linewidth]{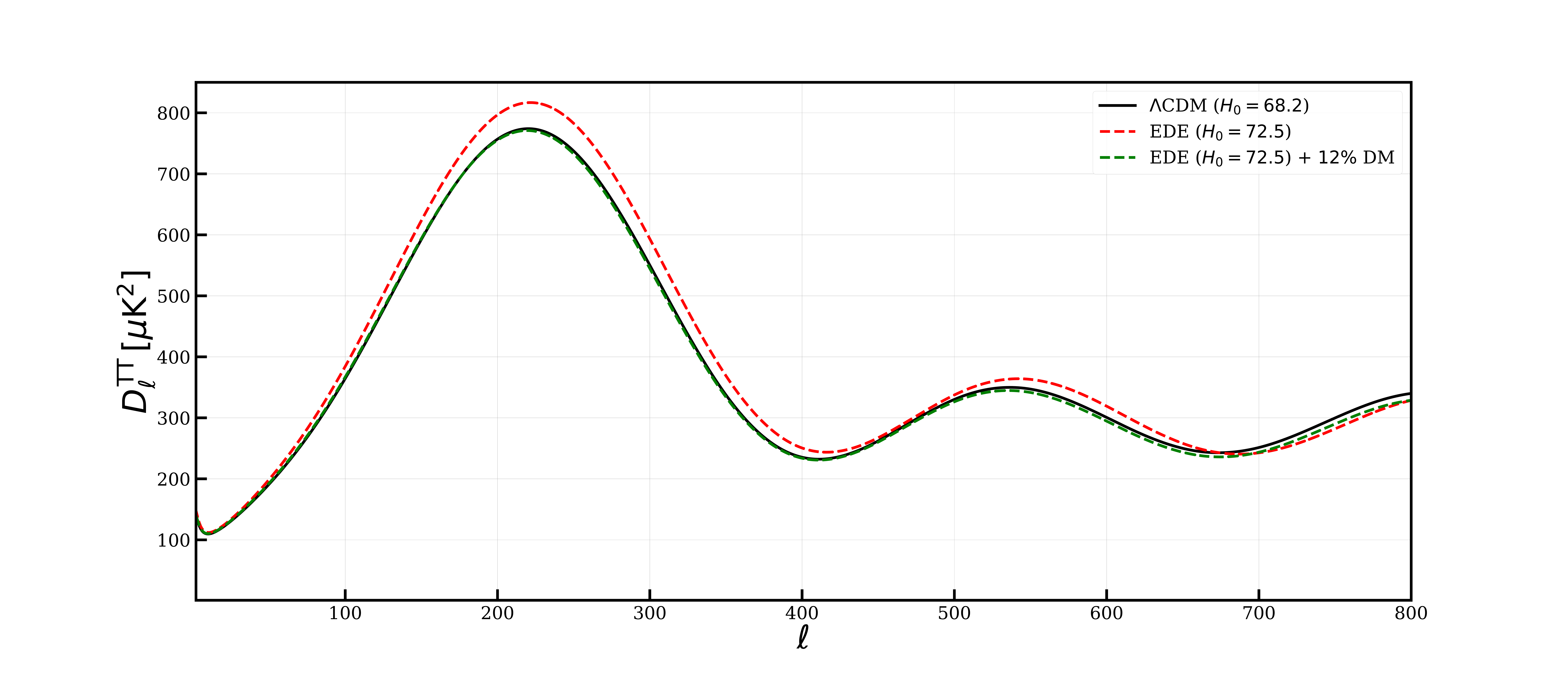}
\caption{ EDE and the first peak of the CMB:  We show a  $\Lambda$CDM cosmology with $H_0=68.16 $ km/s/Mpc (black), EDE with $H_0=72.52$ and $f_{\rm EDE}=14\%$ (red, dashed) but other $\Lambda$CDM parameters unchanged, and the same EDE with an additional $12\%$ dark matter ($\Omega_c h^2$).  The additional dark matter compensates for the effect of EDE on the first peak. The best-fit model has additional compensating parameter shifts, such as the spectral index $n_s$, discussed in detail in \cite{Hill:2020osr}.}
\label{fig:CMB}
\end{figure}

However, the situation is more nuanced than presented thus far: The precision of the \emph{Planck} 2018 data demands a delicate balance between the additional dark energy component and shifts in the standard $\Lambda$CDM parameters, e.g., to undo the EDE-induced modification to the driving of acoustic oscillations (see, e.g.,~\cite{Lin:2019qug}). Notably, the $\Lambda$CDM parameters change as follows in the $H_0$-resolving EDE cosmology (here for concreteness we compare the best-fit $\Lambda$CDM and EDE cosmologies from Tab.~I of \cite{McDonough:2021pdg}): 
\begin{enumerate}
    \item Dark Matter density:  $\gtrsim 10\%$ increase in the physical dark matter density $\Omega_{c}h^2$, corresponding to a shift of $\sim 20 \sigma$ in units of the $\Lambda$CDM error bar. 
    
    \item Spectral Index:  $\gtrsim 5 \sigma$ ($\approx 2\%$) increase in the spectral index $n_s$ relative to that in $\Lambda$CDM, in units of the error bar in the $\Lambda$CDM error bar. 
    
    \item Amplitude of perturbations:  $\sim 2\sigma$ increase in the amplitude of scalar perturbations $A_s$. 
\end{enumerate}
The precise $\sigma$ values depend on the analysis; these parameter shifts refer to difference in best-fit values from Tab. I of \cite{McDonough:2021pdg}, namely the fit to {\it Planck} PR3, BAO, SNIa, and SH0ES.

These shifts drive the tight constraints from large-scale structure data that will be presented in Secs.~\ref{sec:constraintsWL}, \ref{sec:constraintsBOSS}, and \ref{sec:constraintsLymanalpha}.

Of particular interest is the dark matter density. The amplitude of the first acoustic peak of the CMB temperature power spectrum provides a measure of the amount of dark matter in the Universe \cite{Mukhanov:2003xr}. 
As explained in \cite{Lin:2019qug}, the early dark energy component changes the evolution of the Weyl potential and hence the driving of acoustic oscillations (see Ref.~\cite{Hu:1994uz} for a review), leading to an overall enhancement of the first acoustic peak of the CMB. This can alternately be interpreted as an overall enhancement of the early integrated Sachs-Wolfe effect (eISW) \cite{Hill:2020osr,Vagnozzi:2021gjh}, arising from the EDE-induced suppression of perturbation growth prior to recombination. This is illustrated in Fig.~\ref{fig:CMB}, where we show the first peak of the CMB temperature anisotropy power spectrum in $\Lambda$CDM with $H_0=68.2$ km/s/Mpc, and in Early Dark Energy with the same $\Lambda$CDM parameters but with $f_{\rm EDE}=0.14$, namely 14\% of the Universe in an EDE component at a redshift $\log_{10}(z_c)=3.58$. Further parameter adjustments are needed to fit the damping tail of the CMB. In particular, a significant shift in the spectral index $n_s$ towards unity is needed, to compensate for the scale-dependent suppression of growth due to the EDE.

\subsection{UV Completions of EDE}

The canonical EDE potential, Eq.~\eqref{eq:EDE_V} with $n=3$, is puzzling from the perspective of the conventional theory constructions of axion-like particles, where axions arise  as the pseudo-Nambu Goldstone bosons of a spontaneously broken global U(1) symmetry. The assumed lightness of the axion field is justified on the basis of the underlying symmetry, which protects the field from perturbative corrections, leaving non-perturbative effects, such as instantons, as the dominant source of mass generation. 
In conventional quantum field theory and string theory setups, the axion potential is the leading-order term in an expansion of the form,
\begin{equation}
    V\simeq V_0 \sum_n e^{-  S_n} \cos\left( \frac{n \varphi}{f}\right)
    \label{eq:V_inst}
\end{equation}
where $S_n$ is the $n$-th order instanton action,  $S_n \simeq n S$, with $S$ the leading instanton action. Explicit string theory computations (e.g.,~\cite{Svrcek:2006yi, Conlon:2006tq, Cicoli:2012sz}) as well as general principles, e.g., the weak gravity conjecture applied to axions \cite{Arkani-Hamed:2006emk, Rudelius:2015xta, Brown:2015iha, Hebecker:2015zss}, suggest that $S \simeq \lambda M_{pl}/f$ where $\lambda$ is an $\mathcal{O}(1)$ constant.  

In comparison, the canonical EDE potential, Eq.~\eqref{eq:EDE_V}, can  be expanded out as 
\be
V(\varphi) = V_0\left[\frac52-\frac{15}{4}\cos\left(\frac{\varphi}{f}\right)+\frac32\cos\left(\frac{2\varphi}{f}\right)-\frac14\cos\left(\frac{3\varphi}{f}\right)\right],
\label{eq:EDE_V_expanded}
\ee
which resembles the expansion in Eq.~\eqref{eq:V_inst} but truncated at $n=3$.  It is not obvious from the above whether this peculiar hierarchy of harmonics can be realized via a controlled instanton expansion of the form Eq.~\eqref{eq:V_inst}. Moreover, the requirements from data that $f 
\lesssim M_{pl}$ suggests a tension with the weak gravity conjecture, and the accompanying Planckian field excursions $|\Delta \varphi|\sim f$ runs the risk of exponentially lightening other fields, as per the Swampland Distance Conjecture. 

\subsubsection{Multifield Models}

Canonical EDE relies upon the addition of a single scalar field beyond the standard concordance model. However, there are several well-motivated reasons to consider the addition of other fields. 

A solid field theory explanation for a light scalar field in the early Universe is from a string axion. However, axions only represent half the scalar field content in such models. That is, the axion corresponds to the phase of a complex scalar field. The modulus of this field is the so-called dilaton and is related to the string coupling $g_s = e^\chi$. Together, the axion and the dilaton comprise the complex axio-dilaton field. The existence of moduli fields is ubiquitous in string theory (and other higher dimensional theories) and one of the few model-independent predictions. However, it is typically assumed that the moduli fields are stablized and can consequently be ignored as they become non-dynamical. Nevertheless, one can also consider the case where both the axion and dilaton are dynamical. 

An interesting application of the axio-dilaton model is to consider the case where the dilaton is initially stabilized, in this case endowed with an exponential potential. In this scenario it is the dilaton that behaves as the source of ``dark energy'' for an EDE model. The canonical evolution of the axion field will then lead to a destabilization of the dilaton causing it enter into fast-roll. This has the desirable effect of quickly dissipating the energy density of the EDE scalar (in this case the dilaton) and providing an alternative mechanism for EDE.

An offshoot of the axio-dilaton model is where one considers the axion with a non-canonical kinetic term \cite{Alexander:2022own}. This results in a coupling between the axion and the dilaton in the axion kinetic term. This Kinetic MIXing (KMIX) is a generic feature from the effective theory for string compactifications of complex moduli fields. The Lagrangian takes on the form,
\begin{equation}
    \mathcal{L} = - \frac{1}{2} (\partial \chi)^2  - \frac{1}{2} f(\chi) (\partial \varphi)^2 - V(\chi,\varphi),
\end{equation}
where $\chi$ is the dilaton and $f(\chi) \equiv e^{\lambda \chi}$ with $\lambda$ an $\mathcal{O}(1)$ number which is dependent on the choice of internal manifolds used for string compactification.  The impact of the kinetic mixing is evident from the equations of motion,
\begin{equation}
    \ddot\varphi + \left(3 H + \frac{f_\chi}{f} \dot\chi \right) \dot\varphi + \frac{1}{f} V_\varphi = 0,
\end{equation}
\begin{equation}
    \ddot\chi + 3 H \dot \chi - \frac{1}{2}f_\chi \dot\varphi^2 + V_\chi = 0.
\end{equation}
The two major changes from the standard Klein-Gordon equations are the presence of a new friction term for the axion and a source term for the dilaton. The latter allows for the energy transfer from the axion field into the dilaton and critically allows for the use of a standard axion-like potential. Note that it is not possible to accommodate an $n=1$ potential in a single-field model as the new scalar will act as an additional form of dark matter post-recombination and, furthermore, its contribution to the cosmic energy budget will be non-negligible. However, in the KMIX model the dilaton sources the energy density of the axion as soon as it starts to fall down its potential, allowing for the total energy density of both fields to redshift as radiation or faster. A fundamental difference compared to canonical EDE is that in the KMIX model it is possible for $\lesssim 1\%$ of dark matter to be comprised of the (light) EDE scalar. That is, it is a generic prediction of this model that a small fraction of dark matter will be an ultra-light axion, which has been well-studied in the context of structure formation \cite{Marsh:2010wq,Amendola:2005ad,Allali:2021azp}. Since the Jeans scale of the EDE scalar is $\sim$ 30~Mpc, this results in a suppression of the matter power spectrum for scales smaller than $k_J$ which naturally has the dynamics to suppress $S_8$.

Further multicomponent models of EDE include the `Early Dark Sector' \cite{McDonough:2021pdg,Lin:2022phm}, wherein the EDE is directly coupled to dark matter,  the closely related `chameleon early dark energy' \cite{Karwal:2021vpk}, the neutrino-assisted EDE \cite{Sakstein:2019fmf,CarrilloGonzalez:2020oac}, and New EDE \cite{Niedermann:2019olb,Niedermann:2020dwg}, wherein the decay of the EDE field occurs via a first-order phase transition triggered by the evolution of a second field.

\subsubsection{Early Dark Energy in String Theory}

An alternative approach to a UV embedding of EDE is to tackle each term in Eq.~\eqref{eq:EDE_V_expanded} individually, with each arising as the leading term in a convergent instanton expansion. Implemented in string theory in \cite{McDonough:2022pku,Cicoli:2023qri}, the desired $1:2:3$ ratio of periodicities corresponds to the rank of $SU(N)$ gauge groups $N_1 : N_2 :N_3$, or other integer quantities such as worldvolume fluxes.

Here we briefly sketch out the mechanics of this. The effective field theory in four dimensions is an ${\cal N}=1$ supergravity theory, where the scalar potential is given by
\begin{equation}
\label{eqn:SUGRApotential}
    V = e^{K}\left( |DW|^2-3|W|^2 \right) \, ,
\end{equation}
where $K$ and $W$ are the Kahler potential and superpotential respectively, each built out of the fields in the model, and $D$ denotes a covariant derivative on field space. A prototypical non-perturbative effect in string theory is gaugino condensation on a stack of $N$ D7 branes, described by a superpotential 
\begin{equation}
W_{\rm D7}  = A \, e^{-\frac{2\pi}{N} f_{\rm D7}}  \,,
\label{eqn:WD7}
\end{equation}
where  $f_{\rm D7}$ depends on the fields and is model dependent. In the case of multiple distinct stacks of branes, wrapped around $4$-cycles which are distinct representatives of the same homology class, the superpotential reads:
\begin{equation}
W_{\rm D7}  = \sum_{i=1}^{N_{\rm stacks}}\,A_i \, e^{-\frac{2\pi}{N_i} f_{{\rm D7},i}} \,.
\label{eqn:WD7Gen}
\end{equation}
This provides multiple avenues to constructing the EDE potential. For example, if $f_{\rm D7}=T$, where $T$ a complex scalar, is the same on each of $N_{\rm stacks}=3$ brane stacks with numbers of branes in the ratios $N_1:N_2:N_3 = 1:2:3$, this leads to a potential whose dominant terms are cosine potentials for Im$(T)$ with the desired relative periodicities in Eq.~\eqref{eq:EDE_V_expanded}. The prefactors in Eq.~\eqref{eq:EDE_V_expanded} are then set by $e^{-\frac{2\pi}{N_i} {\rm Re}(T)}$ which is $\ll 1$ (in Planck units) in order to match the energy scale of the EDE. As detailed in \cite{Cicoli:2023qri}, this can be realized in a convergent instanton expansion, in contrast with naive expectation from Eq.~\eqref{eq:EDE_V}.

%%%%%%%%%%%%%%%%%%%%%%%%%%%%%%%%%%%%%%%%%%%%%%%%%%%%%%%%%%%%%%%%%%%%%%
%%%%%%%%%%%%%%%%%%%%%%%%%%%%%%%%%%%%%%%%%%%%%%%%%%%%%%%%%%%%%%%%%%%%%%

%%%%%%%%%%%%%%%%%%%%%%%%%%%%%%%%%%%%%%%%%%%%%%%%%%%%%%%%%%%%%%%%%%%%%%
%%%%%%%%%%%%%%%%%%%%%%%%%%%%%%%%%%%%%%%%%%%%%%%%%%%%%%%%%%%%%%%%%%%%%%
%\clearpage
\section{Constraints from \emph{Planck} CMB Data}
\label{sec:constraintsCMB}
%%%%%%%%%%%%%%%%%%%%%%%%%%%%%%%%%%%%%%%%%%%%%%%%%%%%%%%%%%%%%%%%%%%%%%
%%%%%%%%%%%%%%%%%%%%%%%%%%%%%%%%%%%%%%%%%%%%%%%%%%%%%%%%%%%%%%%%%%%%%%

To quantitatively understand and assess the EDE model, we turn to data,  beginning with {\it Planck} 2018 primary CMB anisotropy (temperature and polarization) angular power spectra. The first analysis of {\it Planck} data alone in EDE was performed in \cite{Hill:2020osr}. Here we reproduce and expand on these results, including a detailed comparison of more recent likelihoods, including the \texttt{Plik} likelihood for PR3 data, the \texttt{CamSpec} likelihood for PR3 data and for PR4 data, and the \texttt{HiLLiPoP} likelihood for PR4 data.
%We perform MCMC analyses using ***** describe ***** . 

\begin{figure}[h!]
\centering
\includegraphics[width=\textwidth]{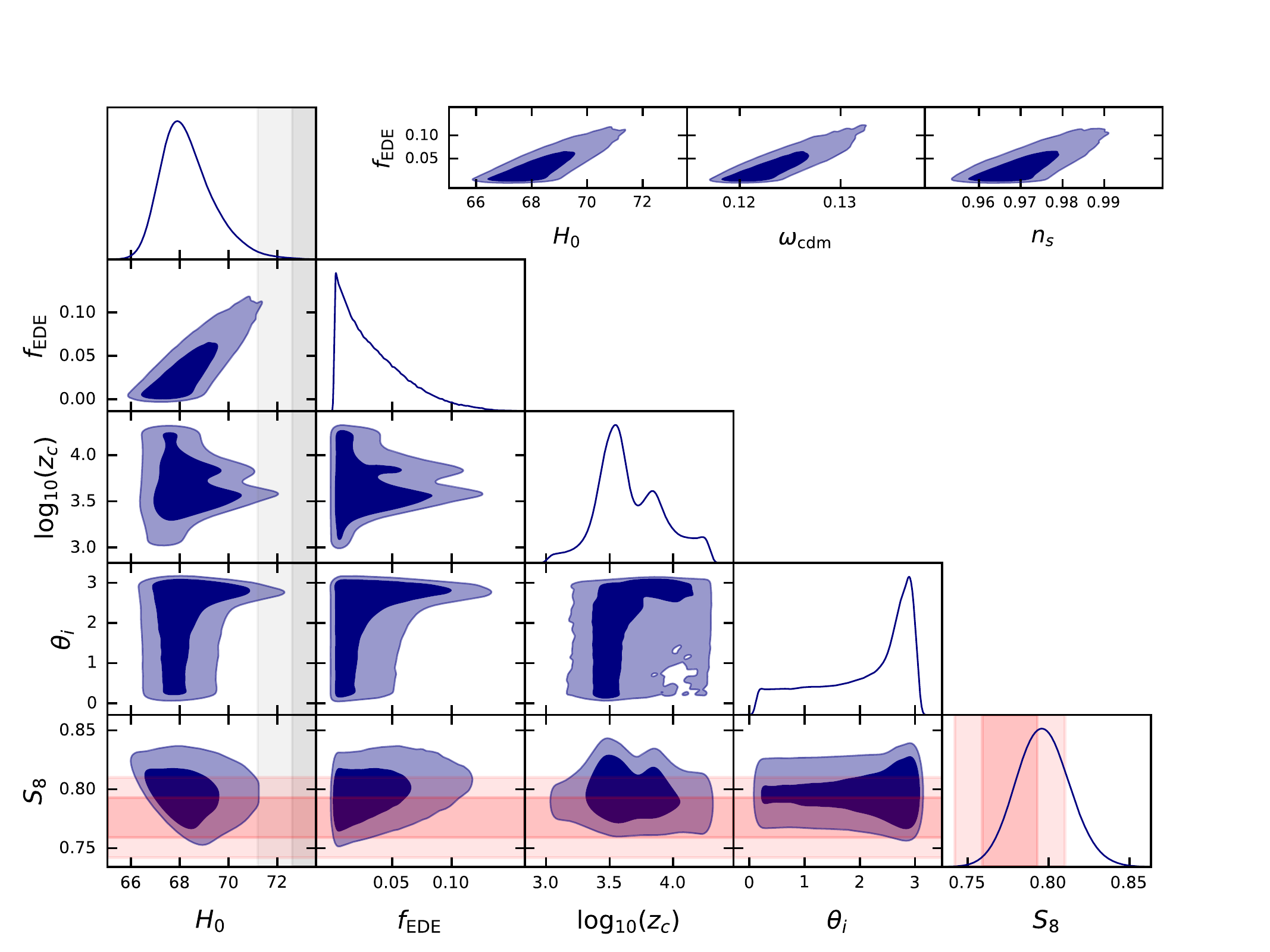}
\caption{Constraints on EDE from {\it Planck} 2018 CMB data (\texttt{Plik} PR3 likelihood). Grey bands show the SH0ES measurement of $H_0$ and pink bands indicate the DES-Y3 measurement of $S_8$.}
\label{fig:EDE-planck-triangle}
\end{figure}

\begin{figure}[h!]
\centering
\includegraphics[width=0.9\textwidth]{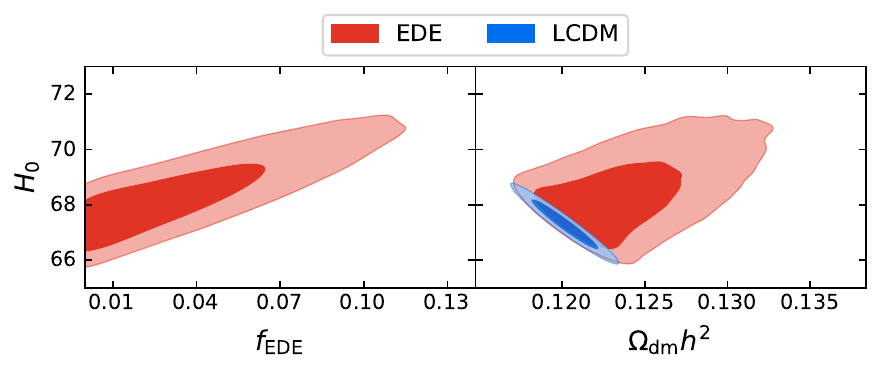}
\caption{
EDE and $\Lambda$CDM posterior distributions in the fit to \emph{Planck} 2018 primary CMB anisotropies. As can be anticipated from Fig.~\ref{fig:CMB}, EDE raises $H_0$ by exploiting a degeneracy in $H_0$ and $f_{\rm EDE}$ (left panel); however, this brings with it additional dark matter (right panel). The $H_0-\Omega_{\rm dm} h^2$ degeneracy direction is orthogonal to that in $\Lambda$CDM; the high $H_0$ region of EDE parameter space has ${\cal O}(10\%)$ more dark matter than the range preferred by $\Lambda$CDM (with $H_0 \sim 67$ km/s/Mpc).}
\label{fig:H0fEDEOmegadmh2}
\end{figure}

We begin with an analysis using the \texttt{Plik} likelihood as presented in \cite{Hill:2020osr}. 
Parameter constraints are given in Tab.~\ref{table:params-P18-only} and posterior distributions are shown in Fig. \ref{fig:EDE-planck-triangle}. The results are consistent with intuition developed in Sec.~\ref{sec:EDECMB}, which can be appreciated from Fig. \ref{fig:H0fEDEOmegadmh2}, where we isolate the 2D posteriors in $H_0-f_{\rm EDE}$ and $H_0 - \Omega_{\rm c}h^2$. As per the discussion of Sec.~\ref{sec:EDECMB}, EDE is able to raise $H_0$ by moving along an $H_0-f_{\rm EDE}$ degeneracy in the fit to data. However, this degeneracy direction also raises the amount  of dark matter $\Omega_{c}h^2$ and the spectral index $n_s$, and to a lesser extent the amplitude $A_s$.

\begin{table}[!h]%[htb!]
%\vspace{-5cm}
%\hspace{2.4cm}
\centering
{Constraints from \emph{Planck} 2018 data only (\texttt{Plik} PR3): TT+TE+EE \vspace{4pt}}
\tbl{The marginalized (mean $\pm1\sigma$) and best-fit constraints on the cosmological parameters in $\Lambda$CDM and in the canonical axion-like EDE scenario with $n=3$, as inferred from \emph{Planck} 2018 PR3 primary CMB data only (TT+TE+EE).  Upper and lower limits are quoted at 95\% CL.  Although there is a small contribution to the constraining power in these data from lensing-induced peak-smearing, the constraints are dominated by information content from the epoch of recombination.  The EDE component is not detected here; a two-tailed limit yields $f_\mathrm{EDE} = 0.033^{+0.027}_{-0.026}$ at 68\% CL, i.e., consistent with zero.}
  %\begin{minipage}{\textwidth}
  %\centering
  {  \begin{tabular}{|l|c|c|c|c|}
    \hline\hline Parameter &$\Lambda$CDM Best-Fit~~&$\Lambda$CDM Marg.~~&~~~EDE ($n=3$) Best-Fit~~~&~~~EDE ($n=3$) Marg.\\ \hline \hline

{\boldmath$\ln(10^{10} A_\mathrm{s})$} & $3.055$ & $3.044\pm 0.016$ & $3.056$ & $3.051\pm 0.017$\\

{\boldmath$n_\mathrm{s}$} & $0.9659$ & $0.9645\pm 0.0043$ & $0.9769$ & $0.9702^{+0.0053}_{-0.0083}$\\

{\boldmath$100\theta_\mathrm{s}$} & $1.04200$ & $1.04185\pm 0.00029$ & $1.04168$ & $1.04164\pm 0.00034$\\

{\boldmath$\Omega_\mathrm{b} h^2$} & $0.02244$ & $0.02235\pm 0.00015$ & $0.02250$ & $0.02250^{+0.00018}_{-0.00022}$\\

{\boldmath$\Omega_\mathrm{c} h^2$} & $0.1201$ & $0.1202\pm 0.0013$ & $0.1268$ & $0.1234^{+0.0019}_{-0.0038}$\\

{\boldmath$\tau_\mathrm{reio}$} & $0.0587$ & $0.0541\pm 0.0076$ & $0.0539$ & $0.0549\pm 0.0078$\\

{\boldmath$\mathrm{log}_{10}(z_c)$} & $-$ & $-$ & $3.75$ & $3.66^{+0.24}_{-0.28}$\\

{\boldmath$f_\mathrm{EDE} $} & $-$ & $-$ & $0.068$ & $< 0.087$\\

{\boldmath$\theta_i$} & $-$ & $-$ & $2.96$ & $> 0.36$\\

    \hline

$H_0 \, [\mathrm{km/s/Mpc}] $ & $67.44$ & $67.29\pm 0.59$ & $69.13$ & $68.29^{+0.73}_{-1.2}$\\

$\Omega_\mathrm{m}         $ & $0.3147$ & $0.3162\pm 0.0083$ & $0.3138$ & $0.3145\pm 0.0086$\\

$\sigma_8                  $ & $0.8156$ & $0.8114\pm 0.0073$ & $0.8280$ & $0.8198^{+0.0090}_{-0.012}$\\

$S_8$                        & $0.8355$ & $0.8331\pm 0.0159$ & $0.8468$ & $0.8393\pm 0.0175$\\

$\mathrm{log}_{10}(f/{\mathrm{eV}})$ & $-$ & $-$ & $26.36$ & $26.57^{+0.26}_{-0.46}     $\\

$\mathrm{log}_{10}(m/{\mathrm{eV}})$ & $-$ & $-$ & $-26.90$ & $-26.94^{+0.39}_{-0.65}    $\\

    \hline
  \end{tabular} 
  
  \label{table:params-P18-only}}
  %\end{minipage}
\end{table}

%\begin{center}
\begin{table}[t!]
\centering
%\scalebox{0.9}{
{$\chi^2$ statistics from {\it Planck} 2018  data (\texttt{Plik} PR3) only: TT+TE+EE \vspace{4pt}}
\tbl{$\chi^2$ values for the best-fit $\Lambda$CDM and EDE models to the primary CMB alone. The reduction in $\chi^2$ is 4.1 for the three-parameter EDE extension of $\Lambda$CDM.}
  {\begin{tabular}{|l|c|c|}
    \hline\hline
    data sets &~~$\Lambda$CDM~~&~~~EDE ~~~\\ \hline \hline
    %CMB TT, EE, TE & &\\
         \textit{Planck} 2018 low-$\ell$ TT & 23.4 & 22.1\\
        \textit{Planck} 2018 low-$\ell$ EE &397.2 & 396.0\\
    \textit{Planck} 2018 high-$\ell$ TT+TE+EE &  2344.9   &  2343.3 \\

    \hline
    Total $\chi^2 $   & 2765.5 & 2761.4\\
     $\Delta \chi^2 $ &  & -4.1 \\ 
    \hline
  \end{tabular}
  %}
  \label{table:chi2_CMB_alone}}
\end{table}
%\end{center}

% \begin{figure}[h!]
% \centering
% \includegraphics[width=0.9\textwidth]{EDE-default-small-Planck.pdf}
% \caption{EDE small triangle}
% \label{fig:EDE-planck-triangle}
% \end{figure}

% \begin{figure}[h!]
% \centering
% \includegraphics[width=0.9\textwidth]{fEDE-row.pdf}
% \caption{EDE row}
% \label{fig:EDE-row}
% \end{figure}

\subsection{Dependence on {\it Planck} Likelihood: \texttt{Plik}, \texttt{CamSpec}, and \texttt{HiLLiPoP} }

To extend this analysis one step further, we investigate constraints on EDE
resulting from alternative choices of the \emph{Planck} likelihood. In particular, we discuss
constraints using the \texttt{CamSpec} likelihood~\cite{efstathiou2021,rosenberg2022}, which uses the
\emph{Planck} 100, 143, and 217 GHz channels in TE and EE, excluding the 100 GHz data from the TT data set, as it
requires a more complex foreground modelling. A dust cleaning step is performed on
the power spectra and residual temperature foregrounds are modelled with a power law
with an amplitude which depends on the cross-frequency. It also includes calibration and
polarization efficiencies (for TE and EE). We also consider the \texttt{Hillipop} likelihood,
described in Ref.~\cite{couchot2017}, containing 100, 143, and 217  GHz spectra with a foreground modelling at the
power spectra level such as \texttt{CamSpec}, but using physically motivated templates to model
Galactic dust, cosmic infrared background (CIB), thermal Sunyaev-Zel'dovich (tSZ) and kinetic Sunyaev-Zel'dovich (kSZ), tSZ-CIB correlations and Poisson-like distributed point sources. While we
use \texttt{Hillipop} only with PR4 data, we use the \texttt{CamSpec} likelihood to
obtain constraints from both the PR3 and PR4 data sets. The PR4 (NPIPE) data set consists in a joint-analysis
of Low Frequency Instrument (LFI) and High Frequency Instrument (HFI) data, including some additional data from re-pointing maneuvers and providing detector splits
instead of half-mission splits to lower the correlation between the two splits. A more detailed description
of the PR4 data processing may be found in Refs.~\cite{planck2020LVII, rosenberg2022}. Cosmological constraints
on $\Lambda$CDM derived from these alternative likelihoods have been shown to be consistent with the \texttt{Plik}
likelihood constraints both for PR3 and PR4 data.  To our knowledge, the analysis presented here contains the first \emph{Planck} PR4-based constraints on EDE.

\begin{figure}[t!]
    \centering
    \includegraphics[width=\textwidth]{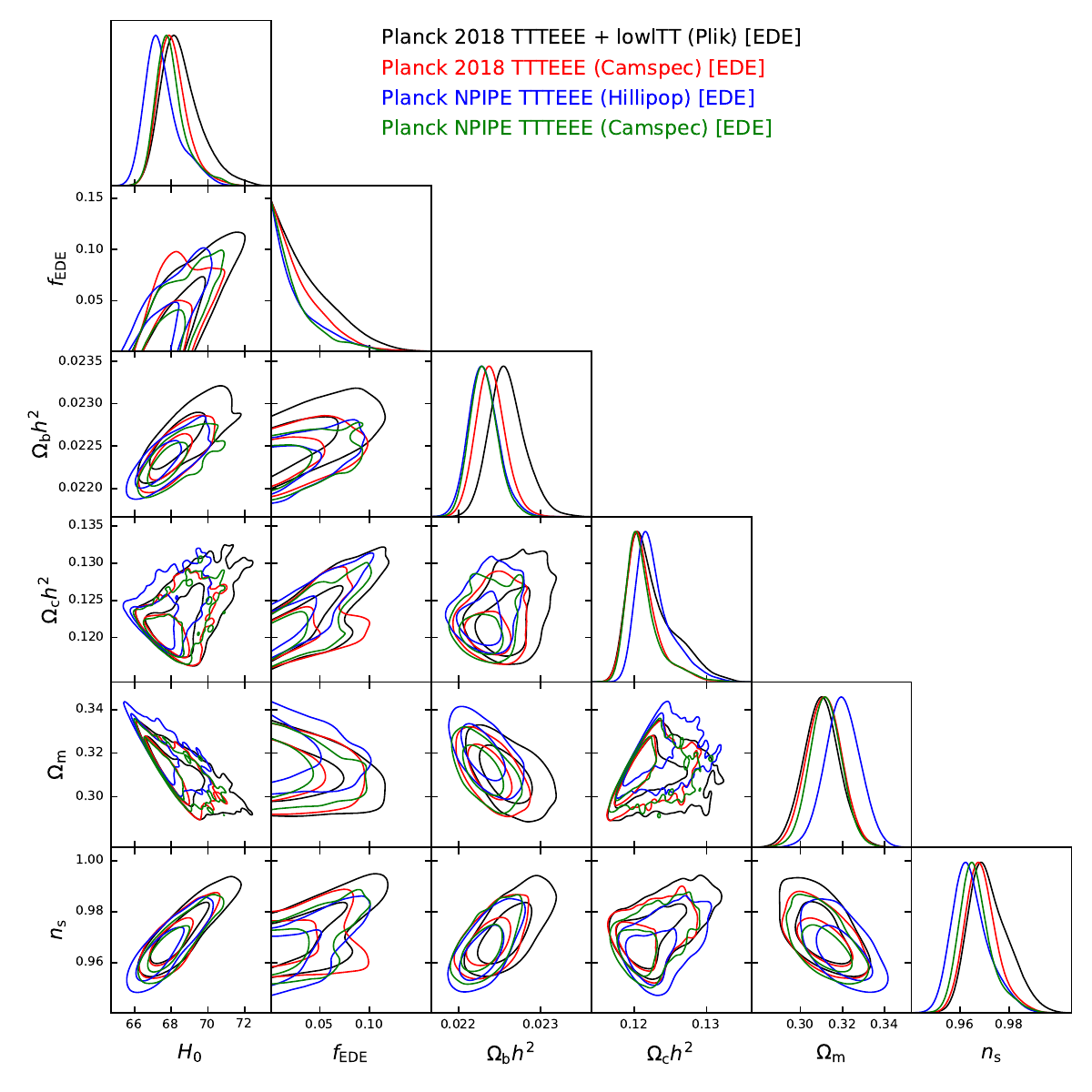}
    \caption{Constraints on EDE from \emph{Planck} CMB data analysed using the \texttt{Plik} PR3 likelihood (black), \texttt{CAMSPEC} PR3 likelihood (red), Hillipop PR4 likelihood (blue), and CAMSPEC PR4 likelihood (green).  We find that the likelihood that has been used in all previous EDE analyses, {\tt Plik} PR3, yields the {\it weakest} constraints on the EDE scenario, with both PR4 (NPIPE) likelihoods yielding notably tighter upper bounds.  }
    \label{fig:planck_ede_comp}
\end{figure}

Figure~\ref{fig:planck_ede_comp} shows the marginalized posterior distributions derived from the \texttt{Plik} likelihood for \emph{Planck} PR3 data, the \texttt{CamSpec} likelihood for both PR3 and PR4 data, and the \texttt{HiLLiPoP} likelihood for PR4 data. As in Ref.~\cite{laposta2022}, we use a wider prior on $\mathrm{log}_{10}(z_c)$ than considered in some previous EDE analyses. The marginalized constraints and best-fit parameter values can be found in Table~\ref{tab:planck_ede_comp}. We used the Gelman-Rubin criterion $R-1 \le 0.03$ to assess the convergence of the chains.

Notice that the overall conclusion regarding EDE does not change when we use the different \emph{Planck} likelihoods. However, we find that the \texttt{Plik} likelihood is the one that imposes the \textit{weakest} constraint on $f_{\rm EDE}$. While this analysis has yielded $f_\mathrm{EDE} < 0.0947\;(95\%~\mathrm{C.L.})$ using the \texttt{Plik} PR3 likelihood, the other constraints are tighter with $f_\mathrm{EDE} < 0.0788\;(95\%~\mathrm{C.L.})$, $f_\mathrm{EDE}<0.0776\;(95\%~\mathrm{C.L.})$ and $f_\mathrm{EDE} < 0.0766\;(95\%~\mathrm{C.L.})$ for \texttt{CamSpec} (PR3), \texttt{CamSpec} (PR4), and \texttt{HiLLiPoP} (PR4), respectively. The maximum-likelihood $f_\mathrm{EDE}=0.0730$ derived from the \texttt{Plik} likelihood is also significantly higher than the best-fit derived from the \texttt{CamSpec} (PR3) likelihood ($f_\mathrm{EDE} = 0.0349$), the {\tt CamSpec} PR4 likelihood ($f_\mathrm{EDE} = 0.0425$), or the {\tt HiLLiPoP} PR4 likelihood ($f_\mathrm{EDE} = 0.0474$). One should note that even if our marginalized constraints are affected with `prior volume effects' due to the broad prior imposed on $\mathrm{log}_{10}(z_c)$, best-fit values are not affected by this type of sampling effect.

\begin{table}[h!]
%\centering
\tbl{Constraints on EDE for various \emph{Planck} likelihoods. Here a wider $z_c$ prior is used than in the previous \emph{Planck} analyses, resulting in weaker upper bounds on $f_{\rm EDE}$ (cf.~Table~\ref{table:params-P18-only}, which uses the {\tt Plik} PR3 likelihood); this arises solely because large amounts of EDE are allowed for very high values of $z_c$, as the EDE field has no discernible impact on cosmological observables in this case. It is noteworthy that the likelihood used in all previous EDE studies ({\tt Plik} PR3) yields the weakest upper bound on $f_{\rm EDE}$ amongst the four likelihoods considered here.}
%\tbl{caption...}
%\centering
%
{\begin{tabular}{|l|c|c|}% \toprule
\hline
\textbf{Parameter} & \texttt{Plik} (PR3) & \texttt{CamSpec} (PR3)\\
\hline\hline
$H_0$ [km/s/Mpc] & $68.55^{+0.75}_{-1.3}$ ($69.88$) & $68.10^{+0.60}_{-0.94}$ ($68.67$)\\
$\Omega_bh^2$ & $0.02260^{+0.00018}_{-0.00024}$ ($0.02279$) & $0.02239^{+0.00016}_{-0.00019}$ ($0.02241$)\\
$\Omega_ch^2$ & $0.1224^{+0.0017}_{-0.0041}$ ($0.1259$) & $0.1213^{+0.0013}_{-0.0027}$ ($0.1226$)\\
$10^9 A_s$ & $2.204 \pm 0.059$ ($2.245$) & $2.188 \pm 0.061$ ($2.189$)\\
$n_s$ & $0.9719^{+0.0062}_{-0.0096}$ ($0.9857$) & $0.9687^{+0.0053}_{-0.0072}$ ($0.9726$)\\
$f_\mathrm{EDE}$ & $< 0.0947$ ($0.0730$) & $< 0.0788$ ($0.0349$)\\
\hline\hline
\textbf{Parameter} & \texttt{CamSpec} (PR4) & \texttt{HiLLiPoP} (PR4)\\
\hline\hline
$H_0$ [km/s/Mpc] & $68.00^{+0.50}_{-0.98}$ ($68.60$) & $67.51^{+0.57}_{-1.1}$ ($68.15$)\\
$\Omega_bh^2$ & $0.02231^{+0.00015}_{-0.00018}$ ($0.02233$) & $0.02231^{+0.00016}_{-0.00020}$ ($0.02245$)\\
$\Omega_ch^2$ & $0.1212^{+0.0011}_{-0.0027}$ ($0.1232$) & $0.1228^{+0.0013}_{-0.0032}$ ($0.1251$)\\
$10^9 A_s$ & $2.165 \pm 0.059$ ($2.137$) & $2.137 \pm 0.058$ ($2.135$)\\
$n_s$ & $0.9664^{+0.0047}_{-0.0076}$ ($0.9728$) & $0.9645^{+0.0054}_{0.0088}$ ($0.9722$)\\
$f_\mathrm{EDE}$ & $< 0.0776$ ($0.0425$) & $< 0.0766$ ($0.0474$)\\
\hline
\end{tabular}
\label{tab:planck_ede_comp}}
\end{table}

%%%%%%%%%%%%%%%%%%%%%%%%%%%%%%%%%%%%%%%%%%%%%%%%%%%%%%%%%%%%%%%%%%%%%%
%%%%%%%%%%%%%%%%%%%%%%%%%%%%%%%%%%%%%%%%%%%%%%%%%%%%%%%%%%%%%%%%%%%%%%
%\clearpage
\section{EDE Meets LSS}
\label{sec:EDExLSS}
%%%%%%%%%%%%%%%%%%%%%%%%%%%%%%%%%%%%%%%%%%%%%%%%%%%%%%%%%%%%%%%%%%%%%%
%%%%%%%%%%%%%%%%%%%%%%%%%%%%%%%%%%%%%%%%%%%%%%%%%%%%%%%%%%%%%%%%%%%%%%

As discussed in the preceding section, Early Dark Energy can successfully accommodate a larger $H_0$ value while maintaining consistency with \emph{Planck} CMB data, although the CMB data on their own do not prefer the existence of this component, with the newest \emph{Planck} likelihoods in fact yielding tighter constraints on the model than those used in previous work. The EDE fit to the CMB is achieved by a delicate balance of the new EDE component and compensating shifts in $\Lambda$CDM parameters, and in particular increases to the physical dark matter density $\omega_{\rm cdm}$, the spectral index $n_s$, and scalar amplitude $A_s$. While the EDE component rapidly decays after the critical redshift $z_c$, the influence of these $\Lambda$CDM parameter shifts on density perturbations, and the subsequent growth of structure, is imprinted across cosmic time.  The critical change comes from the non-negligible contribution of the EDE scalar to the cosmic energy budget at recombination. Unlike perturbations in the matter component, the EDE perturbations are oscillating leading up to last scattering and (leaving $\Lambda$CDM parameters unchanged) results in a suppression of the Weyl potential and also manifests in the CMB as an EDE contribution to the early-ISW effect. In fitting to early-Universe data sets, namely the CMB, the aforementioned $\Lambda$CDM parameter changes are typically able to restore CMB power spectrum to fit data about as well as the concordance model. Given this comes at the cost of a significant increase in the physical dark matter density and that the cosmology hereafter is largely unchanged (canonical EDE redshifts much faster than radiation), one would anticipate possible issues in the late Universe with structure formation.

\subsection{The Matter Power Spectrum}

\begin{figure}[ht]
    \centering
    \includegraphics[width=.49\linewidth]{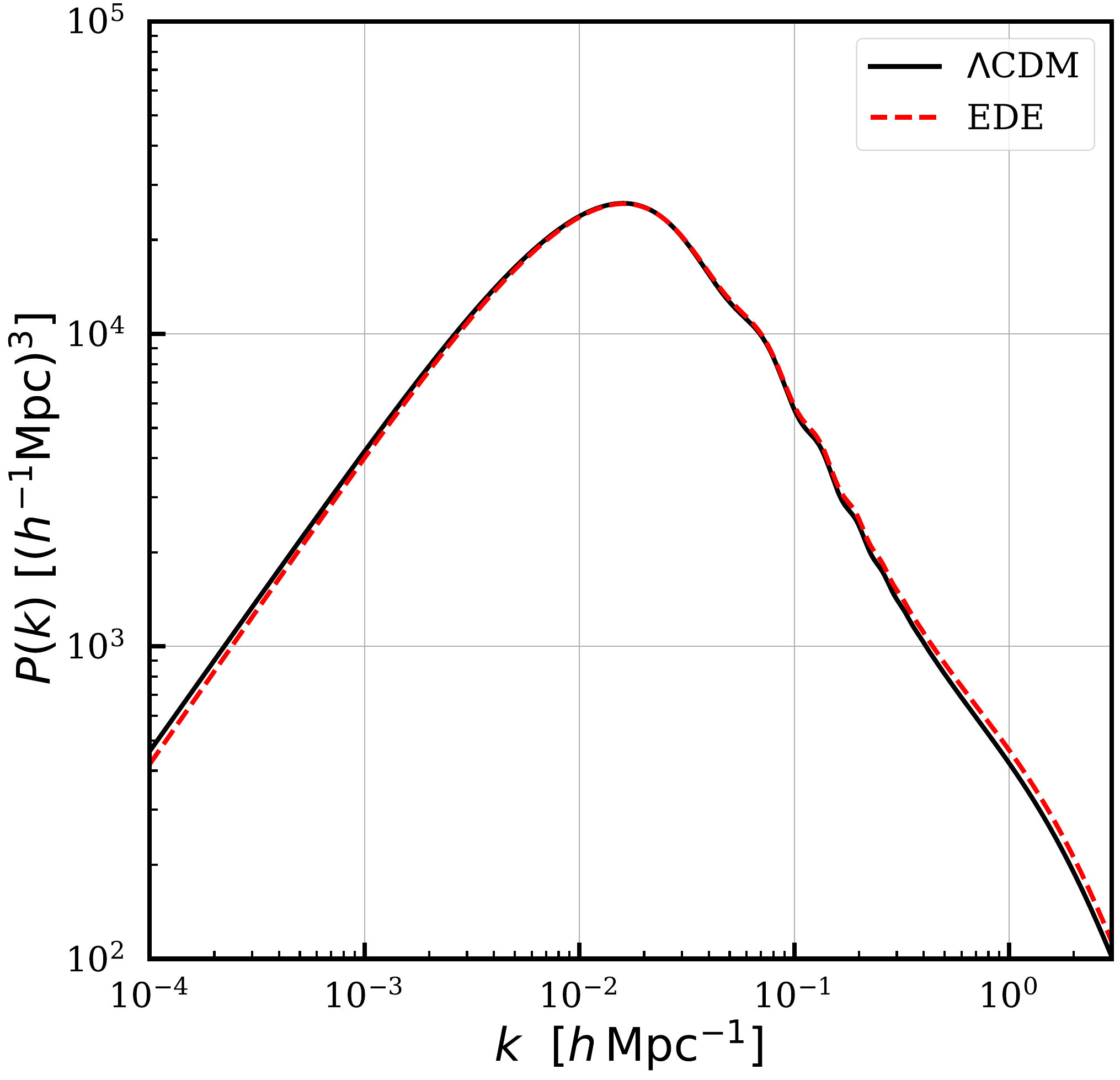}
        \includegraphics[width=0.49\linewidth]{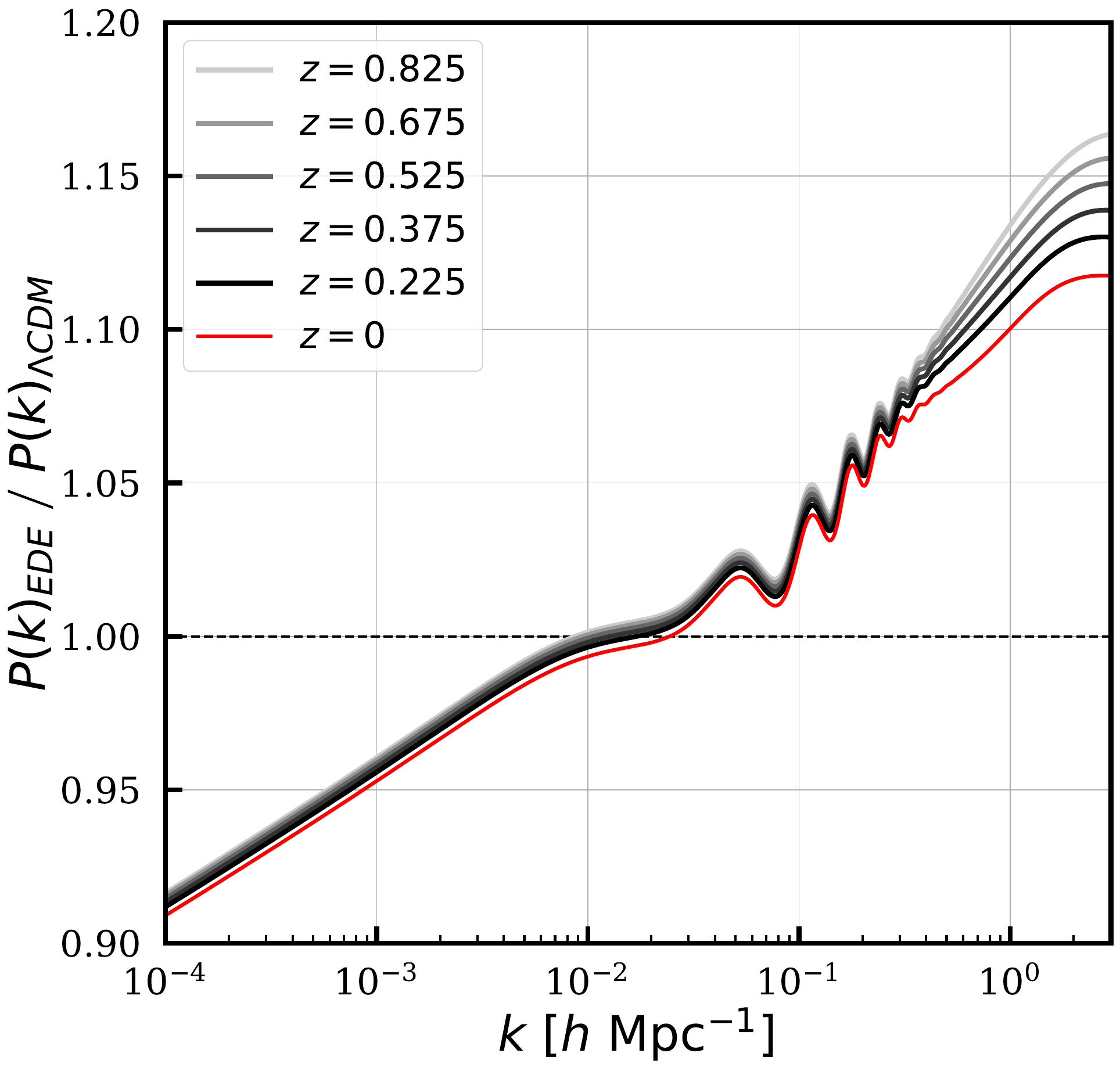}
    \caption{ The non-linear power spectrum $P(k)$ at $z=0$ for EDE and $\Lambda$CDM models that fit the primary CMB, distances, and SH0ES data (left panel).  The change in $\sigma_8$ for the canonical EDE model can be seen here as the relative enhancement to $P(k)$ in the range $0.1 \, h/{\rm Mpc} \lesssim k \lesssim 1 \, h/{\rm Mpc}$ (although $\sigma_8$ is strictly speaking computed from the linear power spectrum). We also show the ratio of the EDE and $\Lambda$CDM non-linear P(k) (right panel) at redshifts chosen to be at the midpoints of bins used in the DES-Y1 analysis (and also $z=0$ in red).  The model parameters used are the same as in Fig.~\ref{fig:CMB_TT}.}
    \label{fig:Pk}
\end{figure}

\begin{figure}[ht]
    \centering
    \includegraphics[width=.49\linewidth]{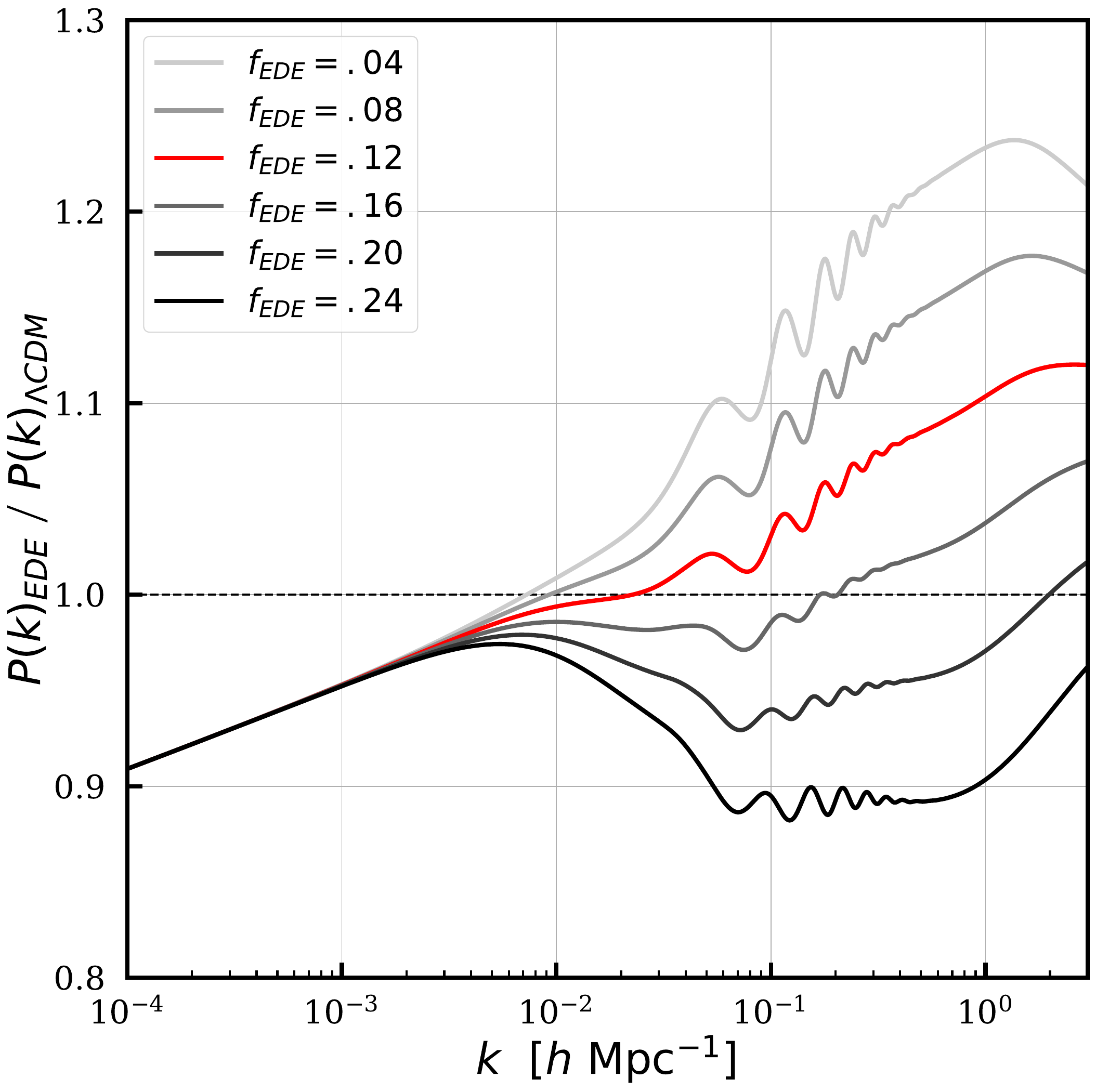}
    \includegraphics[width=.49\linewidth]{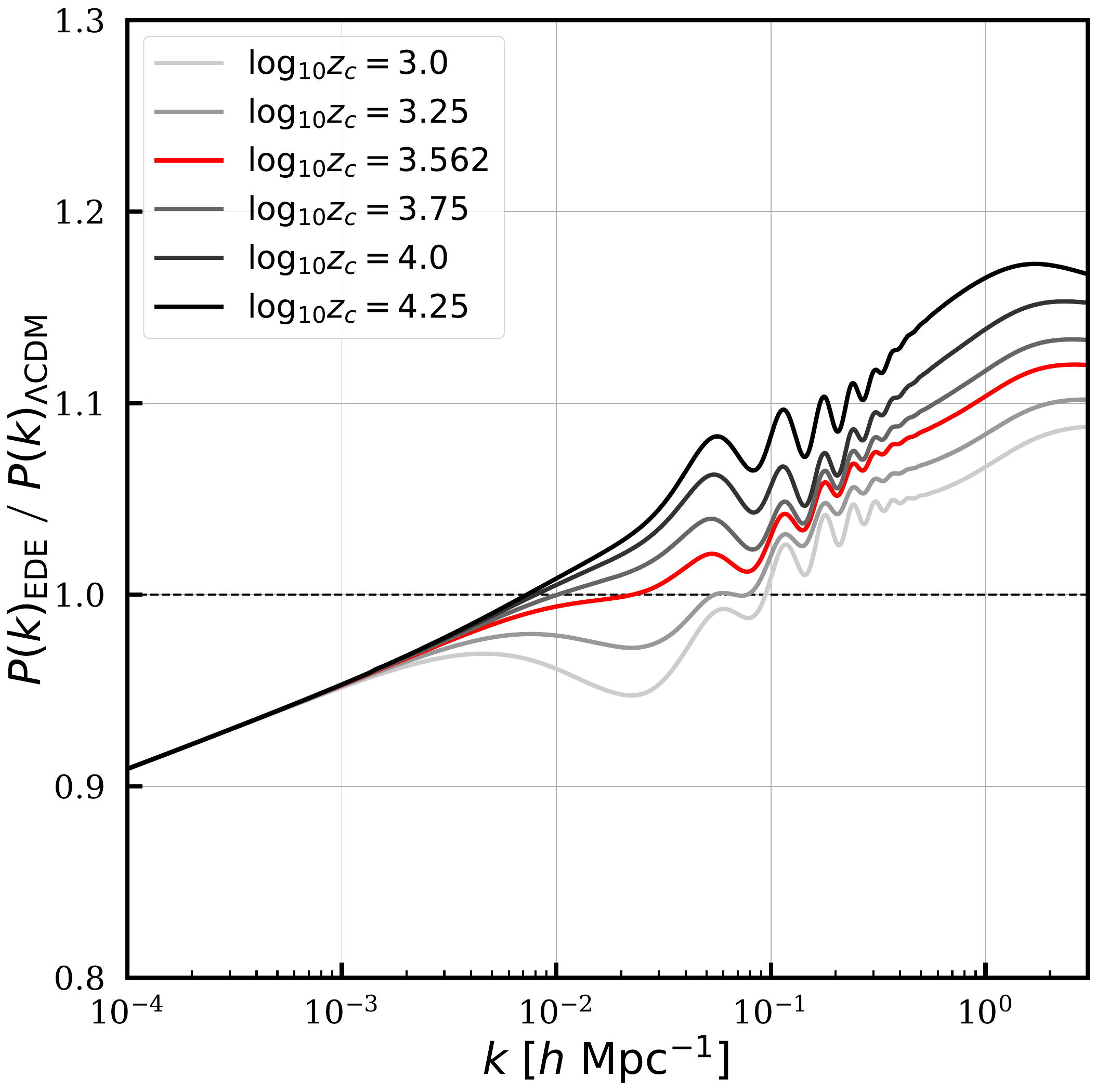}
    \caption{The ratio of $P(k)$ in EDE relative to $\Lambda$CDM as a function of $f_{\rm EDE}$, at fixed $\log_{\rm 10}z_c=3.526$ and $\theta_i = 2.83$ (left panel). 
    When $f_{\rm EDE}$ is increased, the growth of perturbations inside the horizon during the EDE phase is further suppressed.  The red curve depicted here is the same as that shown in Fig.~\ref{fig:Pk}. We also show the ratio of $P(k)$ in EDE to $\Lambda$CDM as a function of $\log_{10}z_{c}$, at fixed $f_{\rm EDE}=0.12$ and $\theta_i=2.83$ (right panel).  
    As critical redshift, $z_c$, is decreased, the suppression of growth due to EDE is pushed to correspondingly later times, and affecting modes on progressively larger scales (lower $k$).}
    \label{fig:fEDEPk}
\end{figure}

The matter power spectrum for EDE and and $\Lambda$CDM is shown in Fig.~\ref{fig:Pk}, where we show the matter power spectrum $P(k)$ in both models and the ratio of the two at varying redshift. The power spectrum in EDE is characterized by an excess of power on small scales $k \gtrsim k_c$, where $k_c$ is the wavenumber of the mode that entered the horizon at the time $z_c$.
The excess is the net result of a competition between an EDE-induced suppression of power and a $\Lambda$CDM-parameter-induced excess. The impact of EDE can be appreciated from Fig.~\ref{fig:fEDEPk}, where we show the imprint of individual shifts in $f_{\rm EDE}$ and $z_c$, with $f_{\rm EDE}$ increased from $0.04$ to $0.24$ (top-to-bottom) and $z_c$ decreased from $4.25$ to $3.0$ (top to bottom). From the left panel of Fig.~\ref{fig:fEDEPk}, one may clearly appreciate that an increase in $f_{\rm EDE}$ causes a dramatic suppression of $P(k)$, e.g., the shift $f_{\rm EDE} =0.04$ to $f_{\rm EDE}=0.24$ converts a $>20\%$ excess in power at $k=1 \, h \, {\rm Mpc}^{-1}$ to a $\sim 10\%$ {\it deficit}. This predominantly impacts modes that are well inside the horizon when the EDE component is active, as can be seen in the right panel, wherein smaller $z_c$ shifts the scale at which EDE-induced changes arise to lower wavenumbers, which enter the horizon later.

The EDE-induced suppression in $P(k)$ is counteracted by both an excess at high-$k$ in the initial conditions for perturbations, namely in the primordial power spectrum as encoded by $n_s$ and $A_s$, and an enhanced growth of structure throughout cosmic time.  The former should not be underestimated; indeed, even just the shift in $n_s$ generates a $6\%$ excess in power at $k=1 h {\rm Mpc}^{-1}$, i.e.,
\begin{equation}
    \frac{{\cal P}_{\cal R}(k=1 h {\rm Mpc}^{-1}, n_s=0.9889)}{{\cal P}_{\cal R}(k=1 h {\rm Mpc}^{-1}, n_s=0.9686)} = 1.06
\end{equation}
where ${\cal P}_{\cal R}(k)$ denotes the primordial power spectrum of comoving curvature perturbations ${\cal R}$. 

It is conventional to parameterize the density perturbations via a small handful of numbers. In particular, the amplitude of perturbations can be characterized by $\sigma_8$, the RMS linear-theory mass fluctuation in a sphere of radius $8 \, {\rm Mpc}/h$ at $z = 0$. This is evaluated in Fourier space as,
\be
\label{sigma8}
(\sigma_8 )^2 = \frac{1}{2 \pi^2}\int {\rm d}\log k \, W^2(kR) \, k^3 P(k) \, .
\ee
where $W(kR)$ is a spherical top-hat filter  of radius $R = 8 \, {\rm Mpc}/h$. In EDE models that fit the CMB, $\sigma_8$ is $\sim 2\%$ larger than in $\Lambda$CDM. This holds across redshift. 

Meanwhile, on the observational side, the parameter that is most well constrained by LSS experiments is $S_8$, namely,
\begin{equation}
    S_8 \equiv \sigma_8 \left( \frac{\Omega_M}{0.3}\right)^{0.5}.
\end{equation}
In the fit to $\Lambda$CDM, the $S_8$ value inferred from CMB experiments is  often larger than that inferred from LSS experiments. For example,  
the {\it Planck} 2018 analysis finds $S_8=0.830 \pm 0.013$ \cite{Aghanim:2018eyx}, while DES-Y3 finds 
$S_8 = 0.776 \pm 0.017$ \cite{DES:2021wwk}, in $2.7\sigma$ tension with the {\it Planck} 2018 CMB result.

\subsection{Anisotropic Galaxy Clustering}

Another sensitive probe of EDE is the anisotropic galaxy power spectrum, which probes the parameter combination $f\sigma_8$, where $f$ is the \textit{logarithmic} growth rate and is defined in terms of the \textit{linear} growth factor $D$, 
\begin{equation}
    f = \frac{{\rm d} \ln{D}}{{\rm d} \ln{a}}.
\end{equation}
Critically, $f$ controls the linear-theory prediction for matter velocity perturbations, $\theta = - f \delta_m$, where $\theta$ is the velocity divergence. The shape of the power spectrum constructed from their perturbations serves as a complementary probe to the CMB; in particular the shape has been shown to be sensitive to the physical dark matter density \cite{Chudaykin:2019ock,DAmico:2019fhj,Ivanov:2019pdj}. In practice, $f\sigma_8$ is probed from measurements of redshift space distortions (RSD).

To really leverage the full power of LSS data sets one has to study the non-linear regime, specifically the non-linear matter power spectrum. However, this is complicated by the non-linear evolution of the matter density field and its connection to luminous tracers like galaxies but is further complicated by baryonic effects. To overcome these challenges one can resort to numerical N-body simulations or alternatively perturbation theory. N-body simulations are advantageous in that it is possible, at least in principle, to model \textit{all} scales. The downside is that it is computationally expensive to run such simulations and one is actually still affected by uncertainties in the modeling of galaxy formation. A typical work around is to construct fitting functions for the non-linear regime (e.g., Halofit~\cite{Takahashi:2012em}) and phenomenologically model the galaxy bias. The downside of the this work-around is that one cannot necessarily trust these fitting functions if there are deviations from $\Lambda$CDM. In some cases, particularly when the new physics affecting LSS is confined to small changes to the high-$z$ linear $P(k)$, it can be argued that the use of Halofit is justified, such as for EDE where the posterior obtained with Halofit is consistent with runs done replacing LSS with an $S_8$ prior (which is a linear-theory quantity).

An alternative to N-body simulations is to tackle the problem of galaxy clustering using the Effective Field Theory of Large-Scale Structure (EFT of LSS)~\cite{Baumann:2010tm,Carrasco:2012cv,Cabass:2022avo,Ivanov:2022mrd}, which serves as a consistent framework to undertake non-linear perturbation theory in the mildly nonlinear regime. While limited to a small window of wavenumbers, $k \lesssim 0.5~h/$Mpc, EFT provides unmatched accuracy in describing clustering on these scales. In the context of spectroscopic surveys, perturbative techniques are useful for reconstructing the 3D distribution of matter. With EFT, all effects contributing to the observed map of LSS are included, including non-linearities in the matter fields, galaxy bias~\cite{Desjacques:2016bnm}, baryonic feedback~\cite{Lewandowski:2014rca}, Fingers-of-God~\cite{Jackson:1971sky}, non-linear redshift space distortions~\cite{Senatore:2014vja,Ivanov:2018gjr,Vlah:2018ygt,Chen:2020zjt}, 
IR resummation for the BAO~\cite{Senatore:2014via,Baldauf:2015xfa,Vlah:2015zda,Blas:2015qsi,Blas:2016sfa,Ivanov:2018gjr,Vasudevan:2019ewf}, etc. The EFT of LSS provides a general large-scale construction of a cosmological fluid which obeys the equivalence principle and has rotational symmetry. Since EDE still possesses these symmetries we are perfectly safe to apply EFT to this extension of $\Lambda$CDM cosmology.

\begin{figure}[h!]
    \centering
    \includegraphics[width=0.49\textwidth]{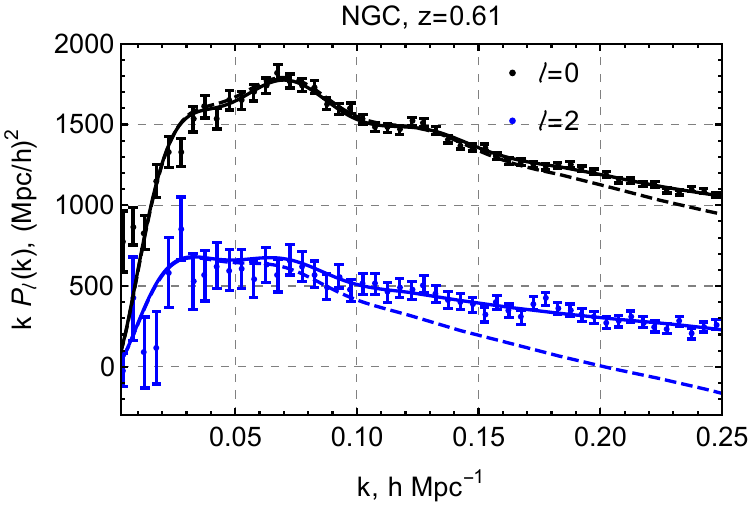}
    \includegraphics[width=0.49\textwidth]{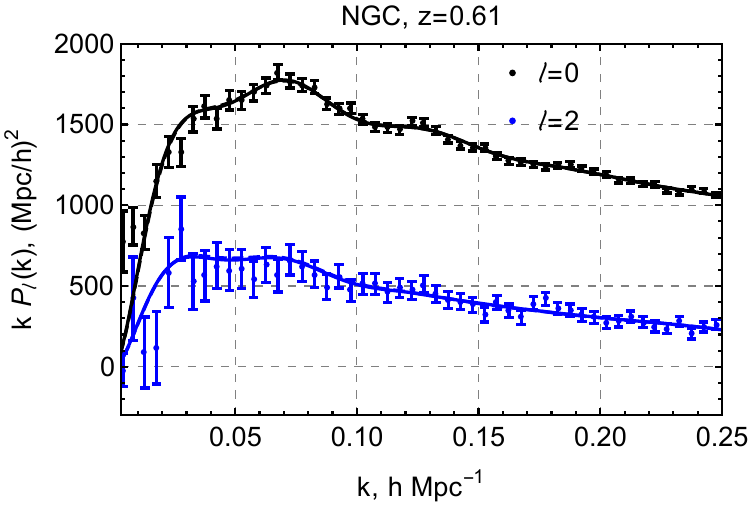}
        \caption{Galaxy power spectrum multipoles for $z=0.61$, before (left panel) and after
    (right panel) the nuisance parameters have been marginalized over, including the high-$z$ NGC BOSS data.
    Results for $\Lambda$CDM are depicted with solid curves, while the EDE model is shown with dashed curves. The right panel (having marginalized over nuisance parameters) shows the curves cannot be distinguished; the fractional difference between these curves is depicted in Fig.~\ref{fig:multipoles}.     }
    \label{fig:spectra}
\end{figure}

\begin{figure}[h!]
    \centering
        \includegraphics[width=0.49\textwidth]{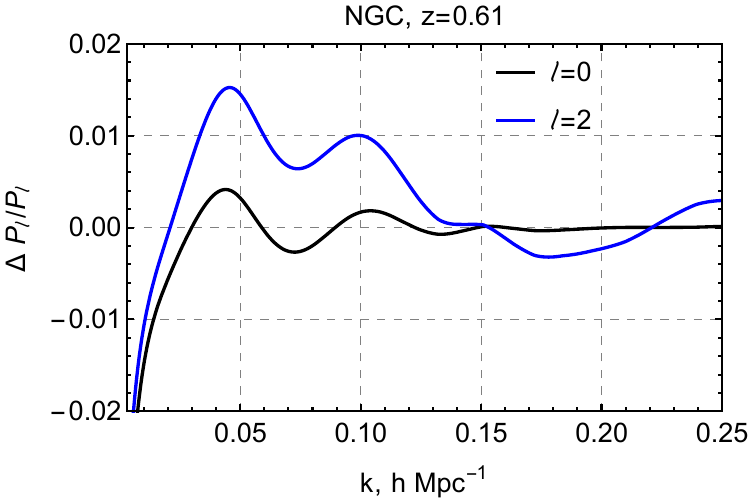}
    \includegraphics[width=0.49\textwidth]{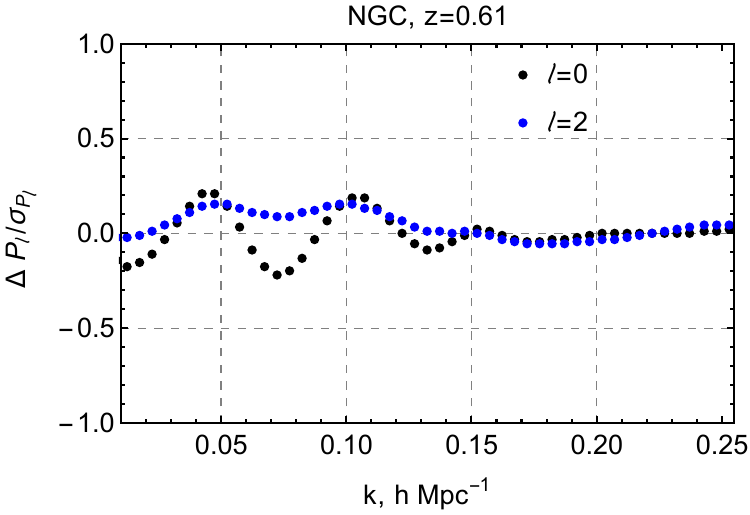}
    \caption{Galaxy power spectrum multipoles for $z=0.61$, where we have marginalized over nuisance parameters as in the right panel of Fig.~\ref{fig:spectra}. {\it Left panel}: The fractional change between EDE and $\Lambda$CDM: $\Delta P/P\equiv(P^{\rm EDE}- P^{\Lambda{\rm CDM}})/P^{\Lambda{\rm CDM}}$.  The monopole has a noticeable $0.3\%$ pattern produced by a mismatch in the shape 
    % and location 
    for the BAO wiggles in the two models. On the other hand, the quadrupole exhibits a $\mathcal{O}(2\%)$ fractional change at low $k$. {\it Right panel}: Fractional change in units of the BOSS error bar for each wavenumber bin: $\Delta P/\sigma_P$. (Keep in mind that neighboring $k$ bins are correlated). The largest discrepancy is observed in the position and shape of the BAO wiggles in the monopole.}
 \label{fig:multipoles}
\end{figure}

In this work we utilize the one-loop EFT of LSS model as laid out in \cite{Ivanov:2019pdj,Chudaykin:2020aoj}. The multipole moments for the redshift-space galaxy power spectrum are given by the one-loop perturbation theory. Collectively, the power spectrum can be written as,
\begin{equation}
    P_{g,\ell} (k) = P^{\rm tree}_{g,\ell} (k) + P^{\rm 1-loop}_{g,\ell} (k) + P^{\rm noise}_{g,\ell} (k) + P^{\rm ctr}_{g,\ell} (k).
\end{equation}
The tree-level piece is simply the linear-theory result, so in this case the Kaiser formula \cite{10.1093/mnras/227.1.1},
\begin{equation}
    P^{\rm tree}_{g} (k,\mu) = (b_1 + f\mu^2)^2 P_{\rm lin}(k),
\end{equation}
where $b_1$ is a scale-independent linear bias and $\mu \equiv \hat{k}\cdot \hat{z}$ is the angle between the Fourier modes and line-of-sight direction $\hat{z}$. The other contributions include the one-loop corrections containing information of the non-linear evolution, stochastic contributions typically modeled as Poisson shot noise, and the last pieces are ultraviolet (UV) counterterms (see, e.g.,~\cite{Ivanov:2019pdj,DAmico:2019fhj,Chudaykin:2020aoj} for more details on the EFT of LSS). In Fig.~\ref{fig:spectra} we show the monopole and quadrupole of the galaxy power spectrum with BOSS data at $z=0.61$ in $\Lambda$CDM and for EDE before and after marginalization over nuisance parameters. We further show the difference between both models in Fig.~\ref{fig:multipoles}. In the first case, the nuisance parameters are kept the same as $\Lambda$CDM, which results in an appreciable deviation between the two models. However, the difference can be fully absorbed by the nuisance parameters, making for a much smaller disagreement between the power spectrum -- changing them by $\sim$ 10$\%$ which is smaller than current precision for these parameters from data or N-body simulations. Focusing on the fractional difference its easy to see the roughly $0.3\%$ deviations in the monopole due to a slight offset in the location of the BAO wiggles. This feature is even more prominent in the quadrupole where it is $\sim2\%$ for small wavenumbers. The deviation in the BAO wiggles is easy to understand physically as it stems directly from the change in the physical dark matter density \cite{Ivanov:2019pdj}.

\begin{figure}[h!]
    \centering
        \includegraphics[width=0.49\textwidth]{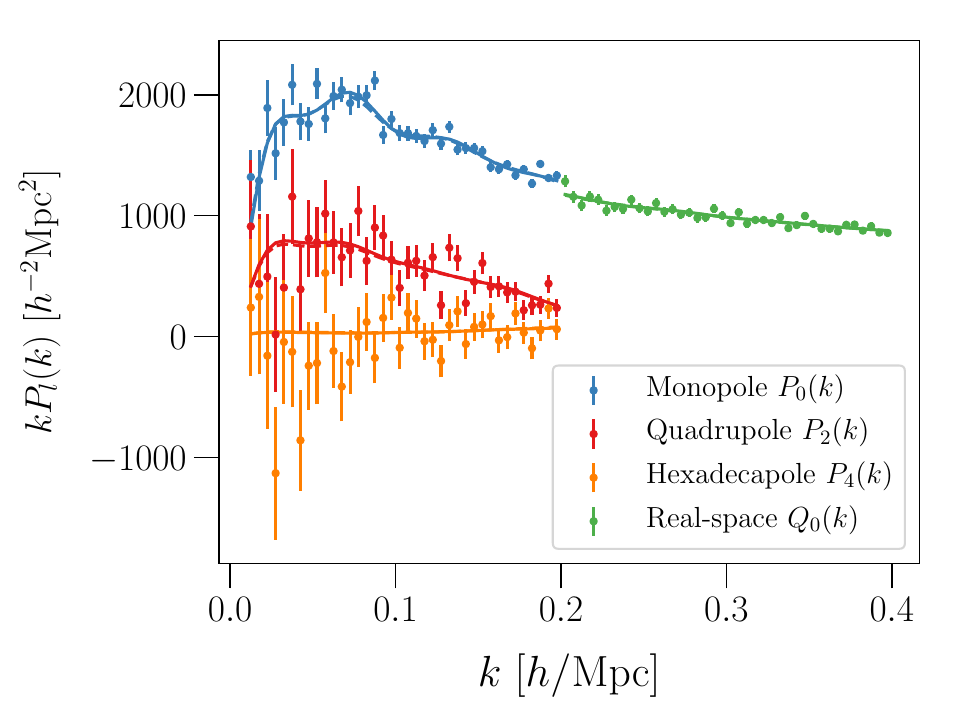}
    \includegraphics[width=0.49\textwidth]{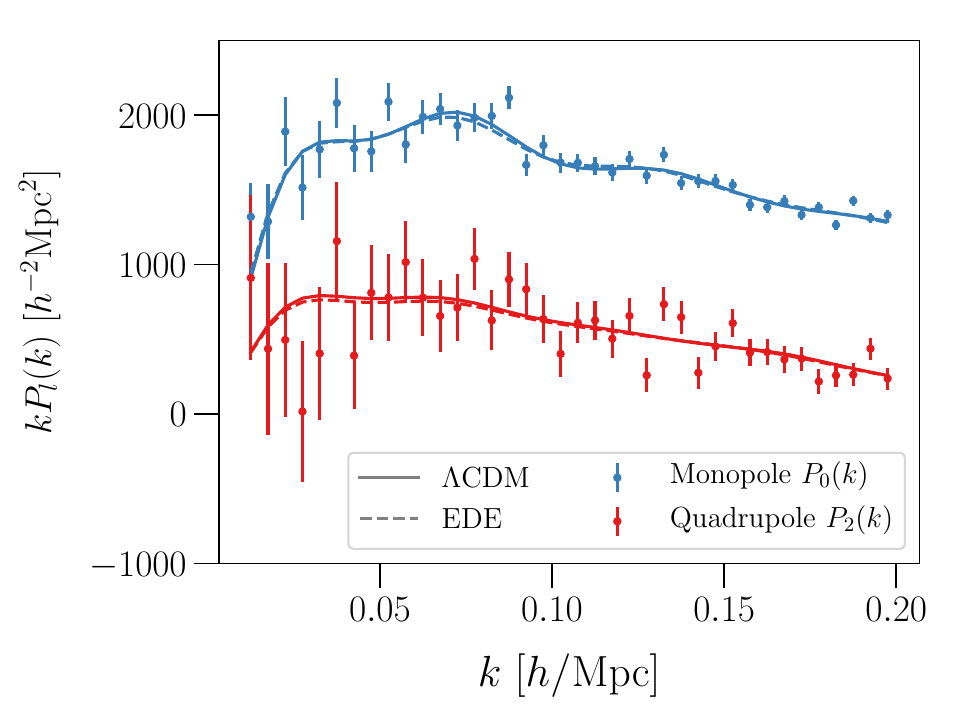}
    \includegraphics[width=0.49\textwidth]{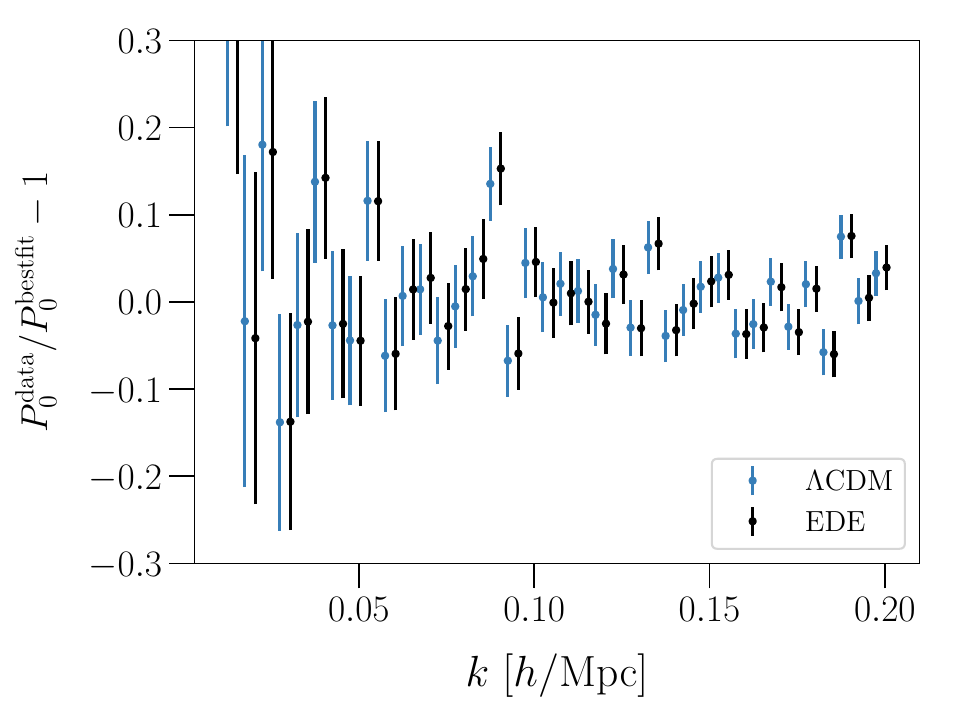}
    \includegraphics[width=0.49\textwidth]{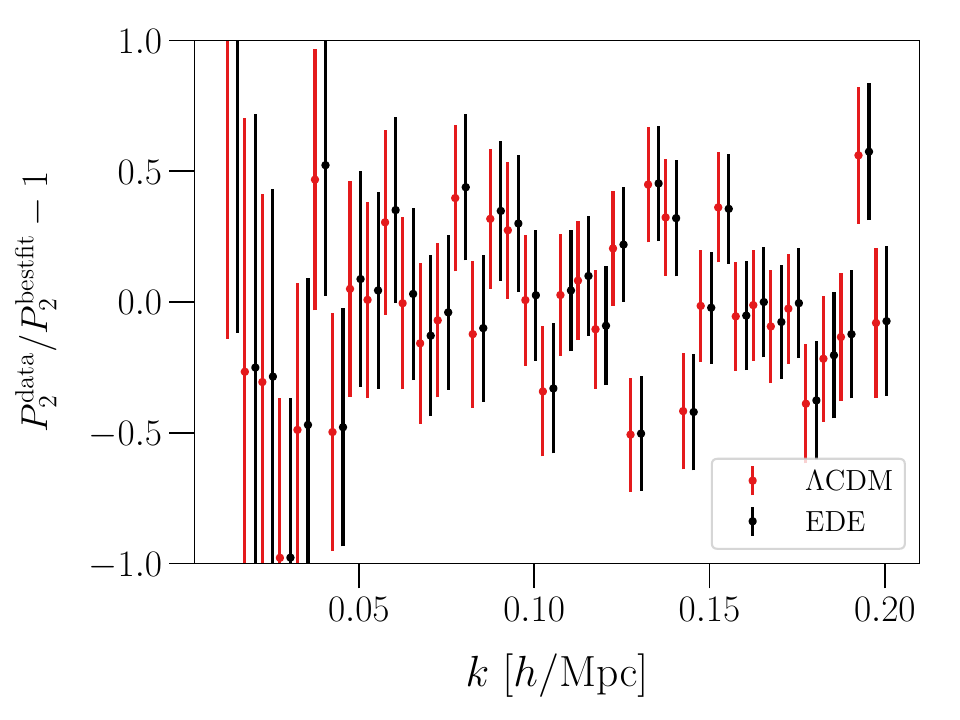}
    \caption{Comparison between the $\Lambda$CDM 
    and EDE best-fit models at the level of the BOSS galaxy power spectrum data. {\it Upper left panel}: the combined power spectrum data set for the NGC low-z ($z=0.38$) slice. 
    {\it Upper right panel}: 
    the monopole and quadrupole 
    moments ($P_0$ and $P_2$) that display the largest difference between EDE and $\Lambda$CDM. 
    {\it Lower left and lower right panels}:
    residuals between the theory and the data 
    for the monopole and quadrupole moments, 
    respectively. 
    }
 \label{fig:boss_new}
\end{figure}

In order to better understand what effects
are most important to distinguish 
between $\Lambda$CDM and the EDE model, 
we compare the corresponding theory curves
against the BOSS DR 12 data from the 
EFT full-shape likelihood version 2021. The theory models 
we consider
are bestfits extracted from the \textit{Planck} + BOSS MCMC
chains for $\Lambda$CDM and for 
the EDE model with $f_{\rm EDE}$ fixed to $0.2$.
The latter value is chosen to be quite large 
in order to clearly 
visualize the impact of the EDE model on the 
galaxy power spectrum. The results 
for our datavector, consisting of 
$\ell=0,2,4$ multipoles, and the 
real-space power spectrum proxy 
$Q_0$~\cite{Ivanov:2021fbu},
are presented 
in Fig.~\ref{fig:boss_new}. 
In our setup, the EDE fit to the 
galaxy power spectrum datavector 
for the NGC $z=0.38$ slice 
is worse than that of 
$\Lambda$CDM by $\Delta\chi^2 = 4$. 
Although this difference is statistically significant,
it accumulates from the entire datavector,
and hence the visual difference between the two models 
is less striking in Fig.~\ref{fig:boss_new}.
In particular, any appreciable discrepancy between the EDE and 
the $\Lambda$CDM models is observed 
only at the level of the monopole and quadrupole
moments. Looking at the corresponding 
residuals in the lower panels of Fig.~\ref{fig:boss_new},
we can identify two most important constraining channels. 
In the monopole, the EDE fit enhances 
power on large scales, $k<0.05~h$Mpc$^{-1}$,
and suppresses power 
on intermediate scales, $k\approx 0.1~h$Mpc$^{-1}$.
At even shorter scales the difference between the EDE and 
$\Lambda$CDM is largely absorbed into the EFT nuisance 
parameters. The scale-dependent modulation 
of $P_0$ can be attributed to 
variations of $\omega_{cdm}$ with fixed $r_s H_0$ (which is
pinned down by the CMB), cf. Fig.~\ref{fig:fEDEPk} and \cite{Ivanov:2019pdj}.
Moving on to the quadrupole, we see that EDE
homogeneously suppresses power on large scales, $k\lesssim 0.1~h$Mpc$^{-1}$, which is 
constrained through RSD, or the $f\sigma_8$
parameter. 
Just like in the monopole case,
the difference 
between $\Lambda$CDM and EDE for wavenumbers
greater than $0.1~h$Mpc$^{-1}$
is absorbed 
by the nuisance parameters in the quadrupole. 
All in all, our analysis suggests
that the galaxy power spectrum parameters most 
relevant for the EDE constraints are $\omega_{cdm}$
and $f\sigma_8$. 
Finally, we note that our results  suggest that better constraints
on EDE can be obtained 
with more restrictive 
priors on the EFT nuisance 
parameters, which can be 
extracted from 
numerical simulations;
see, e.g.,~\cite{Barreira:2021ukk,Lazeyras:2021dar,Cabass:2022epm,
Cabass:2022ymb,Philcox:2022frc,Cabass:2022wjy}.

%\clearpage
\subsection{The Lyman-$\alpha$ forest}
\label{sec:Lyalpha}
Finally, we turn to the high(er)-redshift universe, and in particular the Lyman-$\alpha$ forest. At $2 \lesssim z \lesssim 5$, clouds of neutral hydrogen (HI) gas permeating the intergalactic medium leave imprints in the spectra of photons traveling through, which are visible in the optical region of the spectrum today. Important sources of such photons are luminous quasars whose broad and continuous spectra are susceptible to absorption features. In particular, HI clouds along an observer's line of sight will attenuate flux as absorption features.  This signature manifests as a collection of densely packed Lyman-$\alpha$ absorption features at different frequencies -- this is the Lyman-$\alpha$ forest. Physically, the corresponding optical depth traces the hydrogen column density along the line of sight, so that it acts as a tomographic tracer of large-scale structure. In contrast to other probes, like the galaxy power spectrum, the Lyman-$\alpha$ forest is able to effectively probe scales down to $\sim$ 1 Mpc or even below.  Moreover, at these high redshifts, such small scales are still near the (quasi-)linear regime, thus allowing cosmological inference on scales that are highly polluted by nonlinear and baryonic effects today; see, e.g.,~\cite{Ivanov:2023yla} for further discussion and references.

Current probes of the Lyman-$\alpha$ forest are large spectroscopic surveys of quasars, which include the Baryonic Oscillation Spectroscopic Survey (BOSS) and its successor eBOSS~\cite{Chabanier:2018rga}, as well as the ongoing Dark Energy Spectroscopic Instrument (DESI)~\cite{DESI:2016fyo,DESI:2023pir,DESI:2023xwh,Karacayli:2023afs}.  These surveys have been used to extract the 3D BAO feature, but the Lyman-$\alpha$ forest can also be used to extract information about the correlations along the observer's line of sight via the 1D flux power spectrum. Smaller dedicated surveys using much higher-resolution spectrographs have also been undertaken, including the XQ-100 \cite{Irsic:2017sop} and MIKE/HIRES quasar samples~\cite{Viel:2013fqw}, which allow Lyman-$\alpha$ measurements out to smaller scales, at the cost of higher sample variance.  Interestingly, analyses of Lyman-$\alpha$ forest data have found that there is preference for lower values of both $\Omega_m$ and $n_s$ relative to CMB analyses in $\Lambda$CDM~\cite{Chabanier:2018rga,Palanque-Delabrouille:2019iyz}.  The constraint on $n_s$ is particularly noteworthy given the very small error bars that joint CMB+Ly$\alpha$ analyses obtain on this parameter, suggesting tight constraining power on EDE, given the increase in $n_s$ required to fit EDE to CMB data.  A first such analysis of EDE using Lyman-$\alpha$ forest data was recently performed in Ref.~\cite{Goldstein:2023gnw}, as discussed further below.

\begin{figure}[h!]
\includegraphics[width=\linewidth]{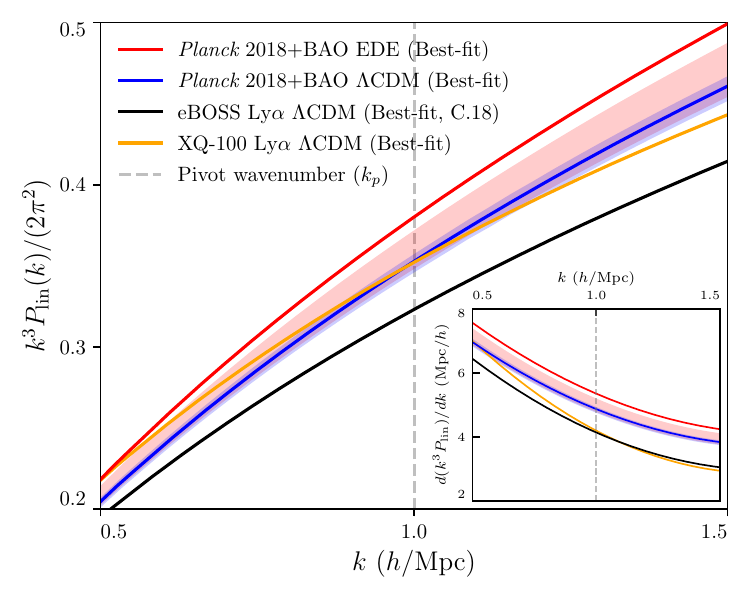}
\caption{
The best-fit (linear) matter power spectrum at $z=3.0$ for EDE (red) compared to $\Lambda$CDM (blue), as fit to the \emph{Planck} PR3 CMB + BAO data, and compared with the best-fit $\Lambda$CDM model for the eBOSS (black)~\cite{Chabanier:2018rga} and XQ-100 (orange)~\cite{Irsic:2017sop} Ly$\alpha$ forest data sets~\cite{Goldstein:2023gnw}. Here, the dashed grey line shows the approximate location of the Ly$\alpha$ pivot wavenumber $(k_p=0.009$ s/km). The shaded bands show the 68\% C.L. uncertainty from the CMB + BAO analyses.  Note that the best-fit model for EDE is outside the 68\% C.L. region due to prior-volume effects in the MCMC analysis. The inset in the lower right shows the derivative $d(k^3P_{\rm lin}(k))/dk$ and the dashed grey line again indicates the approximate location of the Ly$\alpha$ pivot wavenumber.  It is clear that EDE cosmologies with Hubble-tension-resolving parameters that successfully fit CMB + BAO data necessitate an enhancement in the amplitude and an increased slope in $P(k)$ for wavenumbers near the pivot scale, relative to $\Lambda$CDM. These constraints, in particular the steeper derivative, go against the trend from Ly$\alpha$ measurements, and are the origin of the tight upper limits on EDE derived in Ref.~\cite{Goldstein:2023gnw}. Note that this figure is for illustrative purposes and does not include errors for the Ly$\alpha$ data, so as to enhance clarity.}
 \label{fig:ly_alpha_thr_plot}
\end{figure}

To better understand the constraining potential of Lyman-$\alpha$ forest data, in Fig.~\ref{fig:ly_alpha_thr_plot} we compare the best-fit $\Lambda$CDM cosmology in the fit to \emph{Planck} CMB and BAO data to the best-fit $\Lambda$CDM model in the fit to Lyman-$\alpha$ forest data from eBOSS (black)~\cite{Chabanier:2018rga} and XQ-100 (orange)~\cite{Irsic:2017sop}, along with the best-fit EDE model to \emph{Planck} CMB and BAO data.  The shaded bands denote 68\% C.L. regions. The grey dashed line shows the approximate Ly$\alpha$ pivot wavenumber at which the data are most sensitive ($k_p=0.009$ s/km).  All curves are furthermore evaluated at $z=3$, the pivot Ly$\alpha$ redshift.

Of particular interest is the derivative of the matter power spectrum $d(k^3P_{\rm lin}(k))/dk$, which is a proxy for the spectral index $n_s$. This is shown in the lower right inset panel of Fig.~\ref{fig:ly_alpha_thr_plot}. From this one may appreciate that $\Lambda$CDM fit to \emph{Planck} 2018 and BAO data prefers a significantly higher slope at $k_p$ than does either of the Ly$\alpha$ data sets. The best-fit EDE model has a larger slope than that in $\Lambda$CDM, suggesting that EDE, in its effort to resolve the $H_0$ tension, will exacerbate existing moderate tension with Ly$\alpha$ data.

%%%%%%%%%%%%%%%%%%%%%%%%%%%%%%%%%%%%%%%%%%%%%%%%%%%%%%%%%%%%%%%%%%%%%%
%%%%%%%%%%%%%%%%%%%%%%%%%%%%%%%%%%%%%%%%%%%%%%%%%%%%%%%%%%%%%%%%%%%%%%
%\clearpage
\section{Constraints from Weak Lensing}
\label{sec:constraintsWL}
%%%%%%%%%%%%%%%%%%%%%%%%%%%%%%%%%%%%%%%%%%%%%%%%%%%%%%%%%%%%%%%%%%%%%%
%%%%%%%%%%%%%%%%%%%%%%%%%%%%%%%%%%%%%%%%%%%%%%%%%%%%%%%%%%%%%%%%%%%%%%

We now begin our examination of large-scale structure constraints on the Early Dark Energy scenario. We use results from Ref.~\cite{Hill:2020osr}, which utilized data from  DES-Y1, KiDS+VIKING-450 (KiDS), and Hyper Suprime-Cam (HSC). The DES-Y1 data set refers specifically to the ``3x2pt'' likelihood composed of photometric galaxy clustering, cosmic shear two-point correlation functions, and galaxy-galaxy lensing \cite{Abbott:2017wau}.  The KV-450 \cite{Hildebrandt:2016iqg,2020A&A...633A..69H} and HSC \cite{Hikage:2018qbn} surveys are complementary data sets to the DES results, but including them would result in an appreciably more computationally expensive MCMC analysis. To circumvent the added complexity of having to sample from two new likelihoods, we have opted to approximate KV-450 and HSC-Y1 data utilizing a prior on $S_8$. For KiDS and HSC-Y1 this corresponds to a constraint of $S_8 = 0.737^{+ 0.040}_{-0.036}$ and $S_8 = 0.780^{+0.030}_{-0.033}$, respectively. One may be concerned that this approach could be problematic, but since we have the full likelihood for DES-Y1 3x2pt we can test if that data set's $S_8$ constraint is a good approximation of the information content in the full likelihood. A validation test was performed in Ref.~\cite{Hill:2020osr} for both the EDE and $\Lambda$CDM scenarios, where it was found that the parameter constraints and posterior distributions in the fit to a combined data set including \emph{Planck}, BAO, RSD, Pantheon, SH0ES, and the DES-Y1 3x2pt likelihood are effectively identical to the posteriors and constraints from the analysis with the DES-Y1 likelihood replaced by a Gaussian prior on $S_8$.
Since that time, new data has become available from DES-Y3~\cite{DES:2021wwk}, but incidentally the DES-Y3 constraint on $S_8$ is nearly identical to that from the inverse-variance-weighted combination of DES-Y1, KV-450, and HSC-Y1 $S_8$ measurements, and thus the combined DES-Y1+ KiDS + HSC-Y1 analysis presented here can be considered a proxy for an analysis using DES-Y3 in place of these three.  Further improvements could be made using the latest KiDS-1000~\cite{Li:2023azi,Kilo-DegreeSurvey:2023gfr} and HSC-Y3~\cite{Li:2023tui,Dalal:2023olq} data, but we leave such an exploration to future work.

\begin{figure}
\centering
 \includegraphics[trim={0 2cm 0 2cm},clip,width=\textwidth]{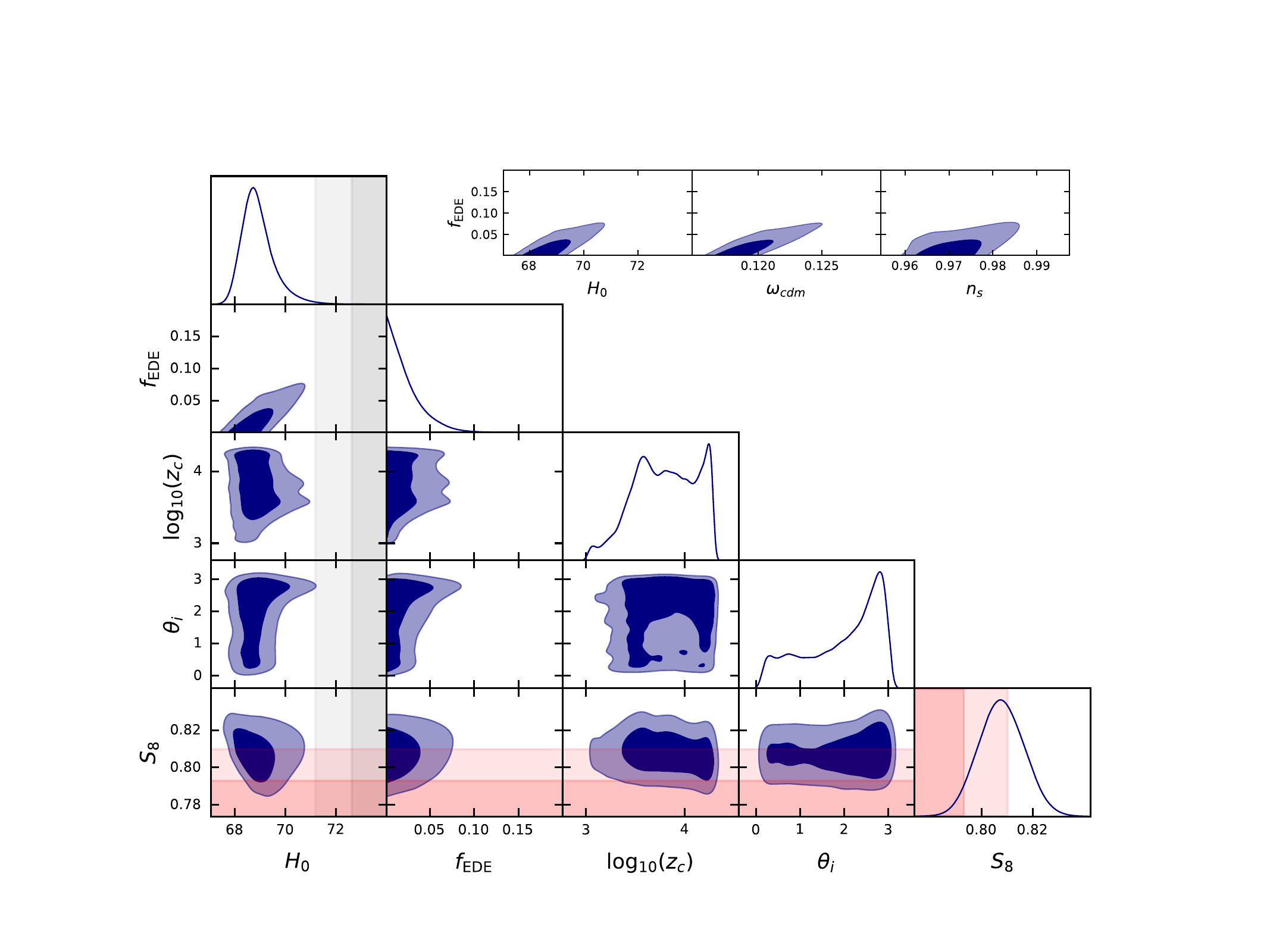}
    \caption{Cosmological parameter constraints on EDE from the combination of DES-Y1 3x2pt data, KiDS and HSC data approximated by an $S_8$ prior, and  \emph{Planck} PR3 CMB and CMB lensing data, BAO data from 6dF, SDSS DR7, and SDSS DR12, Pantheon SNIa data, and SDSS DR12 RSD data. Grey bands indicate the SH0ES $H_0$ measurement while pink bands indicate the DES-Y3 measurement of $S_8$.}
    \label{fig:no-SH0ES}
\end{figure}

To mitigate `prior volume effects' (discussed further in Sec.~\ref{sec:priors}) we have performed the MCMC analysis with a SH0ES $H_0$ prior, which was subsequently removed in post-processing, such that our analysis is independent of SH0ES entirely. The posteriors for canonical EDE or $\Lambda$CDM analyses are presented in Fig.~\ref{fig:no-SH0ES}, and in Table~\ref{table:params-uberlikelihoodKiDSHSCNoSH0ES} we show the mean $\pm 1\sigma$ constraints with a combined data set consisting of \emph{Planck} PR3 {\tt Plik} primary CMB data (TT+EE+TE) and \emph{Planck} PR3 CMB lensing data; data for BAO from SDSS DR7, SDSS DR12, and 6dF; Pantheon SNIa data; RSD data from SDSS DR12; DES-Y1 3x2pt data; and a prior for $S_8$ derived from KV-450 and HSC-Y1 data. Note that we have treated the three weak lensing surveys (DES, KiDS, and HSC) as independent, which is a valid assumption as the sky overlap between them is small for the data releases used here.

After the inclusion of all these data sets, we find that the evidence for EDE is weakened to below the 2$\sigma$ mark, as assessed from the $f_{\rm EDE}$ posterior. Concretely we find for a two-tailed analysis that $f_{\rm EDE} = 0.023 \pm 0.017$ (68\% CL), which is statistically consistent with $f_{\rm EDE} = 0$; as an upper bound we instead find that $f_{\rm EDE} < 0.060$ at $95\%$ CL. Furthermore, we find that $H_0 = 68.92^{+0.40}_{-0.72}$ km/s/Mpc, which is in $\approx 3.5\sigma$ tension with SH0ES. On the other hand, this value for $H_0$ is perfectly consistent with the $\Lambda$CDM value of $H_0$ fit to the same combination of data sets, $H_0 = 68.33 \pm 0.36$~km/s/Mpc. Collectively this corresponds to a non-detection of a new EDE component.

The posteriors after the inclusion of large-scale structure data are consistent with that from the primary {\it Planck} 2018 CMB data alone, although the constraints on $f_{\rm EDE}$ are even stronger after the inclusion of the additional data sets.

\begin{table}[h!]%[ht!]
   %\hspace{-0.24cm} 
\centering
{Constraints from \emph{Planck} 2018 PR3 TT+TE+EE + CMB Lensing, BAO, SNIa, RSD, DES-Y1, KiDS-$S_8$, and  HSC-$S_8$ \vspace{4pt}}
\tbl{The mean $\pm1\sigma$ constraints for parameters in $\Lambda$CDM and in the canonical EDE model, from \emph{Planck} 2018 (PR3) primary CMB data (TT+TE+EE) and CMB lensing data; BAO data from SDSS DR7, SDSS DR12, and 6dF; SDSS DR12 RSD data; DES-Y1 3x2pt data; Pantheon SNIa data; and with priors on $S_8$ from HSC and KiDS data.  Here we have not provided the best-fit parameters because of our use of approximate likelihoods for KiDS and HSC.  In our analysis we find no preference for the EDE component; using a two-tailed analysis we find $f_{\rm EDE} = 0.023 \pm 0.017$, which is consistent with  our  \emph{Planck} CMB-only analysis (Table~\ref{table:params-P18-only}).}
  %\centering
  {
    \begin{tabular}{|l|c|c|}
    \hline\hline Parameter &$\Lambda$CDM Marg.~~&~~~EDE ($n=3$) Marg.\\ \hline \hline

{\boldmath$\ln(10^{10} A_\mathrm{s})$} & $3.041\pm 0.014            $ & $3.044\pm 0.014$\\

{\boldmath$n_\mathrm{s}$} & $0.9691\pm 0.0035          $ & $0.9718^{+0.0041}_{-0.0055}          $\\

{\boldmath$100\theta_\mathrm{s}$} & $1.04200\pm 0.00028       $ & $1.04177^{+0.00038}_{-0.00030}$\\

{\boldmath$\Omega_\mathrm{b} h^2$} & $0.02253\pm 0.00013        $ & $0.02264^{+0.00015}_{-0.00018}$\\

{\boldmath$\Omega_\mathrm{c} h^2$} & $0.11785\pm 0.00077        $ & $0.11956^{+0.00088}_{-0.0020}$\\

{\boldmath$\tau_\mathrm{reio}$} & $0.0552\pm 0.0070$ & $0.0558\pm 0.0070$\\

{\boldmath$\mathrm{log}_{10}(z_c)$} & $-$ & $> 3.28$\\

{\boldmath$f_\mathrm{EDE} $} & $-$ & $ < 0.060 $\\

{\boldmath$\theta_i$} & $-$ & $> 0.35$\\

    \hline

$H_0 \, [\mathrm{km/s/Mpc}] $ & $68.33\pm 0.36            $ & $68.92^{+0.40}_{-0.72}$\\

$\Omega_\mathrm{m}         $ & $0.3021\pm 0.0045          $ & $0.3008\pm 0.0047$\\

$\sigma_8                  $ & $0.8032\pm 0.0053         $ & $0.8064^{+0.0057}_{-0.0073}$\\

$S_8$                        & $0.8060\pm 0.0082         $ & $0.8074\pm 0.0089$\\

$\mathrm{log}_{10}(f/{\mathrm{eV}})$ & $-$ & $26.52^{+0.28}_{-0.44}     $\\

$\mathrm{log}_{10}(m/{\mathrm{eV}})$ & $-$ & $-26.67\pm 0.69            $\\

    \hline
  \end{tabular} 

  \label{table:params-uberlikelihoodKiDSHSCNoSH0ES}}
\end{table}

%%%%%%%%%%%%%%%%%%%%%%%%%%%%%%%%%%%%%%%%%%%%%%%%%%%%%%%%%%%%%%%%%%%%%%
%%%%%%%%%%%%%%%%%%%%%%%%%%%%%%%%%%%%%%%%%%%%%%%%%%%%%%%%%%%%%%%%%%%%%%
%\clearpage
\section{Constraints from BOSS}
\label{sec:constraintsBOSS}
%%%%%%%%%%%%%%%%%%%%%%%%%%%%%%%%%%%%%%%%%%%%%%%%%%%%%%%%%%%%%%%%%%%%%%
%%%%%%%%%%%%%%%%%%%%%%%%%%%%%%%%%%%%%%%%%%%%%%%%%%%%%%%%%%%%%%%%%%%%%%

We now turn our attention to the role of BOSS data in constraining EDE. We consider BOSS data in conjunction with {\it Planck} 2018 (PR3) CMB data and these two together in conjunction with $S_8$ measurements from DES, HSC, and KV450.

In the CMB analysis, we adopt the \texttt{Plik} likelihood for the final {\it Planck} 2018 TT+TE+EE+low $\ell$+lensing
data \cite{Planck2018likelihood}. Employing the standard analysis procedure, we adjust all necessary nuisance parameters to accommodate observational and instrumental uncertainties.

When analysing BOSS we use a number 
of joint EFT-based full-shape+BAO likelihoods 
extracted from the final release, DR12~\cite{Alam:2016hwk}. 
Namely, our result presented first 
will be for the 
FS+BAO likelihood from 
Ref.~\cite{Philcox:2020vvt}, based on the 
FS likelihood version 2019~\cite{Ivanov:2019pdj}.
Our final results will be presented for the 
FS+BAO likelihood version 2021 from
Ref.~\cite{Philcox:2021kcw}; see Refs.~\cite{Ivanov:2019pdj,Ivanov:2019hqk,Philcox:2020vvt,Chudaykin:2020aoj,Wadekar:2020hax,Chudaykin:2020ghx,Philcox:2021kcw} for more detail. 
Note that the FS part of the EFT-based BAO+FS 
likeilihood has been upgraded twice
since the original 2019 release~\cite{Ivanov:2019pdj}. 
All subsequent FS+BAO likelihoods have the 
same BAO part, so in what follows we will 
focus only on the FS part when discussing
different BOSS DR12 likelihoods, and distinguish between them by using the notations
FS 2019, FS 2020, and FS 2021. 
FS 2020 is the intermediate version used in (e.g.) Refs.~\cite{Wadekar:2020hax,Chudaykin:2020ghx}.

The FS+BAO likelihood
from Ref.~\cite{Philcox:2020vvt} (FS-2019)
was used in the first BOSS EDE constraints presented in Ref.~\cite{Ivanov:2020ril}. It
includes the monopole ${\ell}=0$ and dipole $\ell=2$ multipole moments of both the pre- and post-reconstruction anisotropic galaxy power spectra, at two redshifts, $z=0.38$ (low-$z$) and $z=0.61$ (high-$z$), observed in both the North Galactic Cap (NGC) and South Galactic Cap (SGC), for a cumulative volume $\simeq 6~(h^{-1}{\rm Gpc})^3$. 
The four data subsets (low-$z$ vs. high-$z$, NGC vs. SGC) each have different selection functions, and therefore in our analysis we endow each with an independent set of nuisance parameters. 
We adopt broad priors as in Ref.~\cite{Ivanov:2019pdj}
for FS-2019 and 
more restrictive priors from~\cite{Philcox:2021kcw}
in our final analysis based on the FS-2021 likelihood. 
Our final FS+BAO likelihood will be based on
the FS-2021 BOSS DR12 likelihood, 
which includes the 
monopole ${\ell}=0$, dipole $\ell=2$,
and hexadecapole $\ell=4$ multipole moments of the 
galaxy power spectrum,
the real-space galaxy power spectrum proxy $Q_0$~\cite{Ivanov:2021fbu}, 
and the bispectrum
monopole~\cite{Ivanov:2021kcd} data, for the same BOSS NGC/SGC/high-z/low-z
selections. As far as the measurements are concerned, 
the FS-2021 likelihood is based on the same data, 
but all statistics are extracted with 
the novel optimal 
window-free estimators~\cite{Philcox:2020vbm,Philcox:2021ukg}.
We discuss the differences between FS likelihoods
in more detail shortly.

Finally, we include supplementary LSS data from photometric surveys. As done previously in Sec.~\ref{sec:constraintsWL}, we incorporate DES-Y1 photometric galaxy clustering, galaxy-galaxy lensing, and cosmic shear measurements~\cite{Abbott:2017wau}, as well as weak lensing measurements from KV-450~\cite{Hildebrandt:2016iqg,2020A&A...633A..69H} and HSC-Y1~\cite{Hikage:2018qbn}. The combined measurements, weighted by inverse variance, result in $S_8 = 0.770 \pm 0.017$, nearly identical to the DES-Y3 measurement $S_8 = 0.776 \pm 0.017$ \cite{DES:2021wwk}. Hereafter, we will denote this combination as simply ``$S_8$.''

We compute the CMB and non-linear spectra using the code \texttt{CLASS-PT}  \cite{Chudaykin:2020aoj}, which is a further modification of the publicly available {\tt class\_ede} code \cite{Hill:2020osr}. We perform posterior sampling using the Metropolis-Hastings algorithm as implemented in \texttt{Monte Python}.\footnote{\url{https://github.com/brinckmann/montepython_public/issues/112}} We plot our results using \texttt{GetDist} \cite{GetDist}.\footnote{\url{https://github.com/cmbant/getdist}}

%%%%%%%%%%%%%%%%%%%%%%%%%%%%%%%%%%%%%%%%%%%%%%%%%%%%%%%%%%%%%%%%%%%%%%%%%%%%%%%%%%%%

\subsection{EDE Meets BOSS: circa 2020} %: Constraints on EDE from the CMB and BOSS Full Shape}
\label{sec:BossFS}

\begin{figure}[ht!]
\begin{center}
\includegraphics[width=0.8\textwidth]{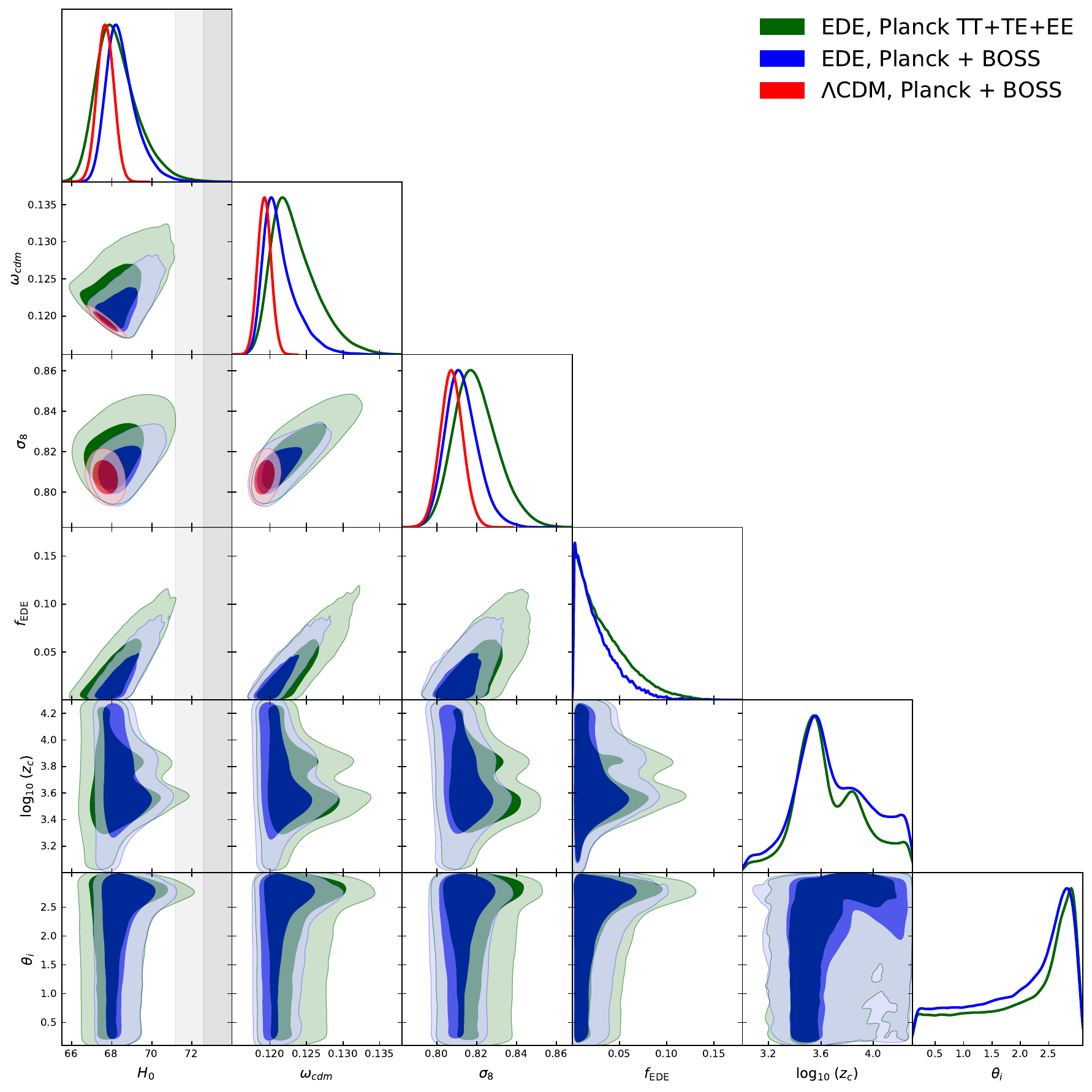}
\end{center}
\caption{Constraints on cosmological parameters for a joint fit to {\it Planck} 2018 TT+TE+EE+low $\ell$+lensing + BOSS DR 12 (FS+BAO) data sets. For comparison, we also show the posteriors from the {\it Planck} 2018 primary CMB analysis obtained in \cite{Hill:2020osr}. 
Also depicted is the SH0ES $H_0$ measurement (gray) where dark- and light-shaded regions represent the $1\sigma$ and $2\sigma$ intervals.\label{fig:planck+BOSS}}
\end{figure}

We begin with an analysis of {\it Planck} 2018 CMB data (PR3 \texttt{Plik}  likelihood) combined with BOSS BAO data and the BOSS DR12 FS-2019 likelihood. The results are given in Table~\ref{table:params-P18+BOSS} and Fig.~\ref{fig:planck+BOSS}. Concordant with the analysis of {\it Planck} data alone, we find no evidence for EDE in this analysis. The constraint from {\it Planck} is strengthened by $\approx 20\%$, to $f_\mathrm{EDE} < 0.072$ at 95\% C.L., when BOSS is included.

To understand these results, it is useful to first review the constraining power of BOSS full shape and BAO data in the context of $\Lambda$CDM. In particular, the shape of the galaxy power spectrum provides a direct measurement of the physical density of dark matter $\omega_{\rm cdm}$ \cite{DAmico:2019fhj,Ivanov:2019pdj}. This leads to a relative decrease in $S_8$, with {$S_8=0.824\pm0.011$} when BOSS is included versus $S_8=0.833\pm 0.016$ in the fit to the CMB alone. Meanwhile there is a relative {\it increase} in $H_0$ relative to {\it Planck}, up to $H_0=67.70 \pm 0.42$ km/s/Mpc, in agreement with the CMB-independent measurement $H_0 = 68.6\pm 1.1$km/s/Mpc from BOSS BAO data with a BBN prior imposed on $\omega_b$ \cite{Philcox:2020vvt}. The slight increase in $H_0$ is a well-known impact of BAO data (see Refs.~\cite{Alam:2016hwk,Aghanim:2018eyx}).

The parameter shifts in $\Lambda$CDM are mirrored by those in EDE. This can be appreciated from the posterior distributions shown in Fig.~\ref{fig:planck+BOSS}. In particular, the addition of BOSS data causes downward shifts in both $\sigma_8$ and $\omega_{\rm cdm}$ (and correspondingly $\Omega_m$; see Table~\ref{table:params-P18+BOSS}). These two effects combine to drive a relative shift downward in $S_8$:  we find $S_8=0.822 \pm 0.010$ when BOSS is included, compared to $S_8=0.839 \pm 0.017$ for the CMB alone,  similar to $\Lambda$CDM. Again, like in $\Lambda$CDM, 
there is a small relative increase in $H_0$ when BOSS is included, to $H_0=68.54 ^{+0.52}_{-0.95}$ km/s/Mpc. Likewise, the constraints on the other standard $\Lambda$CDM parameters, such as the spectral index $n_s$, track the $\Lambda$CDM constraints when BOSS is included. The spectral index is reduced to $n_s=0.9696 ^{+0.0046} _{-0.0068}$ in EDE, compared to $n_s=0.9656 \pm 0.0037$ in $\Lambda$CDM. That the BOSS-induced shifts in parameter constraints in EDE so closely match those in $\Lambda$CDM reflects the lack of an overall preference for the EDE component, and indicates that the EDE constraints on $H_0, n_s$, etc.,  are primarily driven by the degeneracies of the underlying $\Lambda$CDM components and not the physics of EDE {\it per se}.

\begin{table}[htb!]
%\vspace{-5cm}
% \hspace{2.4cm} 
\centering
{Constraints from \emph{Planck} 2018 data  + BOSS DR12 (FS-2019) \vspace{4pt}}
\tbl{Constraints on parameters in $\Lambda$CDM and the canonical EDE model with mean (best-fit) $\pm1\sigma$ constraints from a combined data set of \emph{Planck} 2018 TT+TE+EE+low $\ell$+lensing and BOSS FS+BAO. Our analysis does not detect an EDE component; a $68\%$ confidence limit gives $f_\mathrm{EDE} = 0.025_{-0.025}^{+0.0061}$, which is statistically consistent with zero. The upper limit on the EDE component, $f_{\rm EDE} < 0.072$, is quoted at 95\% CL. The EDE value of $H_0$ is in $3.9\sigma$ tension with the latest SH0ES measurement  ($H_0=73.04\pm1.04$ km/s/Mpc)\cite{Riess:2021jrx}.}
  %\begin{minipage}{\textwidth}
  %\centering
  {\begin{tabular}{|l|c|c|}
    \hline\hline Parameter &$\Lambda$CDM~~&~~~EDE ($n=3$) ~~~\\ \hline \hline

    {\boldmath$\ln(10^{10} A_\mathrm{s})$} & $3.043 \, (3.034) \, \pm 0.014$ & $3.047 \, (3.049) \, \pm 0.014$ \\

    {\boldmath$n_\mathrm{s}$} & $0.9656 \, (0.9655) \, \pm 0.0037 $ & $0.9696 \, (0.9717)^{+0.0046}_{-0.0068}$ \\

    {\boldmath$100\theta_\mathrm{s}$} & $1.04185 \, (1.04200) \, \pm 0.00029 $ & $1.04172 \, (1.04126) \, \pm 0.00032$\\

    {\boldmath$\Omega_\mathrm{b} h^2$} & $0.02241 \, (0.02233) \, \pm 0.00014 $ &  $0.02255 \, (0.02245) \, \pm 0.00018$ \\

    {\boldmath$\Omega_\mathrm{cdm} h^2$} & $ 0.1192\, (0.1191)_{-0.00095}^{+0.00087}$ & $0.1215 \, (0.1243)^{+0.0013}_{-0.0029}$ \\

    {\boldmath$\tau_\mathrm{reio}$} & $0.0546 \, (0.0503)_{-0.0072}^{+0.0065} $ & $0.0553 \, (0.0543)_{-0.0075}^{+0.0069}$\\

    {\boldmath$\mathrm{log}_{10}(z_c)$} & $-$ & $3.71 \, (3.52)^{+0.26}_{-0.33}$ \\

    {\boldmath$f_\mathrm{EDE} $} & $-$ & $< 0.072 \, (0.047)$\\

    {\boldmath$\theta_i$} & $-$ & $2.023(2.734)_{-0.34}^{+1.1} $\\

    \hline

    $H_0 \, [\mathrm{km/s/Mpc}]$ & $67.70 \, (67.56) \, \pm 0.42$ & $68.54 \, (68.83)^{+0.52}_{-0.95}$ \\

    $\Omega_\mathrm{m}$ & $0.3105 \, (0.3112)_{-0.0058}^{+0.0053}$ & $0.3082 \,(0.3120)_{-0.0057}^{+0.0056}$ \\

    $\sigma_8$ & $0.8077 \, (0.8039) \, \pm 0.0058$ & $0.8127 \, (0.8195)_{-0.0091}^{+0.0072}$ \\
    
    $S_8$ & 
    $0.822 \, (0.819) \, \pm 0.010$
    & 
    $0.824 \, (0.827)\, \pm 0.011$
    \\
    \hline
  \end{tabular} 
  \label{table:params-P18+BOSS}}
  %\end{minipage}
\end{table}

%%%%%%%%%%%%%%%%%%%%%%%%%%%%%%%%%%%%%%%%%%%%%%%%%%%%%%%%%%%%%%%%%%%%%%%%%%%%%%%%%%%%
%\clearpage
\subsection{Full combination of CMB and LSS data: circa 2020}
\label{sec:S8}
%%%%%%%%%%%%%%%%%%%%%%%%%%%%%%%%%%%%%%%%%%%%%%%%%%%%%%%%%%%%%%%%%%%%%%%%%%%%%%%%%%%%

%DES \cite{Abbott:2017wau}, the Kilo-Degree Survey (KiDS) \cite{Hildebrandt:2016iqg,2020A&A...633A..69H}, and the Subaru Hyper Suprime-Cam (HSC) survey \cite{Hikage:2018qbn}

 The constraints on EDE significantly tighten when additional large-scale structure data are included in the analysis. In particular we supplement \emph{Planck} and BOSS data with additional LSS data from the Dark Energy Survey Y1 \cite{Abbott:2017wau}, the Kilo-Degree Survey (KiDS) \cite{Hildebrandt:2016iqg,2020A&A...633A..69H}, and the Subaru Hyper Suprime-Cam (HSC) survey \cite{Hikage:2018qbn}. Following \cite{Hill:2020osr}, we parameterize these additional data sets via a prior on $S_8$ corresponding to the measurement of each. This analysis comes with important caveats, as follows:
\begin{enumerate}
    \item The LSS data sets are treated as independent. The approximation is justified on the basis of the small relative sky overlap between BOSS and the weak lensing surveys considered here ($1\%$, $2\%$, and $1.5\%$ of BOSS for DES-Y1, KV450, and HSC \cite{Lee:2019pfw,Troster:2019ean,Kondo:2019ind}), respectively, as well as the different survey depths and photo-$z$ calibrations. We note that 
    the combined constraint of DES-Y1+KV450+HSC-Y1 is near identical to the DES-Y3 constraint~\cite{DES:2021wwk}, and thus this caveat can easily be avoided by replacing the former with the latter.
        \item We approximate the weak lensing data by a Gaussian prior on $S_8$, which is justified on the basis of the validation check performed in~\cite{Hill:2020osr} for DES-Y1, for both the EDE and $\Lambda$CDM models. This allows the inclusion of of HSC-Y1 and KV-450 measurements, which do not have publicly available likelihoods. However, this comes with another caveat that the HSC and KV-450 measurements are based on non-linear biasing models and not an EFT-based framework as is implemented here for BOSS. We expect a joint EFT treatment of weak lensing and BOSS to lead to a stronger breaking of degeneracies between cosmological and nuisance parameters, thereby improving the constraints on the former.
\end{enumerate}
With these caveats in mind, we proceed with our analysis.

\begin{figure}[ht!]
\centering
 \includegraphics[width=0.8\textwidth]{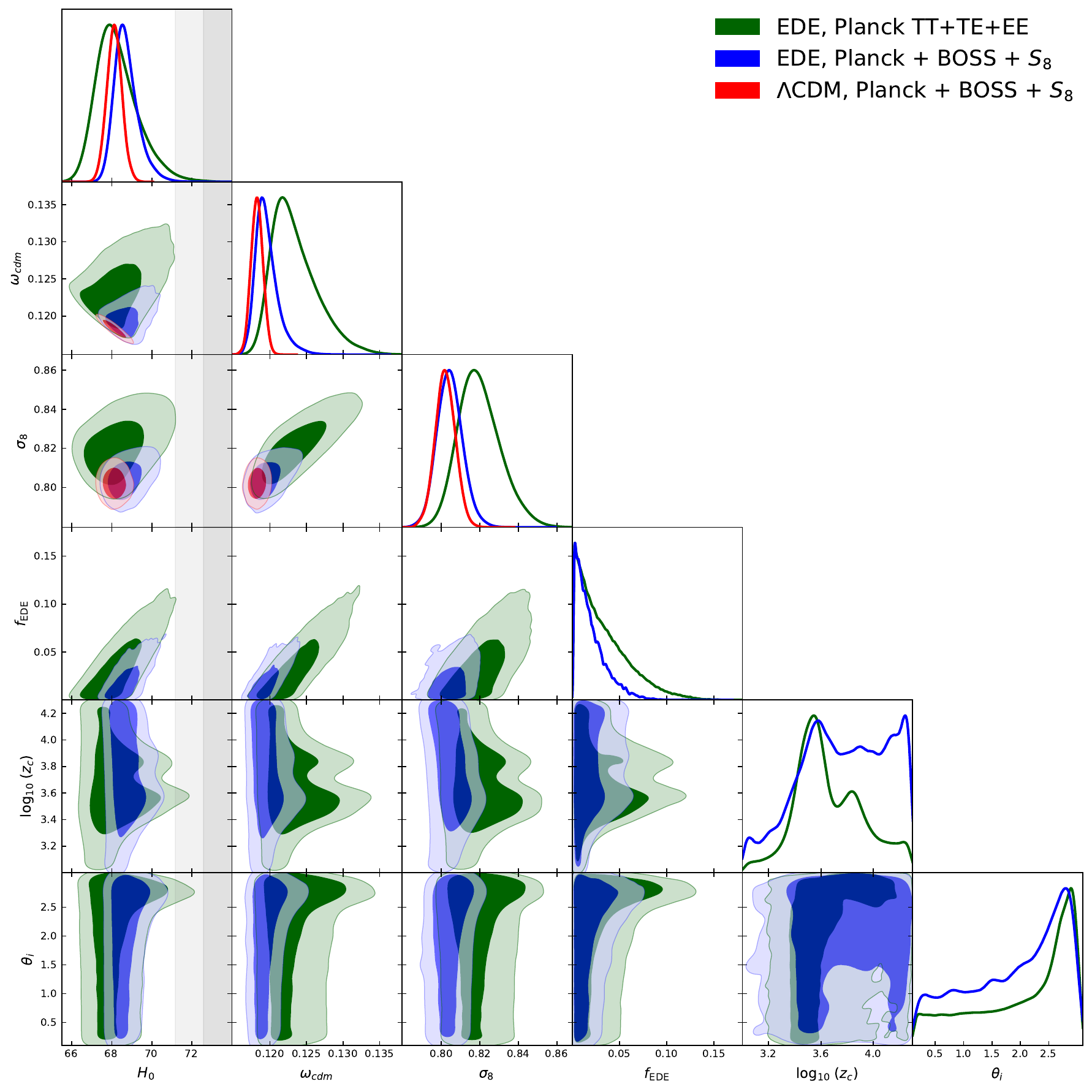}
    \caption{Constraints on cosmological parameters from a joint fit to a data set comprising {\it Planck} 2018 TT+TE+EE+low $\ell$+lensing + BOSS DR12 (FS+BAO) + $S_8$ (DES-Y1 + KV-450 + HSC-Y1) on parameters in the EDE (blue) model and $\Lambda$CDM (red) models. For comparison, we also show the results for the EDE model with just the {\it Planck} 2018 primary CMB data (green) from \cite{Hill:2020osr}. 
    We also include the SH0ES $H_0$
    measurement (gray); the dark- and light-shaded contours represent 
$68\%$ and $95\%$ confidence intervals, respectively.
    }
    \label{fig:BOSS-S8}
\end{figure}

The results are shown in Fig.~\ref{fig:BOSS-S8} and Table~\ref{table:params-P18+BOSS-S8}. Consistent with the analyses of {\it Planck} alone and of {\it Planck} and BOSS DR12, we find an upper bound $f_{\rm EDE}<0.053$ at $95\%$ CL, a $26\%$ improvement from the analysis with {\it Planck} and BOSS DR12. The Hubble constant is nearly identical in EDE and $\Lambda$CDM, with the two constraints given by $H_0=68.73^{+0.42}_{-0.69}$ km/s/Mpc and $H_0=68.13 \pm 0.38$ km/s/Mpc, respectively, and both in significant tension with the SH0ES measurement. As in the {\it Planck}+BOSS DR12 analysis, the constraints in EDE very nearly match those in $\Lambda$CDM, indicating that LSS data has broken the degeneracies between EDE parameters and $\Lambda$CDM parameters (e.g., $f_{\rm EDE}$ and $H_0$), leaving only the standard $\Lambda$CDM degeneracies to drive the cosmological constraints.

\begin{table}[htb!]
% \hspace{2.4cm} 
\centering
{Constraints from \emph{Planck} 2018 data  + BOSS DR12 (FS-2019) + $S_8$ from DES+KV-450+HSC\vspace{4pt}}
\tbl{The mean (best-fit) $\pm1\sigma$ constraints for parameters in $\Lambda$CDM and the canonical EDE model for a combination of BOSS FS+BAO, DES+KV-450+HSC $S_8$, and \emph{Planck} 2018 TT+TE+EE+low $\ell$+lensing data sets.  Here we report the upper limit at 95\% C.L. for $f_{\rm EDE}$. In this scenario EDE is not detected; we find the $68\%$ confidence limit is given by $f_\mathrm{EDE} = 0.019_{-0.019}^{+0.0040}$ which is statistically consistent with zero.}
  {\begin{tabular}{|l|c|c|}
    \hline\hline Parameter &$\Lambda$CDM~~&~~~EDE ($n=3$) ~~~\\ \hline \hline

    {\boldmath$\ln(10^{10} A_\mathrm{s})$} & $3.036 \, (3.039) \, \pm 0.014$ & $3.038 \, (3.034) \, \pm 0.014$ \\

    {\boldmath$n_\mathrm{s}$} & $0.9674 \, (0.9727) \, \pm 0.0037 $ & $0.9696 \, (0.9621)^{+0.0042}_{-0.0051}$ \\

    {\boldmath$100\theta_\mathrm{s}$} & $1.041945 \, (1.041966) \, \pm 0.00030 $ & $1.04178 \, (1.04176) \, \pm 0.00035$\\

    {\boldmath$\Omega_\mathrm{b} h^2$} & $0.02249 \, (0.02273) \, \pm 0.00013 $ &  $0.02259 \, (0.02243)^{+0.00016}_{-0.00018}$ \\

    {\boldmath$\Omega_\mathrm{cdm} h^2$} & $ 0.1182\, (0.1157)\pm 0.00081$ & $0.11958 \, (0.11951)^{+0.00096}_{-0.0018}$ \\

    {\boldmath$\tau_\mathrm{reio}$} & $0.0527 \, (0.0591)\pm 0.0067 $ & $0.0535 \, (0.0521)_{-0.0075}^{+0.0069}$\\

    {\boldmath$\mathrm{log}_{10}(z_c)$} & $-$ & $3.77 \, (4.24)^{+0.51}_{-0.33}$ \\

    {\boldmath$f_\mathrm{EDE} $} & $-$ & $< 0.0526 \, (0.0115)$\\

    {\boldmath$\theta_i$} & $-$ & $1.91(1.55)_{-0.47}^{+1.2} $\\

    \hline
    $H_0 \, [\mathrm{km/s/Mpc}]$ & $68.13 \, (69.28) \, \pm 0.38$ & $68.73 \, (67.92)^{+0.42}_{-0.69}$ \\

    $\Omega_\mathrm{m}$ & $0.3046 \, (0.2859)\pm 0.0049$ & $0.3024 \,(0.3091)\pm 0.0050$ \\

    $\sigma_8$ & $0.80204 \, (0.7947) \, \pm 0.0053$ & $0.8044 \, (0.8023)_{-0.0069}^{+0.0060}$ \\
    
    $S_8$ & 
    $0.8082 \, (0.7810) \, \pm 0.0086$
    & 
    $0.8075 \, (0.8143)\, \pm 0.0092$ 
    \\
    \hline
  \end{tabular} 
  \label{table:params-P18+BOSS-S8}}
\end{table}

\begin{figure}[ht!]
\centering
 \includegraphics[width=0.99\textwidth]{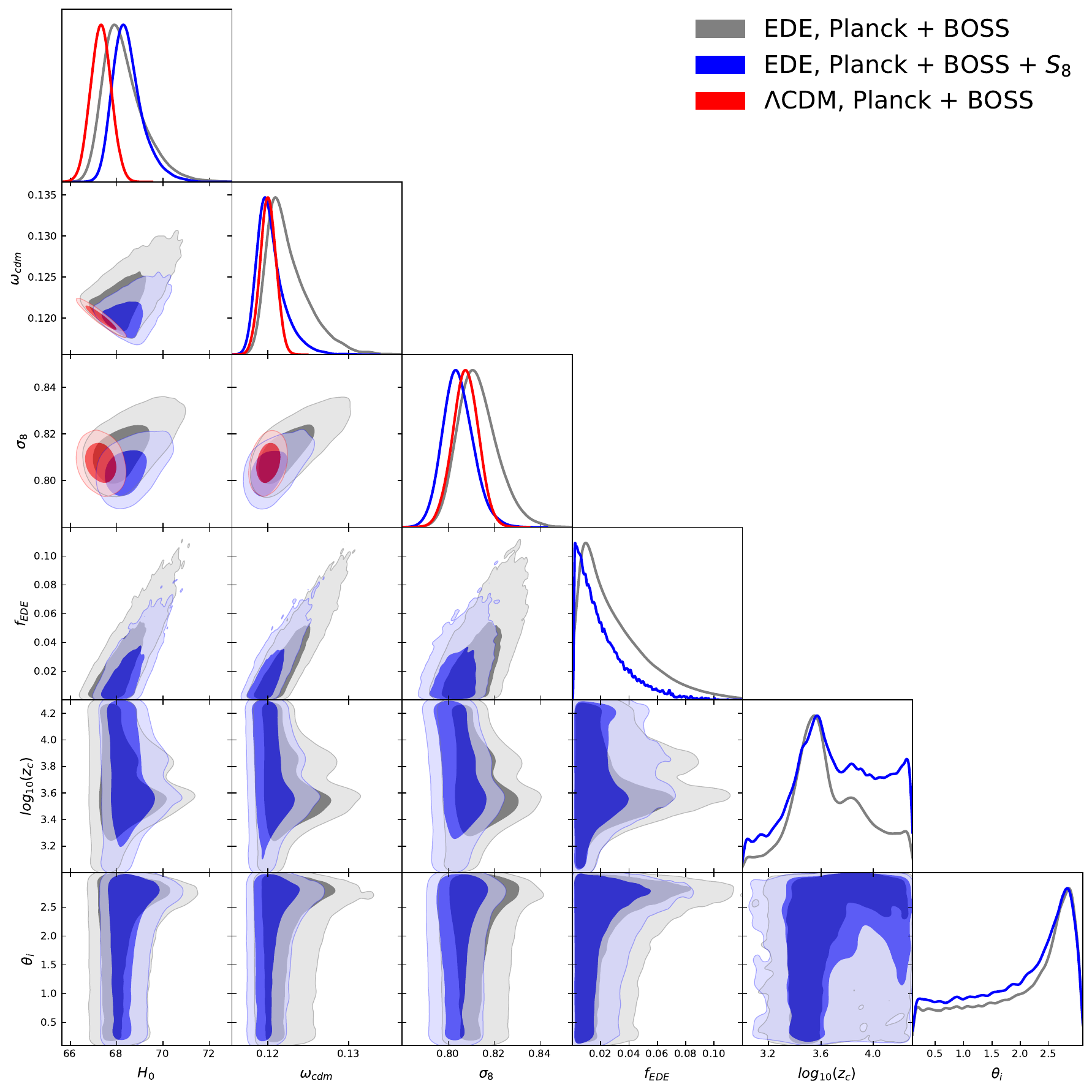}
    \caption{Cosmological parameter constraints from the joint {\it Planck} 2018 TT+TE+EE+low $\ell$+lensing + BOSS DR12 BAO+FS-2021 + eBOSS ELG + $S_8$ (DES-Y3) likelihood. We present combinations 
    with (blue) and without (grey) the $S_8$ DES-Y3 prior. 
    In addition, we display the {\it Planck} + BOSS $\Lambda$CDM posteriors (red).
    }
    \label{fig:BOSS-S8_2021}
\end{figure}

\begin{table}[htb!]
\centering{Constraints on EDE from \emph{Planck} 2018 data  + BOSS DR12 (FS-2021), with and without $S_8$ from DES-Y3 \vspace{4pt}}
\tbl{Mean (best-fit) $\pm1\sigma$ constraints on the cosmological parameters in the EDE scenario from the combination of \emph{Planck} 2018 TT+TE+EE+low $\ell$+lensing data with with EFT-based  BOSS BAO+FS-2021  and eBOSS ELG BAO+FS (suppressed in notation for brevity) measurements, with and without a DES-Y3 $S_8$ prior. The upper limit on $f_{\rm EDE}$ is quoted at 95\% CL.}
  {\begin{tabular}{|l|c|c|}
    \hline\hline Parameter & \emph{Planck}  + BOSS DR12 BAO+FS-2021 &~~~ $+$ DES-Y3 $(S_8)$ ~~~\\ \hline \hline

    {\boldmath$\ln(10^{10} A_\mathrm{s})$} & $3.04 \, (3.041) \, _{-0.015}^{+0.015}$ &  $3.033 \, (3.041 )\, _{-0.014}^{+0.015}$ \\

    {\boldmath$n_\mathrm{s}$} & $0.9693 \, (0.9746) \,_{-0.0076}^{+0.0049}$ & $0.9692 \, (0.9716)\,_{-0.0058}^{+0.0044}$ \\

    {\boldmath$100\theta_\mathrm{s}$} & $1.042 \, (1.042) \, _{-0.00032}^{+0.00035}$ &  $1.042 \, (1.042) \,_{-0.00031}^{+0.00036}$\\

    {\boldmath$\Omega_\mathrm{b} h^2$} &  $0.0225 \, (2.265) \,_{-0.00022}^{+0.00018}$ & $0.02254 \, ( 0.02276)\,_{-0.00019}^{+0.00016}$  \\

    {\boldmath$\Omega_\mathrm{cdm} h^2$} & $0.1225 \, ( 0.1236)\,_{-0.0033}^{+0.0015}$ & $0.1204 \, (0.1211 )\,_{-0.0021}^{+0.001}$ \\

    {\boldmath$\tau_\mathrm{reio}$} & $0.05119 \, (0.05306) _{-0.0071}^{+0.0071}$ & $0.05009 \, ( 0.05435)\,_{-0.0069}^{+0.0076}$ \\

    {\boldmath$\mathrm{log}_{10}(z_c)$} & $3.663 \, (3.83) \, _{-0.31}^{+0.25}$ & $3.732 \, (3.905)\,_{-0.33}^{+0.39}$ \\

    {\boldmath$f_\mathrm{EDE} $} & $ < 0.08028\,(0.04702)$ & $<0.05893\,(0.0387)$ \\

    {\boldmath$\theta_i$} &  $2.0 \, (2.903) \,_{-0.39}^{+1.1}$& $1.945 \, (2.763)\,_{-0.433}^{+1.13}$\\

    \hline
    $H_0 \, [\mathrm{km/s/Mpc}]$ & $68.33 \, (68.99) \, _{-1.1}^{+0.58}$ &  $68.5 \, (69.2)\,_{-0.77}^{+0.46}$ \\

    $\Omega_\mathrm{m}$ & $0.3121 \, (0.3087) \, _{-0.0063}^{+0.0063}$  & $0.306 \,0.3018 \,_{-0.0055}^{+0.0055}$ \\

    $\sigma_8$ &  $0.8129 \, (0.815)\,_{-0.01}^{+0.0072}$ & $0.8044 \, ( 0.8067)\,_{-0.0074}^{+0.0063}$  \\
    
    $S_8$ &  $0.829 \, (0.809)\, _{-0.012}^{+0.012}$ &  $0.812 \,(0.809) \,_{-0.009}^{+0.009}$
    \\
    \hline
  \end{tabular} \label{tab:BOSS_2021}}
 \end{table}

\subsection{EDE Meets LSS: circa 2023} 
\label{sec:BossFSNEW}

The EDE {\it Planck}+BOSS 2020 results 
were derived 
from the 
BOSS FS-2019 likelihood. This 
likelihood 
was based 
on the public BOSS DR12 products~\cite{Alam:2016hwk}: 
power spectrum measurements, 
the window function kernels, 
and the covariance matrices. 
This likelihood was gradually 
updated with two major versions:
2020 and 2021. The 2020 version, released in Fall 2020, 
included
the new analytic covariance,\footnote{Based on Refs.~\cite{Wadekar:2019rdu,Wadekar:2020hax};
also see~\cite{Philcox:2020zyp,Chudaykin:2020hbf}
for covariance matrix treatment in the 
context of the EFT-based FS likelihood. 
} 
and new measurements
of the power spectrum 
from a custom-made estimator. 
This likelihood also assumed 
more restrictive priors on
nuisance parameters, following 
the discussions in~\cite{Chudaykin:2020aoj}.
In order to simplify the comparison between 
FS likelihoods,  
we will focus on the CMB-free
$\Lambda$CDM cosmological parameters
extracted from analyses with a fixed 
spectral tilt $n_s$ and without the BAO in what follows. 

At the $\Lambda$CDM level, 
the FS-2020 likelihood
results are consistent 
with those from FS-2019 \cite{Ivanov:2019pdj},
with the biggest change being 
a $2\sigma$ shift of $\omega_{cdm}$:
$0.1113 \pm 0.0046$ in FS-2019
versus $0.1203 \pm 0.0054$ in FS-2020.
We stress that the likelihood 
update did not noticeably 
affect
the optimal value of $\sigma_8$, 
$0.721 \pm 0.041$ in FS-2019 \cite{Ivanov:2019pdj} vs. 
$0.734 \pm 0.043$ in FS-2020~\cite{Wadekar:2019rdu}.
In the context of 
the EDE model, {\it Planck}+BOSS BAO+FS-2020 yields
the upper limit 
$f_{\rm EDE}<0.088$ (95\% C.L.).
The small degradation compared to 
the {\it Planck} + BOSS FS 2019 results 
is mainly due to $\omega_{cdm}$, which is slightly higher in the  BOSS FS 2020
likelihood, and which confirms
the original arguments 
of Ref.~\cite{Ivanov:2020ril}
that this is the main parameter
responsible for the 
constraints from the BOSS
FS side.

A new update
of the BOSS full-shape 
likelihood took place in 
December 2021. 
As mentioned above, the BOSS FS-2021 likelihood
includes the BOSS galaxy 
power spectrum monopole, quadrupole,
and hexadecapole moment,
the real-space proxy $Q_0$,
as well as the bispectrum
monopole. The power
spectra and bispectra are extracted from the BOSS data using new 
optimal window-free estimators. 
The power spectrum scale cuts 
of FS-2021  
are somewhat more conservative
than before: $k_{\rm max}=0.2~h$Mpc$^{-1}$.
In the $\Lambda$CDM context, the new likelihood gave
$\sigma_8=0.722^{+0.032}_{-0.036}$, consistent 
with both BOSS FS-2019 and BOSS FS-2020.
We stress that this result 
is derived with optimal estimators
that appropriately include the BOSS
window function normalization. 
We also note that the apparent mismatch 
in the official BOSS window function 
normalization~\cite{Chen:2021wdi,Philcox:2021kcw} did not  
affect the results of the 
the FS-2019 and BOSS FS-2020
likelihoods, which had a custom 
window function implementation that 
mitigated the normalization error 
of the BOSS collaboration. 

Another important development 
was the creation of the eBOSS 
full-shape likelihood 
for the emission line galaxy (ELG) sample~\cite{Ivanov:2021zmi}. This sample probes
higher redshifts $z\simeq 0.85$ than BOSS, and hence 
gives an additional lever on  
large-scale structure growth. 
The final results for the EDE model 
fit to the {\it Planck} 2018 (PR3 {\tt Plik}), BOSS BAO+FS-2021, 
and eBOSS BAO+FS ELG data are presented in Table~\ref{tab:BOSS_2021}
and Figure~\ref{fig:BOSS-S8_2021}.
The net result is $f_{\rm EDE}<0.080$ (95\% C.L.) (best-fit 
$f_{\rm EDE}=0.047$). 
This constraint
is marginally weaker than 
the original {\it Planck}+BOSS BAO+FS-2019 
result~\cite{Ivanov:2020ril}. At the same time,  
it is tighter than the {\it Planck}+BOSS BAO+FS-2020
constraint, which can be attributed to 
additional BOSS and eBOSS statistics
present in that likelihood. 
All in all, we see
that {\it Planck} + BOSS EDE results
together with the new likelihood based 
on the \texttt{CLASS-PT}
code and the baseline prior choices of Ref.~\cite{Chudaykin:2020aoj}
are fully consistent
with those from the 2020 EDE {\it Planck}+BOSS constraints~\cite{Ivanov:2020ril}. 

Finally, we combine BOSS BAO+FS with the 
weak lensing data. 
Adding the $S_8$ prior
from DES-Y3, $S_8=0.776\pm 0.017$~\cite{DES:2021wwk}, we  
obtain our final {\it Planck} 2018+BOSS+
$S_8$ results presented in Table~\ref{tab:BOSS_2021}
and Figure~\ref{fig:BOSS-S8_2021} (we omit the eBOSS ELG data in the caption for brevity.)
We obtain an upper limit
$f_{\rm EDE}<0.059$ (95\% C.L.) with a best-fit 
$f_{\rm EDE}=0.038$. These results 
are fully consistent with 
our previous 
estimate $f_{\rm EDE}<0.053$~\cite{Ivanov:2020ril}. 
At face value, the 
Hubble constant constraint, $H_0=68.50^{+0.46}_{-0.77}$
km/s/Mpc, is strongly 
inconsistent with the SH0ES 
measurements.

Note that DES-Y3 data cannot be 
easily combined with other weak lensing 
measurements due to 
significant sky overlaps; see Ref.~\cite{Kilo-DegreeSurvey:2023gfr} for more detail. 
This is the reason we use only DES-Y3 in our 
final LSS+CMB analysis. Also, note that 
there are data sets that have 
become available while 
this work was being completed: 
eBOSS EFT-based full-shape likelihood for quasars~\cite{Chudaykin:2022nru}, 
and the BOSS bispectrum
quadrupole and hexadecapole 
moments~\cite{Ivanov:2023qzb}. It would be interesting to 
include these data sets in a future analysis.

%%%%%%%%%%%%%%%%%%%%%%%%%%%%%%%%%%%%%%%%%%%%%%%%%%%%%%%%%%%%%%%%%%%%%%%%%%%%%%%%%%%%
%\clearpage
\subsection{EFT-based vs. Standard  BOSS likelihoods}
\label{app:fs8}

As a final check on the intuition developed in the previous analyses, we perform an analysis with the EFT-based BOSS likelihood replaced by the  ``consensus'' BOSS DR12 FS+BAO likelihood \cite{Alam:2016hwk}. The latter is a highly compressed representation of the data, generated from a fit of $f\sigma_8$ and the BAO parameters to BOSS data, using a fixed power spectrum template computed for a fiducial $\Lambda$CDM cosmology consistent with {\it Planck}.\footnote{Note that the template for the galaxy power spectrum used in fixed shape analyses is based on approximate phenomenological models (i.e., not the EFT of LSS), and hence may be affected by uncontrollable 
systematic errors.
}

\begin{figure}[ht!]
\begin{center}
\includegraphics[width=0.8\textwidth]{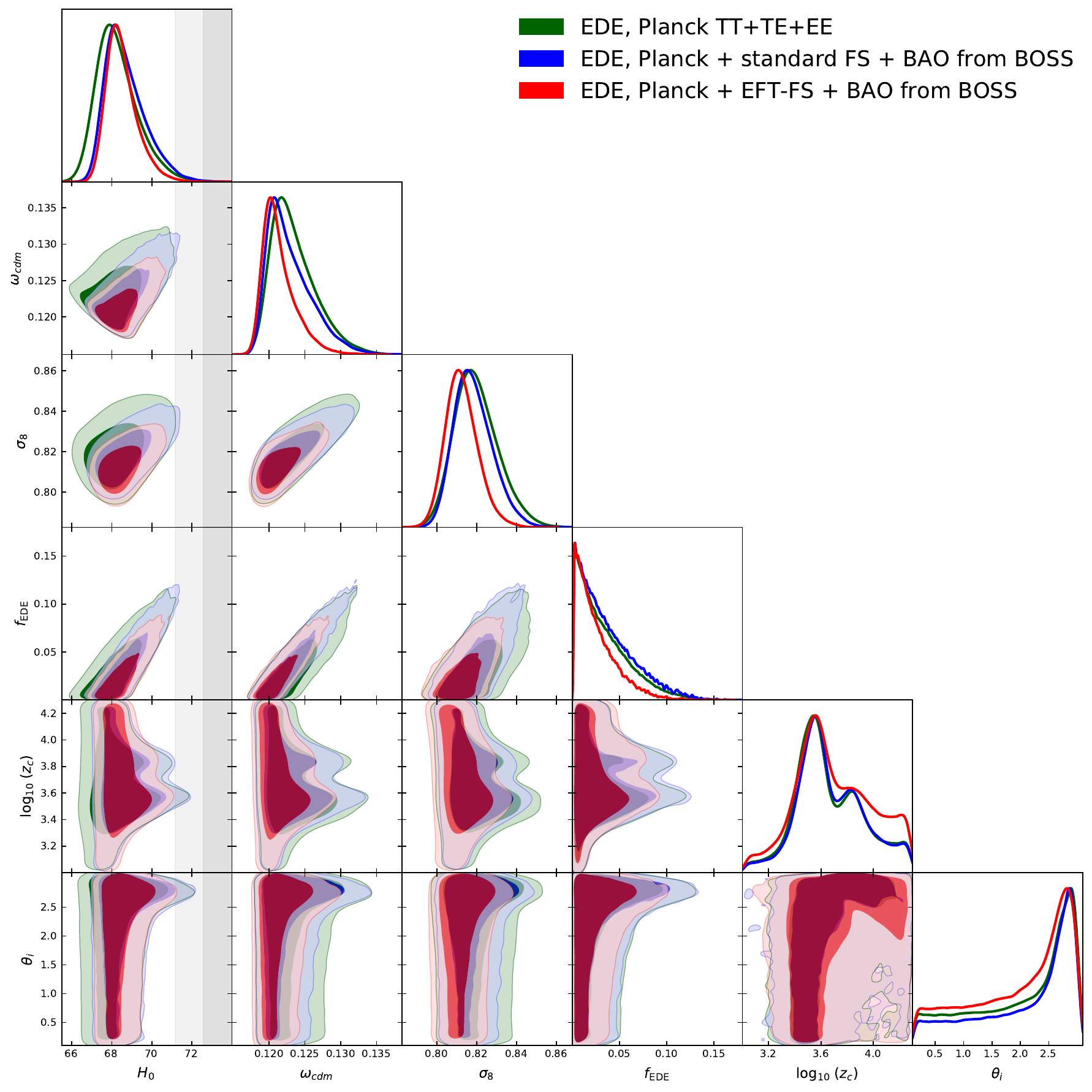}
\end{center}
\caption{
Posterior distributions for the
parameters extracted from the joint {\it Planck} 2018 TT+TE+EE+low $\ell$+lensing + BOSS FS+BAO data.
We show the results obtained
using the standard FS+BAO likelihood (in blue)
and the EFT-based likelihood (in red).
For reference, we also display the  
 constraints from the {\it Planck} 2018 primary CMB data alone (TT+TE+EE), obtained in \cite{Hill:2020osr}. 
The gray band shows the $H_0$ measurement from SH0ES, for comparison.
The dark-shaded 
and light-shaded 
contours mark 
$68\%$ and $95\%$ confidence intervals, respectively.
\label{fig:planck+BOSS-fs8} } 
\end{figure}

The results are shown in Fig.~\ref{fig:planck+BOSS-fs8} and Table~\ref{table:params-P18+BOSS-eft_vs_stand}. From Fig.~\ref{fig:planck+BOSS-fs8} one may appreciate a significantly improved constraint on larger values $\omega_{\rm cdm}$ when the EFT-based likelihood is used, evidenced by a significant suppression of the tail of the 1D posterior of $\omega_{\rm cdm}$. This contributes to the breaking of the degeneracy between $f_{\rm EDE}$ and $\omega_{\rm cdm}$, which in turn restricts the $f_{\rm EDE}$-$H_0$ degeneracy and the ability of the model to be even mildly concordant with the SH0ES $H_0$ measurement.

\begin{table}[htb!]
%\vspace{-5cm}
% \hspace{2.4cm} 
%\centering
\centering{Constraints from \emph{Planck} 2018 data  + BOSS DR12 for EDE ($n=3$)\vspace{4pt}}
\tbl{Mean (best-fit) $\pm1\sigma$ constraints on the cosmological parameters in the EDE scenario with $n=3$, as inferred from the combination of \emph{Planck} 2018 TT+TE+EE+low $\ell$+lensing data with the standard (left column) and EFT-based (right column) BOSS FS+BAO measurements. The upper limit on $f_{\rm EDE}$ is quoted at 95\% CL.}
  %\begin{minipage}{\textwidth}
  {\begin{tabular}{|l|c|c|}
    \hline\hline Parameter &Standard
    BOSS likelihood~~&~~~EFT
    BOSS
    likelihood~~~\\ \hline \hline

    {\boldmath$\ln(10^{10} A_\mathrm{s})$} & $3.053\,(3.060)_{-0.016}^{+0.015}$ & $3.047 \, (3.049) \, \pm 0.014$ \\

    {\boldmath$n_\mathrm{s}$} & $0.9713\,(0.9790)_{-0.0085}^{+0.0055}$ & $0.9696 \, (0.9717)^{+0.0046}_{-0.0068}$ \\

    {\boldmath$100\theta_\mathrm{s}$} & $1.04185 \, (1.04200)_{-0.00032}^{+0.00035}$ & $1.04172 \, (1.04126) \, \pm 0.00032$\\

    {\boldmath$\Omega_\mathrm{b} h^2$} & $0.02256 \,(0.02269)_{-0.00022}^{+0.00018}$ &  $0.02255 \, (0.02245) \, \pm 0.00018$ \\

    {\boldmath$\Omega_\mathrm{cdm} h^2$} & $0.1230\,(0.1269)_{-0.004}^{+0.0018}$ & $0.1215 \, (0.1243)^{+0.0013}_{-0.0029}$ \\

    {\boldmath$\tau_\mathrm{reio}$} & $0.05688\,(0.05755)_{-0.0078}^{+0.0071}$  & $0.0553 \, (0.0543)_{-0.0075}^{+0.0069}$\\

    {\boldmath$\mathrm{log}_{10}(z_c)$} & $3.67\,(3.813)_{-0.27}^{+0.22}$ & $3.71 \, (3.52)^{+0.26}_{-0.33}$ \\

    {\boldmath$f_\mathrm{EDE} $} & $<0.096\,(0.076)$ & $< 0.072 \, (0.047)$\\

    {\boldmath$\theta_i$} & $2.122\,(2.934)_{-0.29}^{+0.98}$ & $2.023(2.734)_{-0.34}^{+1.1} $\\

    \hline

    $H_0 \, [\mathrm{km/s/Mpc}]$ & $68.71\,(69.66)_{-1.2}^{+0.69}$ & $68.54 \, (68.83)^{+0.52}_{-0.95}$ \\

    $\Omega_\mathrm{m}$ & $0.3097\,(0.3096)_{-0.0063}^{+0.0062}$ & $0.3082 \,(0.3120)_{-0.0057}^{+0.0056}$ \\

    $\sigma_8$ & $0.8187\,(0.8286)_{-0.011}^{+0.008}$ & $0.8127 \, (0.8195)_{-0.0091}^{+0.0072}$ \\
    
    $S_8$ & $0.8316 \, (0.8363) \, \pm 0.012$ & $0.8237 \, (0.8275)\, \pm 0.011$ \\
    \hline
  \end{tabular} 
  \label{table:params-P18+BOSS-eft_vs_stand}}
  %\end{minipage}
\end{table}

%%%%%%%%%%%%%%%%%%%%%%%%%%%%%%%%%%%%%%%%%%%%%%%%%%%%%%%%%%%%%%%%%%%%%%%%%%%%%%%%%%%%

%%%%%%%%%%%%%%%%%%%%%%%%%%%%%%%%%%%%%%%%%%%%%%%%%%%%%%%%%%%%%%%%%%%%%%
%%%%%%%%%%%%%%%%%%%%%%%%%%%%%%%%%%%%%%%%%%%%%%%%%%%%%%%%%%%%%%%%%%%%%%
%\clearpage
\section{Constraints from the Lyman-$\alpha$ Forest}
\label{sec:constraintsLymanalpha}
%%%%%%%%%%%%%%%%%%%%%%%%%%%%%%%%%%%%%%%%%%%%%%%%%%%%%%%%%%%%%%%%%%%%%%
%%%%%%%%%%%%%%%%%%%%%%%%%%%%%%%%%%%%%%%%%%%%%%%%%%%%%%%%%%%%%%%%%%%%%%

As a final investigation of LSS constraints on Early Dark Energy, 
we focus on Lyman-$\alpha$ forest data, namely the 1D Ly$\alpha$ flux power spectra from SDSS DR14 BOSS and eBOSS quasars \cite{Chabanier:2018rga}, referred to as eBOSS Ly$\alpha$ in what follows.  We also consider a similar analysis using a smaller sample of high-resolution Ly$\alpha$ forest measurements extracted from XQ-100~\cite{Irsic:2017sop} and MIKE/HIRES~\cite{Viel:2013fqw}.

The physics of the Ly$\alpha$ forest is reviewed in Sec.~\ref{sec:Lyalpha}. In the context of $\Lambda$CDM, Ly$\alpha$ data prefer lower values of $n_s$ and $\Omega_m h$ than CMB data sets \cite{Chabanier:2018rga,Palanque-Delabrouille:2019iyz,Esposito:2022plo}. One of the traditional hurdles for the use of Lyman-$\alpha$ data in joint cosmological analyses has been the high computational cost. This stems from the need to run costly hydrodynamical simulations so that multiple nuisance parameters can be carefully marginalized over to make robust theory predictions for the Ly$\alpha$ forest power spectrum. Here, the leading challenge arises from carefully accounting for the thermal and ionization history of the intergalactic medium, as well as potential feedback effects at low redshifts. In practice, computing the $\Lambda$CDM prediction for the Ly$\alpha$ forest flux spectra is conventionally done via interpolation over a grid of these hydrodynamical simulations. In light of the significant computational expense, in our analysis of EDE we instead use a two-parameter compressed likelihood, analogous to the compression of BOSS full-shape and BAO data in the standard (non-EFT) likelihood \cite{Alam:2016hwk}. This approach was validated for a range of $\Lambda$CDM and beyond-$\Lambda$CDM cosmological models in Ref.~\cite{Pedersen:2022anu}. The compressed eBOSS Ly$\alpha$ likelihood is characterized by a dimensionless amplitude $\Delta_L^2\equiv k^3P_{\rm lin}(k_p,z_p)/(2\pi^2)$ and spectral index $n_L\equiv d\ln P_{\rm lin}(k_p, z_p)/d\ln k$, both evaluated at a pivot redshift $z_p=3$ and pivot scale $k_p=0.009~{\rm s/km}$ \cite{SDSS:2004aee}. Our compressed likelihood is built from a 2D Gaussian in $\Delta_L^2$ and $n_L$, with an a priori non-zero correlation coefficient, fit to a 30$\times$30 grid of $\{ \Delta_L^2, n_L\}$ values taken from the contour shown in Fig.~20 of Ref.~\cite{Chabanier:2018rga}. The resulting best-fit parameters for the 2D Gaussian are shown in~Table~\ref{tab:ly_alpha_gauss_params}. The log-likelihood is then given by 
\begin{align}
\label{eq:Lyalphalike}
\log\mathcal{L}= -\frac{1}{2(1-\rho^2)}
\bigg\{ \Delta x^2-2\rho\Delta x\Delta y+\Delta y^2\bigg\},
\end{align}
where $\Delta x\equiv (\Delta_{L}^2-\bar{\Delta}_{L}^{2})/\sigma_{\Delta_L^2}$ and $\Delta y\equiv (n_{L}-\bar{n}_{L})/\sigma_{n_L}$, overbar $(\, \bar{\,}\,)$ and $\sigma$ denote the mean and errors of the 2D Gaussian, and $\rho$ is the correlation coefficient between $\sigma_{\Delta_{L}^2}$ and $\sigma_{n_L}$. The best-fit parameters describing the eBOSS and XQ-100 Ly$\alpha$ data sets are shown in Table~\ref{tab:ly_alpha_gauss_params}. For further details we refer the reader to Ref.~\cite{Goldstein:2023gnw}.

%%%%%%%%%
% Results 
%%%%%%%%%

\begin{figure}[h!]
\includegraphics[width=\linewidth]{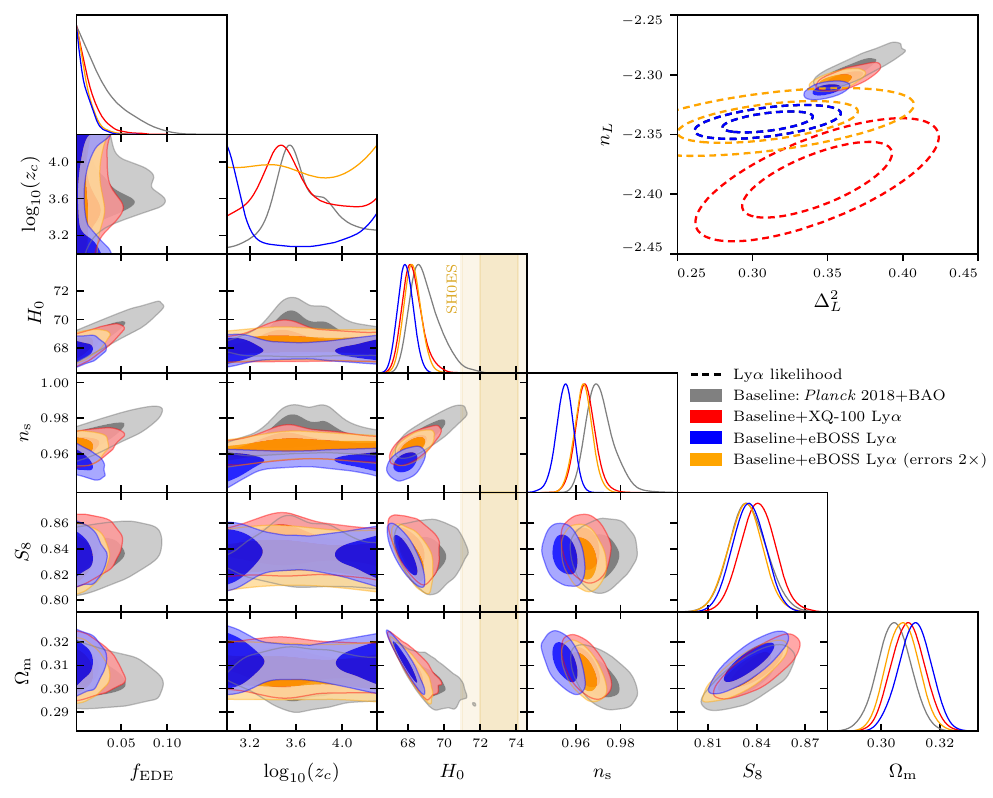}
\caption{ The marginalized posteriors before and after including Ly$\alpha$ forest data for the canonical axion-like EDE model and $\Lambda$CDM parameters of interest. The baseline analysis (grey) consists of \emph{Planck} 2018 high-$\ell$ (TT+TE+EE) and low-$\ell$ (TT+EE), in addition to BOSS DR12, SDSS MGS, and 6dFGS BAO data. The inclusion of eBOSS (blue) or XQ-100 (red) Ly$\alpha$ data sets reduces the upper bound on $f_{\rm EDE}$ significantly. In particular, both analyses which include Ly$\alpha$ data strongly exclude $H_0$ values capable of resolving the Hubble tension. The panel in the upper right depicts constraints for these data sets in the $\Delta_L^2$ -- $n_L$ plane. Ly$\alpha$ likelihoods alone (dashed lines) are also included. Although both Ly$\alpha$ likelihoods are in noticeable tension with the baseline analysis (in EDE or in $\Lambda$CDM), EDE cosmologies that can reduce the Hubble tension are inconsistent with the direction of this tension, and in fact exacerbate it. Even if the Ly$\alpha$ errors on $\sigma_{n_L}$ and $\sigma_{\Delta_L^2}$ are doubled, our conclusion with respect to the eBOSS Ly$\alpha$ likelihood (orange) is unchanged. See Ref.~\cite{Goldstein:2023gnw} for further details.
}  
 \label{fig:ly_alpha_marg_post_and_likelihood}
\end{figure}

The expectation from Sec.~\ref{sec:Lyalpha} is indeed borne out in the joint analysis of CMB, BAO, and Lyman-$\alpha$ forest data~\cite{Goldstein:2023gnw}.  Results of this analysis are shown in Fig.~\ref{fig:ly_alpha_marg_post_and_likelihood} and Table~\ref{tab:posterior_param_limits}.   Jointly analyzing the eBOSS Lyman-$\alpha$ forest data with \emph{Planck} PR3 ({\tt Plik}) CMB and BAO data lowers the 95\% C.L. upper bound on $f_{\rm EDE}$ from 0.07 (CMB+BAO) to 0.03 (CMB+BAO+Ly$\alpha$) and yields $H_0=67.9_{-0.4}^{+0.4}~{\rm km/s/Mpc}$ (68\% C.L.), with maximum \textit{a posteriori} value $H_0=67.9~{\rm km/s/Mpc}$. Similarly tight constraints are obtained using the MIKE/HIRES and XQ-100 Lyman-$\alpha$ forest data.

Moreover, from the $n_L - \Delta_L^2$ panel of Fig.~\ref{fig:ly_alpha_marg_post_and_likelihood} (top right) one may appreciate that the parameter constraints (solid contours) are in significant tension with both the XQ-100 (red dashed) and eBOSS (blue dashed) Ly$\alpha$ data sets. While a similar tension exists in $\Lambda$CDM \cite{Chabanier:2018rga,Esposito:2022plo}, the parameter space opened up in the EDE fit to \emph{Planck} and BAO data is in the exact opposite direction to that required to solve this ``Ly$\alpha$ tension'': $H_0$-resolving EDE requires larger values of both $n_L$ and $\Delta_L^2$. As a heuristic check on the robustness of this tension, we repeat the analysis with the error bars $\sigma_{n_L}$ and $\sigma_{\Delta_{L}^2}$ artificially doubled in the Ly$\alpha$ likelihood in Eq.~\eqref{eq:Lyalphalike} (orange contours), finding little change in our results.

Taken at face value, these results thus indicate that canonical axion-like EDE is effectively ruled out by the Lyman-$\alpha$ forest as a resolution of the Hubble tension.  Due to the complexity of modeling this observable (e.g.,~\cite{Fernandez:2023grg}), however, further work will be necessary to cement this conclusion.  The upcoming DESI Lyman-$\alpha$ data will also play a key role in this effort.

\begin{table}[h!]
    \tbl{Best-fit parameter values for the compressed eBOSS Ly$\alpha$ likelihood used in this work.}
    {\begin{tabular}{ |c|c|c|c|c|c| } 
     \hline
    Ly$\alpha$ data set & $\Delta_{L^2}^0$ & $n_{L}^0$ & $\sigma_{\Delta_L^2}$ & $\sigma_{n_L}$ & $\rho$ \\ 
    \hline
     eBOSS &  0.310 & -2.340 & 0.020 &  0.006 & 0.512  \\ 
     \hline
   
     XQ-100/MIKE-HIRES & 0.343 &  -2.388 &  0.033 & 0.021 & 0.694  \\
    \hline
    \end{tabular}
   \label{tab:ly_alpha_gauss_params}}
\end{table}

\begin{table}[!t]
    %\centering
    \tbl{Marginalized constraints on cosmological parameters for the canonical axion-like EDE model from the Lyman-$\alpha$ data sets considered in this work. For each data set, we report the posterior mean and 68\% upper and lower limits for all parameters, except for $f_{\rm EDE}$, for which we report 95\% upper limits.  Maximum \textit{a posteriori} values are shown in parentheses.}
    {\begin{tabular}{ |c|c|c|c|} %c|c|c| } 
      \hline
      & $f_{\rm EDE} $ &    $n_s$ &   $H_0$ [km/s/Mpc] 
      \\ 
      \hline
      \emph{Planck} + BAO     &  $<0.079~(0.059)$ &   $0.9710_{-0.0072}^{+0.0045}~(0.9778)$ &  $68.98_{-1.02}^{+0.55}~(70.11)$ 
      \\ 
      \hline
      +eBOSS Ly$\alpha$        &  $<0.027~(0.015)$ &    $0.9549_{-0.0035}^{+0.0040}~(0.9525)$ &  $67.86_{-0.43}^{+0.43}~(67.89)$  
      \\ 
      \hline
      +XQ-100 Ly$\alpha$       &  $<0.041~(0.012)$ & 
      $0.9643_{-0.0045}^{+0.0040}~(0.9633 )$ &  $68.24_{-0.63}^{+0.45}~(68.21)$ 
      \\ 
      \hline
    \end{tabular}
   \label{tab:posterior_param_limits}}
\end{table}

%%%%%%%%%%%%%%%%%%%%%%%%%%%%%%%%%%%%%%%%%%%%%%%%%%%%%%%%%%%%%%%%%%%%%%
%%%%%%%%%%%%%%%%%%%%%%%%%%%%%%%%%%%%%%%%%%%%%%%%%%%%%%%%%%%%%%%%%%%%%%
%\clearpage
\section{Priors and Prior Volume Effects}
\label{sec:priors}
%%%%%%%%%%%%%%%%%%%%%%%%%%%%%%%%%%%%%%%%%%%%%%%%%%%%%%%%%%%%%%%%%%%%%%
%%%%%%%%%%%%%%%%%%%%%%%%%%%%%%%%%%%%%%%%%%%%%%%%%%%%%%%%%%%%%%%%%%%%%%

\subsection{Choice of Priors}

Finally, we return to the cosmological aspects of the EDE model, with an eye towards data analysis. A Bayesian analysis of the model, e.g., posterior sampling via Markov chain Monte Carlo (MCMC) -- the staple of the field 
-- necessitates a choice of priors on model parameters. While there is some arbitrariness in the choice of priors, the standard approach to analyses of extensions to $\Lambda$CDM has been broad uniform priors. 

Analyses of EDE, including those presented here, typically impose uniform priors on the {\it derived parameters} $f_{\rm EDE}$ and $z_c$, instead of the particle physics model parameters $m$ and $f$ appearing in the potential $V(\varphi)$. MCMC analyses can then be performed by sampling over $f_{\rm EDE}$ and $z_c$. The motivation for this approach is to reduce the computational expense of the MCMC, since directly sampling in the derived parameters to which the data is most sensitive tends to dramatically improve the time to convergence of MCMC chains.  This approach is justified by a lack of any prior knowledge of either $f $ and $m$ or $f_{\rm EDE}$ and $z_c$, and thus the broad uniform (`uninformative') priors may be imposed in whatever parametrization is most convenient to do so.

However, string theory does provide priors on axion model parameters: The mass and decay constant of axions in the string axiverse \cite{Arvanitaki:2009fg,Cicoli:2012sz} are both log distributed \cite{Broeckel:2021dpz}. In particular, Planck-scale axion decay constants ($f\approx M_{\rm pl}$) are in conflict with both the general expectations of any theory of quantum gravity and with string theory specifically \cite{Banks:2003sx, Rudelius:2014wla, Bachlechner:2015qja, Conlon:2016aea, long:2016jvd,Heidenreich:2015nta, Hebecker:2016dsw, Hebecker:2018ofv}.
The resulting implications for EDE have been studied in Refs.~\cite{Rudelius:2022gyu} and \cite{Cicoli:2023qri}.
The statistical predictions of the string landscape for axion model parameters is an area of active research; see, e.g., \cite{Broeckel:2021dpz,Halverson:2019cmy,Mehta:2020kwu,Mehta:2021pwf,Demirtas:2021gsq}.  We note axiverse constructions can also emerge in field theory, e.g.,~\cite{Alexander:2023wgk,Maleknejad:2022gyf}, where the distribution of axion masses has also been computed \cite{Alexander:2023wgk}.

\begin{figure}[h!]
    \centering
    \includegraphics[width=0.75 \textwidth]{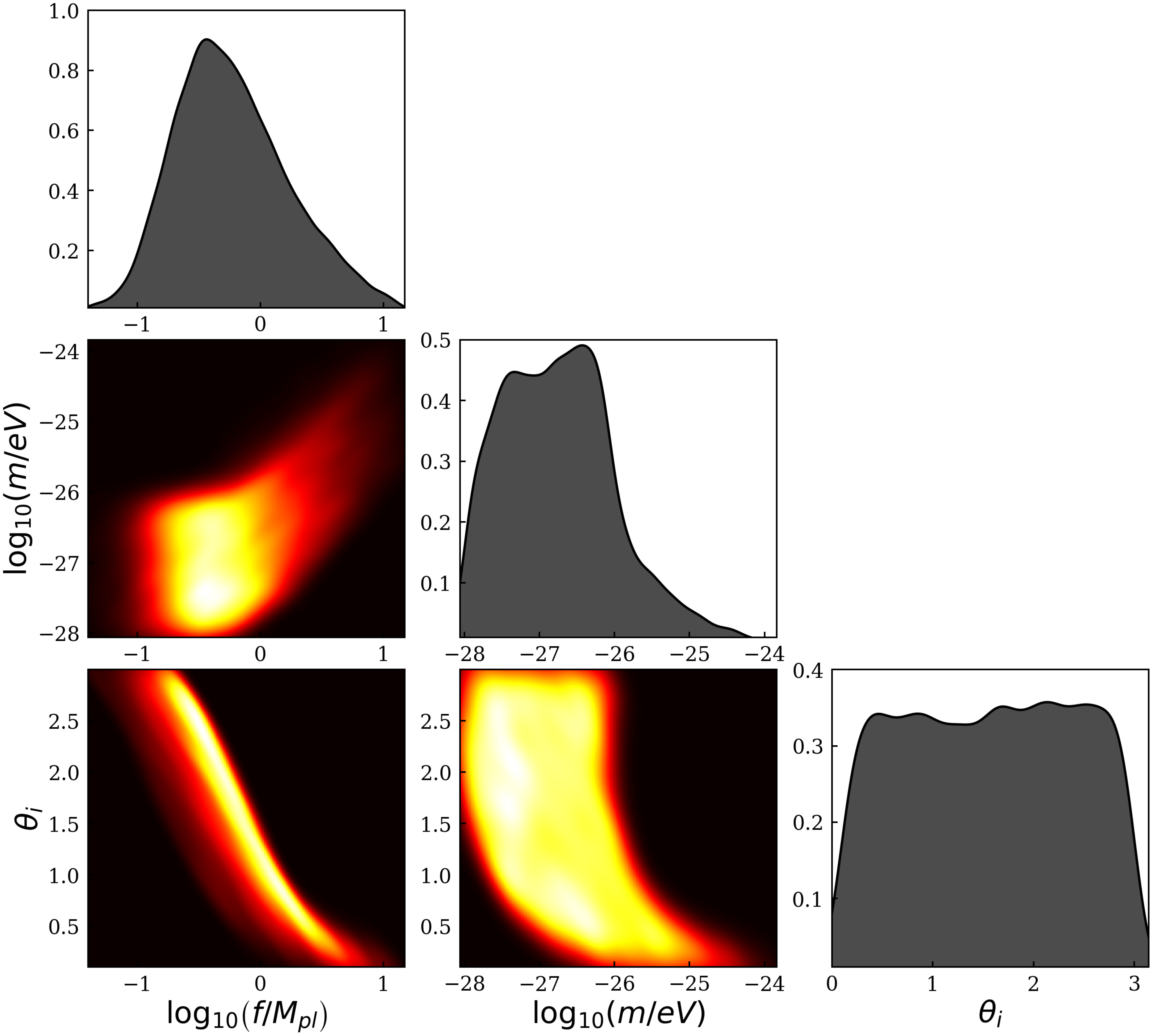}
    \caption{The effective priors on EDE particle physics parameters $f$ and $m$ implied by uniform priors on $f_{\rm EDE}$, $\log_{10}(z_c)$, and $\theta_i$. We see that the distribution of $f$ is peaked at $f=0.59 M_{pl}$ and extends well above the Planck scale. }
    \label{fig:smithpriors}
\end{figure}

To assess the consistency of theoretical priors and the priors typically used in MCMC analyses of EDE, we generate 10,000 realizations of EDE for $f_{\rm EDE}$, $\log_{10}(z_c)$, and $\theta_i$ drawn from uniform distributions in the range $[0.01,0.25]$, $[3.1,4.2]$, and $[0.1,3.0]$ respectively. The results are shown in Fig.~\ref{fig:smithpriors}. From this one may appreciate that the commonly used uniform priors on EDE parameters are an implicit prior on the axion decay constant that is peaked at a near-Planckian value $\approx 0.59 M_{pl}$, in conflict with the prior from theory \cite{Banks:2003sx, Rudelius:2014wla, Bachlechner:2015qja, Conlon:2016aea, long:2016jvd,Heidenreich:2015nta, Hebecker:2016dsw, Hebecker:2018ofv}. Meanwhile the effective prior on the mass is that it should lie in the range $\log_{10}(m/{\rm eV}) \approx [-28,-25]$. This is an odd choice given that axions conceivably explain one if not both of dark matter and dark energy, and this mass range covers neither of those possibilities (see Ref.~\cite{Marsh:2015xka} for a review of axion cosmology).

\begin{figure}[h!]
\centering
 \includegraphics[width=\textwidth]{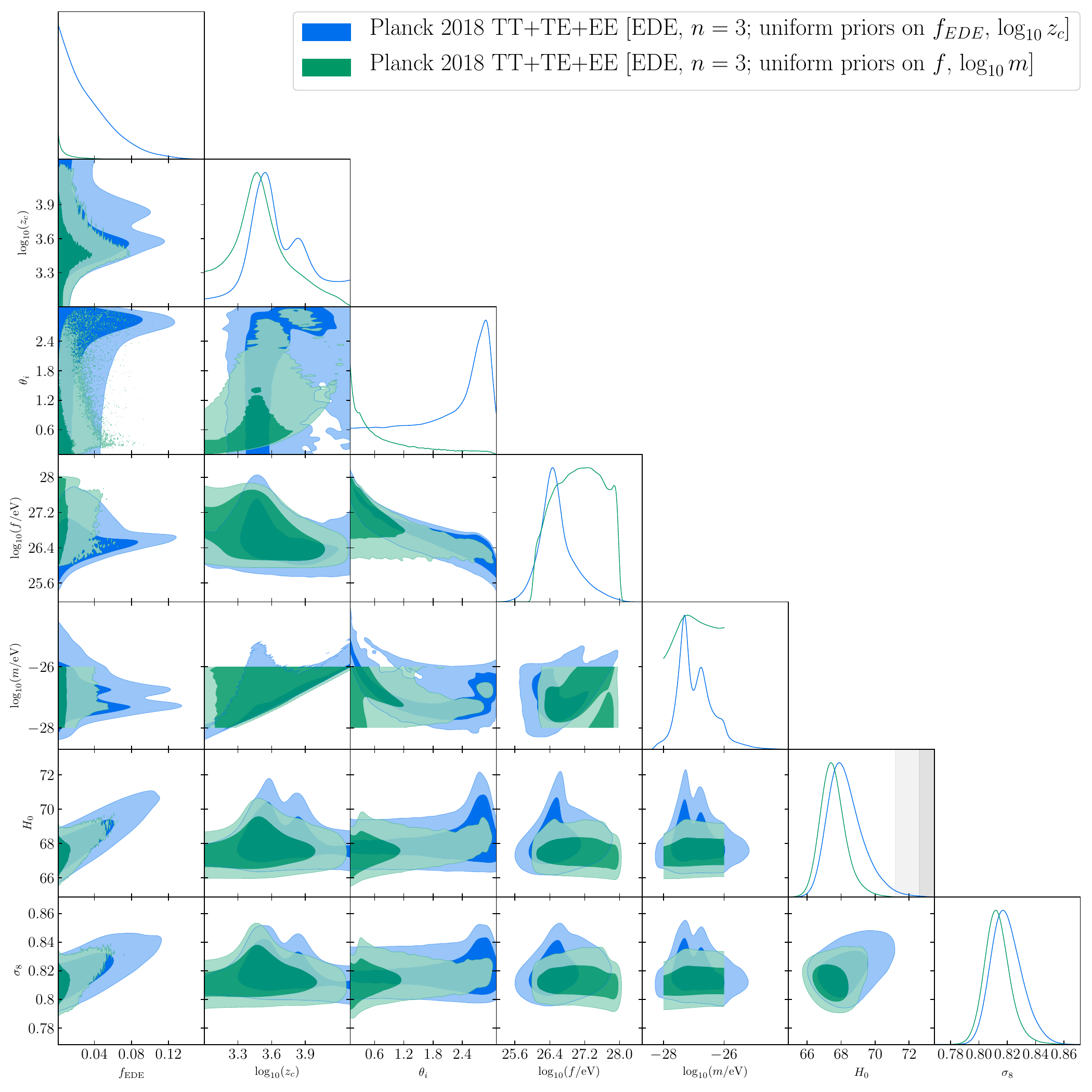}
    \caption{ Constraints on EDE from \emph{Planck} PR3 primary CMB data alone, analyzed with the \texttt{Plik} likelihood, with priors imposed on the particle physics parameters $f$ and ${\rm log}_{10}(m)$ (green contours) or on the effective EDE parameters $f_{\rm EDE}$ and $\log_{10}(z_c)$ (blue contours).}  
    \label{fig:EDEconstraints_priors}
\end{figure}

With this in mind, we now repeat the analysis of {\it Planck} CMB data (TT+EE+TE) using theory-motivated priors on the axion mass and decay constant. Specifically, we make a choice of uniform priors on $f$ and $\log_{10}{(m)}$. We display the results for this analysis in Fig.~\ref{fig:EDEconstraints_priors},
which are compared to an equivalent analysis with uniform priors on the \textit{derived} EDE parameters. We find an even more restrictive upper bound,  $f_{\rm EDE} <  0.041$
at 95\% C.L.

The take-away from this result is that, despite using dramatically different priors, the conclusion for EDE is \textit{unchanged}. Both parametrizations of the model, whether using particle physics or derived parameters results in the same conclusion. This strongly suggests that the Bayesian preference of CMB and LSS data sets for $\Lambda$CDM over EDE is not merely an artifact of the choice of priors.

\subsection{Prior Volume Effects \& The Inclusion of SH0ES in the combined data set}

A built-in feature of the canonical EDE model  is that the parameters $z_c$ and $\theta_i$ are unconstrained in the $\Lambda$CDM limit, $f_{\rm EDE}\rightarrow 0$. This property is shared by {\it any} multi-parameter extension to $\Lambda$CDM where one parameter predominantly controls the magnitude of the departure from $\Lambda$CDM. Meanwhile, the parameters $z_c$ and $\theta_i$ 
have to be fine-tuned in the $H_0$-tension-resolving regime, $f_{\rm EDE} \approx 0.1$, 
which creates the possibility for a potential pitfall in the MCMC analysis. That is, it could happen that an MCMC analysis may cross a threshold Gelman-Rubin statistic and be deemed converged, but not have been run long enough to fully explore finely-tuned minima of the likelihood, such as the high-$H_0$ regime of EDE ($f_{\rm EDE}\approx 0.1$).
One might therefore be concerned that the marginalized parameter constraints of the MCMC do not accurately reflect the true Bayesian constraints.

One mitigation strategy (advocated for explicitly in, e.g.,~Refs.~\cite{Niedermann:2020dwg,Berghaus:2022cwf}, and utilized in nearly all works on EDE) is to include the SH0ES $H_0$ measurement as a prior in the MCMC analysis. Such an analysis cannot possibly test concordance of SH0ES with other data sets, since concordance is by definition the compatibility of constraints arising from independent analyses of differing data sets.

To test concordance while still ensuring the $H_0$-tension-resolving regime is fully explored, one may simply include the SH0ES $H_0$ measurement as a prior in sampling, and then afterwards remove this prior, and recompute the posterior probability of every sample in the chain (so-called ``importance reweighting'' of MCMC samples). This is precisely the strategy of Sec.~\ref{sec:constraintsWL}, with results shown in Fig.~\ref{fig:no-SH0ES} and Table \ref{table:params-uberlikelihoodKiDSHSCNoSH0ES}. The exponential fall-off of the $f_{\rm EDE}$ posterior is consistent with that seen in the analyses of Secs.~\ref{sec:constraintsCMB}, \ref{sec:constraintsBOSS}, and Sec.~\ref{sec:constraintsLymanalpha}, where the SH0ES measurement was not taken as a prior, and indicates that the posterior distributions in these analyses are an accurate representation of the Bayesian constraints on the model.

\subsection{Mean Likelihood Profile}

In order to understand the nature of our 
CMB+LSS
constraints on $f_{\rm EDE}$, it 
is useful to compare the Bayesian 
posteriors 
with the \textit{mean
likelihood profile} \cite{Lewis:2002ah}
for this parameter. 
The mean likelihood profile has been used in past CMB analyses in, e.g., Refs.~\cite{LewisBridle2002,Montier:2014bra,Martin:2010kz,Audren:2012wb,Planck2015likelihood}.
The mean likelihood curve is obtained by
averaging the likelihood 
in the MCMC chain for a given 
binned value of $f_{\rm EDE}$ over all 
other 
cosmological parameters. 
It is a smeared version 
of the usual \textit{profile likelihood}
curve, in which one instead maximizes the 
likelihood for $f_{\rm EDE}$  over 
other cosmological parameters. 
The mean likelihood
curve, however, has the important property 
that it shows how much the 
other cosmological parameters 
need to be tuned in order to provide a 
good fit. If the mean likelihood 
is bigger with non-zero $f_{\rm EDE}$, 
it suggests that this parameter indeed
improves the fit on average. 
The mean likelihood 
is a cheap statistic that suppresses 
marginalization 
(prior volume) bias~\cite{Lewis:2002ah}.\footnote{Strictly speaking, the calculation of the mean likelihood 
depends on the prior, e.g., through the parameter range of the samples in the case of a flat prior on a relevant parameter. The bulk of the prior dependence
is, however, cancelled out in the 
mean likelihood estimator. 
} 
In particular, 
in the limit $f_{\rm EDE}\to 0$
the mean likelihood will simply 
be a constant function of 
$\log_{10}(z_c)$
and $\theta_i$,
which  fully  takes into 
account the fact that the EDE 
model itself will not depend on these 
parameters in that limit. 

\begin{figure}[ht!]
\begin{center}
\includegraphics[width=0.5\textwidth]{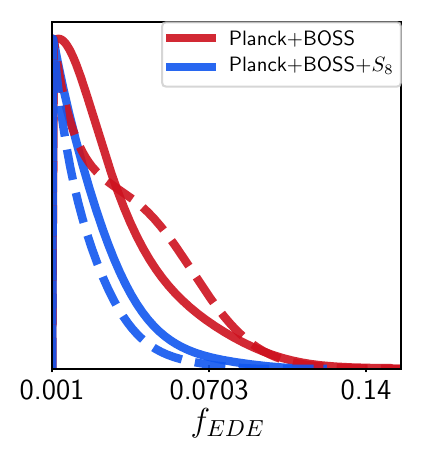}
\end{center}
\caption{1D posterior distributions (solid lines) and 
mean likelihood profiles (dashed lines) for $f_{\rm EDE}$
derived from MCMC chains for the {\it Planck}+BOSS (in red)
and {\it Planck}+BOSS+$S_8$ (in blue) data sets. 
\label{fig:mean_prof} } 
\end{figure}

The mean likelihood profiles 
for $f_{\rm EDE}$ extracted from the MCMC chains for 
{\it Planck} + BOSS and {\it Planck} + BOSS + $S_8$
are displayed in Fig.~\ref{fig:mean_prof}, 
where solid lines represent 
the marginalized Bayesian posteriors, while the dashed
lines depict the binned mean likelihood
profiles. The mean likelihood 
curves closely follow the 
1D marginalized posterior curves: they monotonically decrease 
as $f_{\rm EDE}$ increases. 
There are no noticeable flat regions 
in the $f_{\rm EDE}$ mean likelihood, which 
suggests that the posterior
is indeed determined by the likelihood, rather 
than the prior volume. 
We see that on average, the EDE 
provides a \textit{worse} fit 
to the data than $\Lambda$CDM. 
Hence, the shape of the posterior
does genuinely reflect the behavior of the 
likelihood. 
As we will discuss in detail
later, this behavior
is fully consistent with the 
profile likelihood results of Ref.~\cite{Herold:2021ksg}. 
On the one hand, EDE can
fit the {\it Planck}+BOSS BAO+FS
data slightly better than $\Lambda$CDM
($\Delta \chi^2\approx -4$) 
for $f_{\rm EDE}\approx 0.072$
provided that $\theta_i$
and $\log_{10}(z_c)$
are strongly fine-tuned.
On the other hand, 
as Fig.~1 of 
~\cite{Herold:2021ksg} shows,
most of the allowed volume of the 
$\theta_i-\log_{10}(z_c)$ parameter space
returns $\chi^2$-statistics that prefer
$f_{\rm EDE}\approx 0$, i.e., 
$\Lambda$CDM over EDE. This is 
confirmed by the mean likelihood analysis
of~\cite{Ivanov:2020ril} and our Fig.~\ref{fig:mean_prof}: 
the probability of the EDE model
is lower than that of $\Lambda$CDM,
provided that one averages the likelihood over 
possible parameters $\theta_i,\log_{10}(z_c)$ 
instead of maximizing it over them. 
In other words, if one picks values of 
$\theta_i,\log_{10}(z_c)$ at random, 
instead of fine-tuning them,
the {\it Planck}+BOSS BAO+FS (+$S_8$) likelihood would be
maximized at $f_{\rm EDE} \approx 0$ ($\Lambda$CDM) for most of the $\theta_i,\log_{10}(z_c)$ choices. All in all, the mean likelihood 
profile clearly illustrates that the {\it Planck} + LSS 
constraints on EDE are not driven by 
prior volume effects.

%\clearpage
%\pagebreak 
\subsection{EDE away from the $\Lambda$CDM limit: Analysis with a high lower bound on $f_{\rm EDE}$}
\label{app:prior}

\begin{figure}[h!]
\centering
 \includegraphics[width=0.8\textwidth]{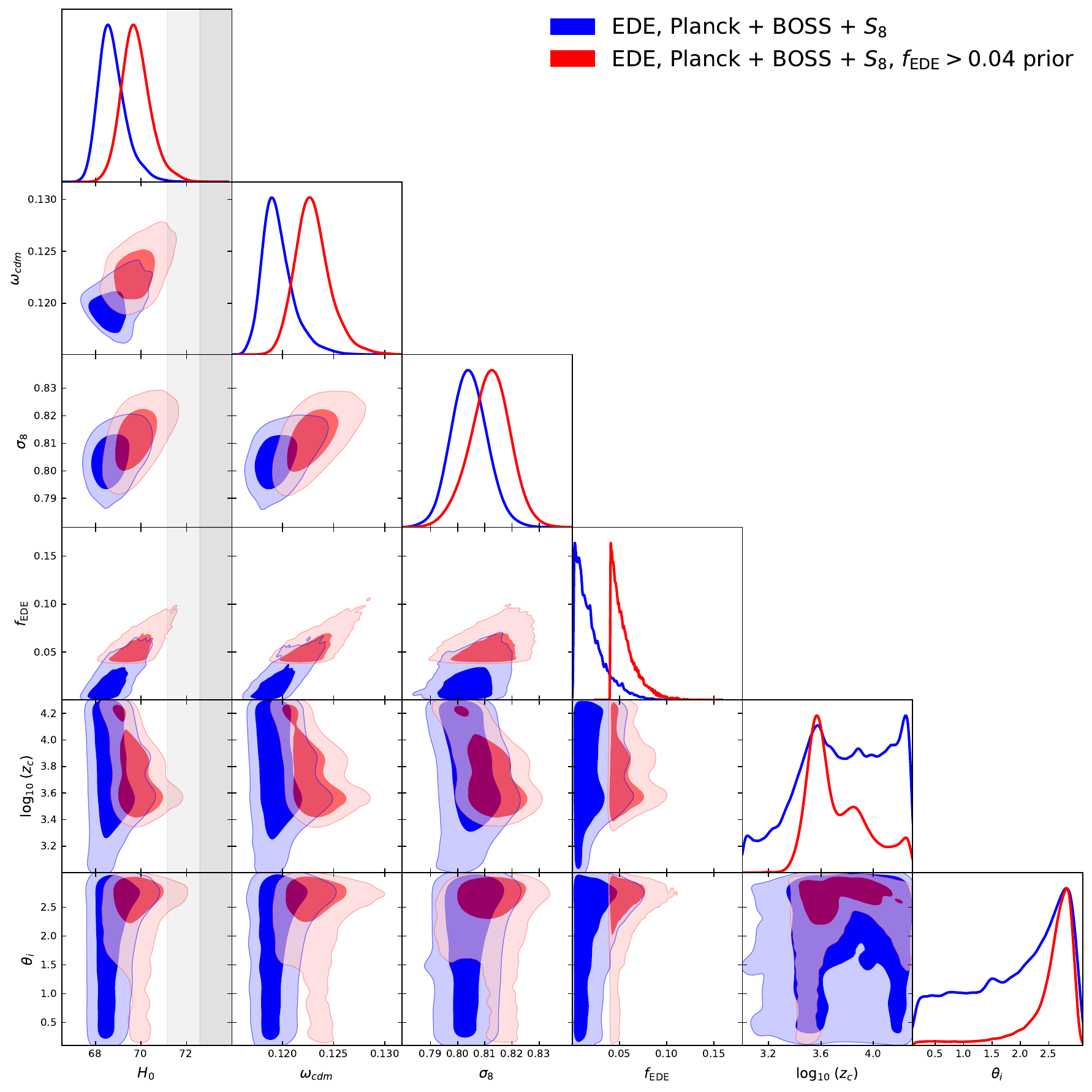}
    \caption{Constraints from \emph{Planck}+BOSS+$S_8$ data on cosmological parameters in EDE for two different choices of prior on $f_{\rm EDE}$: $f_{\rm EDE}>0.04$ (red) and $f_{\rm EDE}>0.001$ (blue). There is substantial support at the lower bound of the prior in the posterior for $f_{\rm EDE}$ in \textit{both} scenarios. Interestingly, this even holds in the $f_{\rm EDE}>0.04$ case which by construction excludes the $\Lambda$CDM limit ($f_{\rm EDE} \approx 0$). This signifies that there is no preference for an EDE component given either choice of prior and precludes `prior volume effects.' Also depicted is the SH0ES measurement of $H_0$ (gray) where the dark- and light-shaded contours represent the $68\%$ and $95\%$ confidence intervals, respectively.
    }
    \label{fig:fEDE-prior}
\end{figure}

To further mitigate the finite prior volume at $f_{\rm EDE}\rightarrow 0$, we repeat our Bayesian analysis of BOSS data in Sec.~\ref{sec:constraintsBOSS} with an artificial lower bound imposed on $f_{\rm EDE}$, given by $f_{\rm EDE}>0.04$. This acts to force the MCMC sampler away from the $\Lambda$CDM limit ($f_{\rm EDE} \rightarrow 0$). The results are shown in Fig.~\ref{fig:fEDE-prior}. This again results in no preference for an EDE component, and the marginalized posterior for $f_{\rm EDE}$ exhibits an exponential fall-off for values above the imposed lower bound $0.04$.

\begin{table}[h!]%[htb!]
% \hspace{2.4cm}
\centering
{EDE away from the $\Lambda$CDM limit Constraints from\\ \emph{Planck} 2018 data  + BOSS DR12  + $S_8$ from DES+KV-450+HSC \vspace{4pt}}
\tbl{The mean (best-fit) $\pm1\sigma$ constraints on the cosmological parameters in the EDE scenario with $n=3$, as inferred from the combination of BOSS FS+BAO, \emph{Planck} 2018 TT+TE+EE + low $\ell$ + lensing, and DES+KV-450+HSC data.  Upper and lower limits are quoted at 95\% CL. We present the results of two analyses differing by a lower prior bound on $f_{\rm EDE}$: baseline physical choice $f_{\rm EDE}>0$ (right column) and artificial unphysical choice $f_{\rm EDE}>0.04$ (left column).}%\vspace{2pt}} \\
  %\centering
  {\begin{tabular}{|l|c|c|}
    \hline\hline Parameter & $f_{\rm EDE}>0.04$  ~~&~~~$f_{\rm EDE}>0$ ~~~\\ \hline \hline
    {\boldmath$\ln(10^{10} A_\mathrm{s})$} & $3.042~(3.027)_{-0.015}^{+0.014}$ & $3.038 \, (3.034) \, \pm 0.014$ \\
    {\boldmath$n_\mathrm{s}$} & $0.9763\,(0.9742)_{-0.0052}^{+0.0061}$ & $0.9696 \, (0.9624)^{+0.0042}_{-0.0051}$ \\
    {\boldmath$100\theta_\mathrm{s}$} & $1.041945 \, (1.041966) \, \pm 0.00030 $ & $1.04178 \, (1.04176) \, \pm 0.00035$\\
    {\boldmath$\Omega_\mathrm{b} h^2$} & $0.02274 \, (0.02278) \,^{+0.00019}_{-0.00017}$ &  $0.02259 \, (0.022433)^{+0.00016}_{-0.00018}$ \\
    {\boldmath$\Omega_\mathrm{cdm} h^2$} & $ 0.1229\,(0.1219)_{-0.002}^{+0.0014}$ & $0.11958 \, (0.11951)^{+0.00096}_{-0.0018}$ \\
    {\boldmath$\tau_\mathrm{reio}$} & $0.05282\,(0.04781)_{-0.0072}^{+0.0074}$ & $0.0535 \, (0.0521)_{-0.0075}^{+0.0069}$\\
    {\boldmath$\mathrm{log}_{10}(z_c)$} & $3.746(3.67)_{-0.28}^{+0.17}$ & $3.77 \, (4.24)^{+0.51}_{-0.33}$ \\
    {\boldmath$f_\mathrm{EDE} $} & $<0.08384\,(0.04078)$ & $< 0.0526 \, (0.0115)$\\
    {\boldmath$\theta_i$} & $2.522(2.505)_{-0.064}^{+0.46}$ & $1.91(1.55)_{-0.47}^{+1.2} $\\
    \hline
    $H_0 \, [\mathrm{km/s/Mpc}]$ & $69.77\,(69.39)_{-0.72}^{+0.55}$ & $68.73 \, (67.92)^{+0.42}_{-0.69}$ \\
    $\Omega_\mathrm{m}$ & $0.3007~(0.3017)
    \pm 0.0052$ & $0.3024 \,(0.3091)\pm 0.0050$ \\
    $\sigma_8$ & $0.8115\,(0.8040)_{-0.0073}^{+0.008}$ & $0.8044 \, (0.8023)_{-0.0069}^{+0.0060}$ \\
    $S_8$ & $0.8126 \, (0.8063) \, \pm 0.0096$ & $0.8075 \, (0.8143)\, \pm 0.0092$ \\
    \hline
  \end{tabular} 
  \label{table:params-fEDE-prior}}
\end{table}

%\clearpage

\subsection{Profile Likelihood}

Finally, we summarize the conclusions of Ref. \cite{Herold:2021ksg,Herold:2022iib}  based on their results for the profile likelihood, and compare to the Bayesian analyses presented here. While the frequentist and Bayesian analyses answer fundamentally different questions, it is nonetheless instructive to perform model comparison independently in the two approaches.

\begin{figure}[h!]
\centering
 \includegraphics[width=0.8\textwidth]{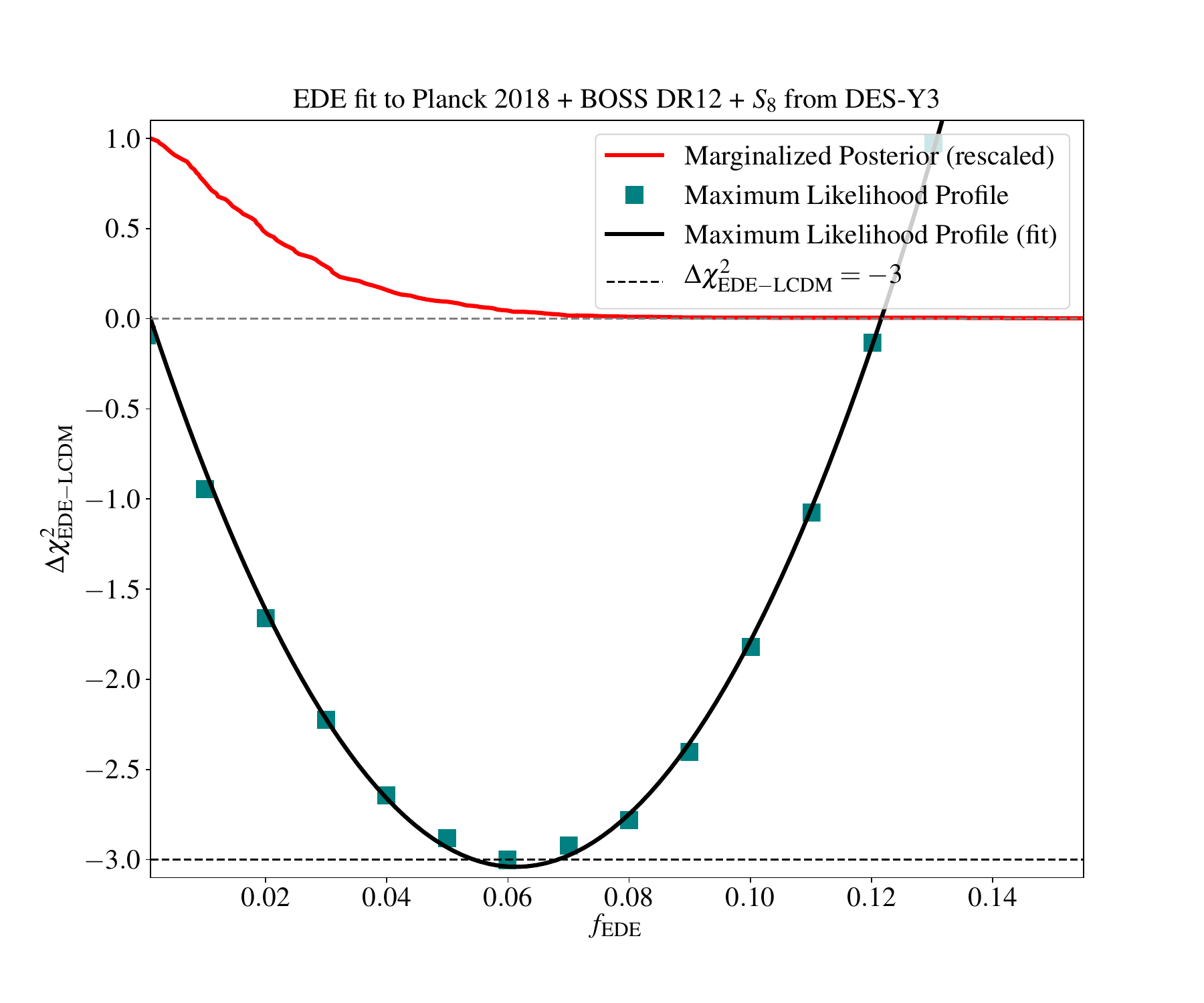}
    \caption{Likelihood profile and marginalized posterior distribution for $f_{\rm EDE}$ in the fit of EDE to {\it Planck} 2018, BOSS DR12 full shape + BAO, and $S_8$ from DES-Y3. 
    }
    \label{fig:profile_likelihood}
\end{figure}

In Fig.~\ref{fig:profile_likelihood} we show the profile likelihood in the fit to {\it Planck} + BOSS (BAO+FS) + $S_8$ from DES-Y3. We highlight the following:
\begin{enumerate}
    \item {\bf Frequentist preference for $\Lambda$CDM}: The best-fit EDE model in the fit to {\it Planck} and BOSS DR12 has $\Delta \chi^2=-5.67$ relative to $\Lambda$CDM, which is decreased to $\Delta \chi^2=-2.93$ when $S_8$ data from DES are included \cite{Herold:2022iib}. This modest improvement in fit comes at the expense of three additional parameters introduced in the EDE model. As a simple frequentist model comparison, one can compute the AIC score \cite{1100705}. One finds an AIC score of $\Delta {\rm AIC} = +0.33$ when $S_8$ data are excluded, which increases to $\Delta{\rm AIC} = +3$ in favor of $\Lambda$CDM once $S_8$ data are included \cite{Herold:2022iib}. This indicates a strong frequentist preference for $\Lambda$CDM.\footnote{Ref.~\cite{Herold:2022iib} uses the full-shape power spectrum at $k_{\rm max}=0.25~h$Mpc$^{-1}$, which is larger than the baseline choice of Ref.~\cite{Philcox:2021kcw}, $k_{\rm max}=0.20~h$Mpc$^{-1}$. 
    The 1-loop EFT model is, strictly
    speaking, not 
    very reliable at $k_{\rm max}=0.25~h$Mpc$^{-1}$, 
    and produces an upward bias
    in the recovered value of $\sigma_8$~\cite{Nishimichi:2020tvu,Chudaykin:2020ghx},
    which
    may lead to an artificially better
    agreement between the data and the 
    EDE model. Thus, the exact value
    of  $\Delta \chi^2$ between $\Lambda$CDM and EDE from \cite{Herold:2022iib} should be viewed with a measure of caution. 
    A more conservative scale cut 
    is expected to produce a smaller
    numerical value of  $|\Delta \chi^2|$ 
    than reported in \cite{Herold:2022iib}, and 
    hence yield an even worse AIC.
    } 
    \item {\bf The marginalized posterior is a monotonically decreasing function:}  This reflects the relative fine-tuning of $\{z_c,\theta_i\}$ in order to maximize the likelihood at $f_{\rm EDE}\sim 0.1$ versus $f_{\rm EDE} \sim 0$. The Bayesian punishment for new finely-tuned parameters, imposed via marginalization, overcomes the modest $\Delta \chi^2$ difference between the $\Lambda$CDM and EDE best-fit minima, leading to the monotonically decreasing behaviour of the marginalized posterior. This indicates a strong Bayesian preference for EDE.
\end{enumerate}
Thus both frequentist and Bayesian model comparison of LSS data lead to an overall preference for $\Lambda$CDM over EDE.

%%%%%%%%%%%%%%%%%%%%%%%%%%%%%%%%%%%%%%%%%%%%%%%%%%%%%%%%%%%%%%%%%%%%%%
%%%%%%%%%%%%%%%%%%%%%%%%%%%%%%%%%%%%%%%%%%%%%%%%%%%%%%%%%%%%%%%%%%%%%%
%\clearpage
\section{Discussion}
\label{sec:Discussion}
%%%%%%%%%%%%%%%%%%%%%%%%%%%%%%%%%%%%%%%%%%%%%%%%%%%%%%%%%%%%%%%%%%%%%%
%%%%%%%%%%%%%%%%%%%%%%%%%%%%%%%%%%%%%%%%%%%%%%%%%%%%%%%%%%%%%%%%%%%%%%

In this work we have performed a comprehensive analysis of the constraints on the canonical Early Dark Energy model from {\it Planck} PR3 and PR4 CMB data and large-scale structure data from a wide array of experiments.  Each of these analyses results in a tight upper bound for the peak EDE energy density fraction, $f_{\rm EDE}$, which is below the benchmark value needed to resolve Hubble tension $f_{\rm EDE} \approx 0.1$. The upper bound is strengthened as more LSS data are added; e.g, the combination of {\it Planck} 2018 (PR3) data, $S_8$ data from DES-Y3 along with BOSS DR12 full shape and BAO data analyzed using the EFT likelihood,  leads to a bound $f_{\rm EDE}< 0.059$ at 95\% C.L.

CMB data play an important role in the model by forcing shifts in the standard $\Lambda$CDM parameters, in particular increases to the dark matter density $\Omega_c h^2$ and spectral index $n_s$, to compensate for the effect of the new EDE component and raised $H_0$. The {\it Planck} data show no preference for an EDE component, but instead yield only an upper bound. We have considered different choices of {\it Planck} likelihood, and found that the \texttt{Plik} PR3 likelihood conventionally used in EDE analyses to date in fact provides the weakest upper bound of all four possibilities considered here, namely, the \texttt{Plik} likelihood for {\it Planck} PR3 data, the \texttt{CamSpec} likelihood for PR3 data and for PR4 data, and the \texttt{HiLLiPoP} likelihood for PR4 data. All likelihoods agree in the non-preference for the EDE component, yielding a set of upper bounds, with {\tt Plik} PR3 being the weakest.

Large-scale structure data break the degeneracies in the fit to the CMB data, leading to significantly tighter constraints on the EDE component. The increases in $\Omega_c h^2$, $n_s$, and $A_s$, that allow EDE to reconcile {\it Planck} and SH0ES, and together drive an increase in the $S_8$ parameter, are reigned in by weak lensing data, including Dark Energy Survey, KiDS-VIKING-450, and HSC. The dark matter density can instead be constrained with anisotropic galaxy clustering data from BOSS DR12, in particular with the full shape and BAO data implemented using an EFT likelihood. Combining {\it Planck} and BOSS DR12 with $S_8$ data results in an upper bound $f_{\rm EDE}< 0.0526$ at 95\% C.L. The model can be further constrained by Lyman-$\alpha$ forest data, which break the $f_{\rm EDE}-n_s$ degeneracy in the fit to CMB data. The combination of {\it Planck}, BAO, and Ly$\alpha$ data leads to upper bounds $f_{\rm EDE}< 0.03$ at 95\% C.L. for eBOSS Ly$\alpha$ data and $f_{\rm EDE}<0.04$ at 95\% C.L. for XQ-100 Ly$\alpha$ data. In all these analyses, the upper bound on $f_{\rm EDE}$ is accompanied by a marginalized constraint on $H_0$ that is in significant tension with the SH0ES measurement.

We have also performed detailed investigations into the impact of the choice of priors on the marginalized parameter constraints in the EDE model, including the imposition of the SH0ES $H_0$ measurement as a prior on the analysis of CMB and LSS data, the choice between theory-motivated and computational-expense-motivated priors, and the impact of a prior $f_{\rm EDE}>0.04$ to restrict the model away from the $\Lambda$CDM limit. All of these analyses lead to a marginalized posterior distribution for $f_{\rm EDE}$ that is a monotonically decreasing function, suggesting a Bayesian preference for $\Lambda$CDM. We compare with a frequentist analysis based on the profile likelihood and find qualitative agreement: the modest $\chi^2$-improvement achieved by EDE is outweighed by the increased number of parameters in EDE (nine parameters versus the six of $\Lambda$CDM), leading to a AIC score comparison that favors $\Lambda$CDM.

Note that in most analyses we did not impose the SH0ES $H_0$ prior. One might argue that 
without this prior,
current CMB+LSS data  
do not have enough precision to detect $f_{\rm EDE}\simeq 0.1$ needed to resolve the Hubble tension. Our results refute such claims 
and suggest that the exclusion 
of $f_{\rm EDE}\simeq 0.1$ with various CMB+LSS measurements
is driven precisely by the statistical power of these data sets. 
We emphasize that combining
LSS and CMB is crucial 
to break degeneracies between the EDE and ``standard'' cosmological parameters
present at the CMB level.

We note there are a significant number of relevant works that have not been discussed or only briefly touched on in this review, in particular, constraints from CMB experiments other than {\it Planck} \cite{Chudaykin:2020acu,Chudaykin:2020igl,Lin:2020jcb,Hill:2021yec,LaPosta:2021pgm,Poulin:2021bjr,Smith:2022hwi,Smith:2023oop}, or extensions to and improvements upon the EDE model, such as those that attempt to address the tension with large-scale structure data \cite{McDonough:2021pdg,Allali:2021azp,Alexander:2022own,Berghaus:2019cls,Berghaus:2022cwf,Clark:2021hlo,Cruz:2023lmn}, let alone proposals other than EDE to resolve the Hubble tension (see, e.g.,~\cite{Abdalla:2022yfr} for 
a recent review).

We close this review on a positive note: The EDE proposal remains an interesting testbed for experimenting with our understanding of the physics implications of data from CMB and LSS experiments. While current data do not exhibit a strong Bayesian preference for the canonical EDE model, it remains possible that EDE or an EDE-like model is the true theory of nature -- this will be tested with future data releases and future experiments, including forthcoming CMB data from the Atacama Cosmology Telescope~\cite{Henderson:2015nzj} and South Pole Telescope~\cite{SPT-3G:2014dbx} and future CMB data from the Simons Observatory \cite{Ade:2018sbj} and CMB-S4~\cite{Abazajian:2019eic}, and LSS experiments such as  \textit{Euclid} \cite{Amendola:2016saw}, DESI \cite{Levi:2019ggs}, WFIRST/Roman \cite{2019arXiv190205569A}, and the Vera Rubin Observatory \cite{Ivezic:2008fe}. Indeed, despite the shortcomings of the canonical EDE model, it is one of the most effective proposals to date that are capable of increasing $H_0$ while simultaneously maintaining a good fit to the CMB. It may well be the case that current or future extensions of EDE are able to avoid, or perhaps overcome, the large deviations from $\Lambda$CDM which seem to preclude the canonical EDE model at present.

\section{Acknowledgments}

EM is supported by a Discovery Grant from the Natural Sciences and Engineering Research Council of Canada. Portions of this work were conducted in MIT’s Center for Theoretical Physics and partially supported by the U.S. Department of Energy under grant Contract Number DE-SC0012567. JCH acknowledges support from NSF grant AST-2108536, NASA grants 21-ATP21-0129 and 22-ADAP22-0145, the Sloan Foundation, and the Simons Foundation.

\bibliography{refs}
\bibliographystyle{ws-ijmpd}

\end{document}